\documentclass[fleqn,usenatbib]{mnras}

\usepackage{newtxtext,newtxmath}
\usepackage[T1]{fontenc}

\DeclareRobustCommand{\VAN}[3]{#2}
\let\VANthebibliography\thebibliography
\def\thebibliography{\DeclareRobustCommand{\VAN}[3]{##3}\VANthebibliography}

\usepackage{graphicx}	% Including figure files
\usepackage{amsmath}	% Advanced maths commands

\usepackage{mdframed}
\usepackage{tikz}
\usetikzlibrary{arrows.meta, positioning}
\usepackage{cleveref}
\usepackage{balance}
\usepackage[normalem]{ulem}
\usepackage{fix-cm}
\usepackage{xcolor}
\definecolor{RED}{named}{red}
\definecolor{BLUE}{named}{blue}
\usepackage{multirow}
\usepackage{orcidlink}

\defcitealias{Elahi2025}{E25}

\title[Bayes-SCF]{Bayes-SCF: A Bayesian filter to mitigate foreground leakage in the 21-cm power spectrum}

\author[K.M.A. Elahi]{Khandakar Md Asif Elahi~\orcidlink{0000-0003-1206-8689}~$^{1,2}$~\thanks{E-mail: asifelahi999@gmail.com}
\\ \\
$^{1}$ Centre for Strings, Gravitation and Cosmology, Department of Physics, Indian Institute of Technology Madras, Chennai 600036, India \\ 
$^{2}$ School of Physical Sciences, National Institute of Science Education and Research, Bhubaneswar 752050, Odisha, India
}

\date{Accepted XXX. Received YYY; in original form ZZZ}

\pubyear{\the\year{}}

\begin{document}
\label{firstpage}
\pagerange{\pageref{firstpage}--\pageref{lastpage}}
\maketitle

\begin{abstract}
Missing channels in radio-interferometric visibility data can introduce systematic artefacts into the estimated 21-cm power spectrum. A common workaround is to first estimate the two-frequency correlation $C(\Delta\nu)$ and then Fourier-transform it to obtain the power spectrum $P(k_\parallel)$. This procedure yields an unbiased estimate when the signal is statistically homogeneous (ergodic) along the line-of-sight, but it fails in the presence of non-ergodic foregrounds. Smooth Component Filtering (SCF) has recently been proposed as a solution to this problem, in which the dominant non-ergodic (spectrally smooth) component is removed prior to estimating $C(\Delta\nu)$. In existing implementations, the smooth component is estimated by convolving the visibilities with a Hann window along the frequency axis. We demonstrate that this Hann-based SCF performs adequately only when foregrounds are extremely spectrally smooth. It breaks down with increased flagging and when foregrounds exhibit spectral structures. We introduce a Bayesian extension, Bayes-SCF, based on Gaussian Process regression, which overcomes these limitations. Bayes-SCF models the smooth component via a covariance function with a fixed correlation length, enabling controlled and data-driven filtering. Using simulated data, we show that Bayes-SCF robustly recovers the input model 21-cm power spectrum in the presence of spectrally unsmooth foregrounds. The filter is demonstrated to work under different flagging patterns, including $80\%$ channels being randomly flagged. Bayes-SCF is also effective in a delay-spectrum approach. The primary trade-off introduced by the Bayesian framework is the increased computational cost; future work will focus on optimizing the algorithm and applying it to real Murchison Widefield Array data. 
\end{abstract}

\begin{keywords}
methods: statistical -- cosmology: observations -- dark ages, reionization, first stars -- techniques: interferometric
\end{keywords}

%%%%%%%%%%%%%%%%%%%%%%%%%%%%%%%%%%%%%%%%%%%%%%%%%%

\section{Introduction}
\label{sec:intro}

The observation of the redshifted 21-cm hyperfine transition of neutral hydrogen (\ion{H}{i}) promises to transform our understanding of the Universe. Several low-frequency radio interferometers, including the Hydrogen Epoch of Reionization Array \citep[HERA;][]{Deboer2017}, the Low Frequency Array \citep[LOFAR;][]{vanharlem2013}, and the Murchison Widefield Array \citep[MWA;][]{Tingay2013}, are currently operating with the goal of detecting the power spectrum of the 21-cm brightness temperature fluctuations from the Epoch of Reionization (EoR). These experiments aim to uncover the properties of the first stars, galaxies, and black holes that transformed the intergalactic medium (IGM) from neutral to ionized \citep[e.g.][]{Madau1997, Bharadwaj2004a, Mondal2017} during redshifts $z \sim 6-15$. Analogous experiments at $z < 6$, such as the Canadian Hydrogen Intensity Mapping Experiment \citep[CHIME;][]{chimelya}, MeerKAT \citep{Carucci2025}, and the Giant Metrewave Radio Telescope \citep[GMRT;][]{Elahi2024}, aim to trace the large-scale structure of the Universe and constrain cosmological parameters. The upcoming SKA Observatory (SKAO) is expected to revolutionize these experiments by providing unprecedented sensitivity and bandwidth.

The EoR 21-cm signal, characterized by brightness temperature fluctuations of order $\sim 10 \ {\rm mK}$ \citep{Madau1997, Zaldarriaga2004, Bharadwaj2004a}, is obscured by astrophysical foregrounds that are brighter by several orders of magnitude \citep{Ali2008, Bernardi2009, Ghosh2012}. The primary discriminant between these components lies in their spectral characteristics. Foregrounds, which primarily consist of Galactic and extragalactic synchrotron radiation, arise from continuum emission and vary `smoothly' with frequency. In contrast, the 21-cm signal is a spectral line; inhomogeneities in the cosmic \ion{H}{i} distribution along the line of sight (LoS) are mapped onto the frequency axis via cosmological redshift, which leads to rapid spectral fluctuations. This distinction in principle allows for a separation in the Fourier domain, where in the cylindrical power spectrum $P(k_\perp, k_\parallel)$ the smooth foregrounds occupy low LoS $(k_{\parallel})$ modes, leaving higher modes accessible for recovering the 21-cm signal \citep{morales2005}. The clean spectral separation is, however, complicated by instrumental chromaticity. The frequency-dependent response of interferometers (often referred to as `baseline migration') couples angular structure into the frequency domain (mode mixing), which causes foreground power to leak into higher $k_{\parallel}$ modes \citep{datta2010, vedantham12, Morales2012}. This results in the measurement of `unsmooth' or spectrally structured foregrounds that contaminate the theoretically clean EoR window \citep{Morales2012, vedantham12, trott1, parsons12, liu14a}.

In low-frequency experiments, it is often necessary to flag specific frequency channels due to Radio Frequency Interference (RFI). Furthermore, instrumental systematics often require flagging specific frequency channels. For instance, the signal processing chain of MWA's Phase II configuration requires a periodic flagging of frequency channels \citep{Prabu2015}. This non-uniform sampling introduces additional spectral structure into the data, further aggravating the challenge of foregrounds. If visibility data are Fourier transformed along the frequency axis (i.e., to the delay space), the spectral discontinuities introduced by flagging result in ringing artefacts, which scatter bright foreground power across the entire $k_{\parallel}$ range. Several algorithms have been proposed to address this, e.g., \cite{Parsons2009} suggested a \textsc{CLEAN}-like nonlinear deconvolution in delay space. \cite{trott16} proposed Least Square Spectral Analysis (LSSA) to handle irregular sampling, a method subsequently used for MWA \citep{Trott2020} and LOFAR \citep{Patil2017, Mertens2020}. More recently, inpainting techniques such as $\mathcal{E}$ppsilon \citep{Barry2019eppsilon} and DAYENU \citep{ewallwice2021} have been developed to fill missing data prior to Fourier transformation \citep[see also][]{Chen2025}. Gaussian Process Regression \citep[GPR;][]{mertens18, Kern2021} and Gaussian Constrained Realizations \citep[GCR;][]{Kennedy2023} have also been employed to reconstruct the signal in the presence of missing channels.

The problem of missing channels can be circumvented by correlating visibilities in the frequency domain \citep{Bharadwaj2001a, Bharadwaj2005} to estimate the multi-frequency angular power spectrum (MAPS) $C_\ell(\Delta\nu)$, before performing a Fourier transform along $\Delta\nu$ to estimate the PS $P(k_\perp,k_\parallel)$. The key advantage of this approach is that, even if specific frequency channels $\nu$ are flagged, the estimated correlation function $C_\ell(\Delta\nu)$ remains continuous in $\Delta\nu$. Consequently, correlation-based estimators, such as the Tapered Gridded Estimator (TGE; \citealt{Bharadwaj2018}), ideally avoid the ringing artifacts caused by spectral discontinuities. \cite{Bharadwaj2018} demonstrated via simulations that the TGE can accurately recover the PS even when $80\%$ channels are randomly flagged from the data, and the estimator has been successfully applied to observational data \citep{Pal2020, Pal2022, Elahi2023, Elahi2023b, Elahi2024}.

However, the robustness of correlation-based estimators relies on the assumption of statistical homogeneity of the signal along the frequency axis. This implies that the statistics of the signal should be translationally invariant, i.e., it should depend only on the frequency separation $\Delta\nu$. Astrophysical foregrounds, however, violate this assumption, as they evolve significantly across the bandwidth (e.g., following a power-law), which renders the signal non-stationary. Additionally, mode mixing due to the instrument further introduces additional non-stationary components in the measured visibility data. As a consequence, simply correlating visibilities and averaging over these non-stationary features leads to residual contamination in the estimated power spectrum (\citealt{Elahi2025}; henceforth \citetalias{Elahi2025}). To address this, \citetalias{Elahi2025} introduced Smooth Component Filtering (SCF), a technique designed to isolate and subtract the dominant, spectrally smooth model from the visibility data prior to power spectrum estimation. SCF has been applied to MWA drift scan data to place upper limits on the EoR ($z = 8.2$) 21-cm power spectrum (\citetalias{Elahi2025}, \citealt{Sarkar2026}), and the bispectrum \citep{Gill2026}. 

While \citetalias{Elahi2025} demonstrated the efficacy of SCF using a fixed Hann window, this linear filtering approach has several inherent limitations. Simulations indicate that when foregrounds exhibit complex spectral structure (e.g., at large $k_\perp$), a fixed window fails to adequately capture the smooth component, which leads to foreground leakage. Such leakage persists even with more aggressive filtering using a window function that has a narrower width. Additionally, the window's finite support requires discarding data at the band edges, where the convolution is ill-defined, thereby reducing the effective bandwidth and sensitivity. Finally, as we demonstrate in this work, fixed-window functions inevitably leak foreground power when strong spectral features are present in the data. To eliminate these shortcomings, we propose an improved strategy that integrates Bayesian probabilistic estimation of the smooth component using Gaussian Process (GP) regression \citep{WR1996, Neal1997, williams1998, Rasmussen2006}. This Bayesian SCF (hereafter, Bayes-SCF) exploits the covariance structure of the data to model the smooth component in a non-parametric and controlled manner. This approach effectively combines the gap-robustness of a correlation-based estimator with the flexibility of Bayesian probabilistic modelling.

The paper is organized as follows: Section~\ref{sec:simulations} describes the simulations of the EoR 21-cm signal and foregrounds, which are the data used in this work. Section~\ref{sec:flagging} provides a description of the different flagging configurations we consider. Section~\ref{sec:pk_from_corr} outlines the methods used for estimating the 21-cm power spectrum in the presence of missing channels. Section~\ref{sec:scf} presents the details of the Hann-based and the Bayesian SCF algorithms and their comparison. We provide a summary of this paper in Section~\ref{sec:conclusions}. Some additional details are deferred to appendices in order to maintain the flow of the paper. 

The values of the cosmological parameters used throughout this work are taken from \citet{Planck2020f}. 

\section{Simulated data}
\label{sec:simulations}

The measured data in 21-cm experiments can be conceptually decomposed into three primary components: the desired cosmological 21-cm signal, bright astrophysical foregrounds, and instrumental noise. Compared to the 21-cm signal, the foregrounds exhibit substantially higher  amplitudes, but are expected to be much smoother in frequency. However, instrumental chromaticity often introduces additional spectral structures into the measured data, thereby disrupting the spectral smoothness. The missing frequency channels further result in additional spectral features in the data. In this work, we focus on the systematic impact of missing frequency channels on estimates of the 21-cm power spectrum. Therefore, we explicitly restrict our analysis to the one-dimensional frequency-dependent behaviour in the data. This reduction in dimensionality also allows us to avoid the complexity of full interferometric simulations and their analysis, which does not provide any additional benefits. We further assume a high-sensitivity observational regime in which instrumental noise is subdominant and focus strictly on the signal and foreground components for the remainder of this analysis.

For the simulated data described in the sections below, we adopt the frequency configuration consistent with the Murchison Widefield Array (MWA). We consider the frequency range $138.9-169.6$~MHz, divided into $N_c = 768$ channels with a width of $\Delta\nu_c = 40$~kHz, thereby yielding a total bandwidth of $B = 30.72$~MHz. The band is centred at $154.2$~MHz, corresponding to the 21-cm signal redshifted from $z = 8.2$. At this redshift, the comoving distance is $r = 9209.7$~Mpc, with its frequency derivative $r' = \mathrm{d}r/\mathrm{d}\nu = 16.99$~Mpc~MHz$^{-1}$. The simulation spans a comoving length of $L \equiv r' B = 522$~Mpc along the line-of-sight (LoS), with a resolution of $r' \Delta\nu_c = 0.68\,{\rm Mpc}$. The LoS wavenumbers probed by this bandwidth cover the range $0 \leq k_\parallel \leq 4.62~\mathrm{Mpc}^{-1}$. We  note that the frequency configuration has been taken to mimic that of \citetalias{Elahi2025}; however, the methodologies developed in this work are equally applicable for a shorter bandwidth of 15~MHz adopted in \cite{Nunhokee2025}.

\subsection{EoR 21-cm signal}
\label{sec:eor_signal}

We first describe the simulation of $T(\nu)$, the LoS fluctuations of the cosmological 21-cm brightness temperature. We assume $T(\nu)$ as a Gaussian random field that follows the $z \approx 8$ EoR 21-cm power spectrum prescribed in \citet{Mondal2017}. Since the frequency structure we consider is sensitive to discrete $k_\parallel$ modes  $k_{\parallel m} = \frac{2\pi m}{L}$, we first interpolate the theoretical power spectrum at these wavenumbers, which serves as the input EoR model $P^{\mathrm{M}}(k_{\parallel m})$. To generate a real-valued realization consistent with this model, we construct complex Fourier coefficients $\tilde{T}_m$ in the delay domain, such that their variance corresponds to the power in the $m$-th delay mode. Explicitly, the coefficients are given by,
\begin{equation}
    \tilde{T}_m =
    \begin{cases}
        0, & m = 0, \\
        \sqrt{\frac{\mathcal{V}_m}{2}}\, (u_m + i v_m), & 1 \le m < N_c/2, \\
        \sqrt{\mathcal{V}_m}\, u_m, & m = N_c/2, \\
        \tilde{T}^{*}_{N_c-m}, & N_c/2 < m \le N_c - 1,
    \end{cases}
\end{equation}
where $u_m$ and $v_m$ are independent random variables drawn from a Gaussian distribution with a mean of 0 and a standard deviation of 1, and $N_c$ is the number of frequency channels. The variance $\mathcal{V}_m$ for each mode is related to the physical power spectrum $P^{\mathrm{M}}$ (in units of $\text{mK}^2\,\text{Mpc}$) by:
\begin{equation}
    \mathcal{V}_m = \frac{N_c^2}{L} P^{\mathrm{M}}(k_{\parallel m}) \,,
\end{equation}
where $L$ is the comoving length in Mpc. For the general modes ($1 \le m < N_c/2$), the power is split equally between the real and imaginary parts. However, for the Nyquist mode ($m=N_c/2$), the coefficient is strictly real-valued, and thus draws the full variance $\mathcal{V}_{N_c/2}$. We set the $m=0$ (DC) mode to zero to obtain a field with zero mean. The definition for $m > N_c/2$ imposes Hermitian symmetry to ensure that the resulting field is purely real-valued.

The simulated 21-cm brightness temperature $T(\nu_n)$ (in mK) for the $n$-th frequency channel is then obtained by performing the inverse discrete Fourier transform,
\begin{equation} 
    T(\nu_n) = \frac{1}{N_c} \sum_{m=0}^{N_c-1} \tilde{T}_m \, \exp\left( i \frac{2\pi m n}{N_c} \right) \,.
    \label{eq:delay_discrete} 
\end{equation}
This construction yields a real-valued stochastic field whose ensemble-averaged power spectrum reproduces the input EoR model. We generate an ensemble of 100 independent realizations using different sets of random variables.

To validate our simulation pipeline, we estimate the power spectrum from the generated ensemble of brightness temperature fields. For each realization $i$, the power spectrum is computed as:
\begin{equation}
    \hat{P}_i(k_{\parallel m}) = \frac{L}{N_c^2} \left| \sum_{n=0}^{N_c-1} T_i(\nu_n) e^{-i \frac{2\pi m n}{N_c}} \right|^2 \,.
    \label{eq:delay_spectrum}
\end{equation}
We then compute the ensemble average $\langle \hat{P}(k_{\parallel}) \rangle$ over 100 realizations. As shown later in the text, the recovered power spectrum matches the input theoretical model $P^{\mathrm{M}}(k_{\parallel})$ within the statistical sampling variance.

\subsection{Foregrounds}
\label{sec:gp_foregrounds}

In contrast to the rapidly fluctuating cosmological 21-cm signal, astrophysical foregrounds are characterised by much larger amplitudes and long-range spectral coherence. We model this behaviour using a Radial Basis Function (RBF) covariance matrix: 
\begin{equation}
    K(\nu,\nu') = \sigma_f^2 \exp \left[-\frac{(\nu-\nu')^2}{2(N_{\rm corr}\,\Delta\nu_c)^2}\right],
    \label{eq:rbf_kernel}
\end{equation}
where $\sigma_f^2$ sets the overall variance, and the product $[N_{\rm corr}\,\Delta\nu_c]$ defines the characteristic spectral correlation length. The dimensionless parameter $N_{\rm corr}$ specifies the effective number of frequency channels over which the signal remains significantly correlated. Large values of $N_{\rm corr}$ are associated with foreground components that are spectrally smooth, whereas smaller values give rise to signals exhibiting pronounced spectral structure. By applying a Cholesky decomposition to the covariance matrix, we generate random realizations of the foreground signals that conform to a prescribed correlation length.   

We simulate two distinct foreground (FG) scenarios typically encountered in 21-cm experiments:
\begin{enumerate}
    \item[I.] Smooth FG ($N_{\rm corr}=10^4$): We model the intrinsically smooth foreground signal using a correlation length that significantly exceeds the observing bandwidth ($N_{\rm corr}\,\Delta\nu_c \gg B$). This produces signals that are highly correlated across the entire band, essentially capturing the generic spectral coherence of power-law synchrotron emission.

    \item[II.] Unsmooth FG ($N_{\rm corr}=10^2$): We simulate a foreground component with a relatively shorter spectral coherence ($N_{\rm corr}\,\Delta\nu_c \simeq 4\,\mathrm{MHz}$). This mimics the behaviour expected in interferometric data at long baselines, where baseline migration couples angular structure into the frequency domain -- a process that shortens the effective correlation length and eventually manifests as the foreground `wedge' in the cylindrical power spectrum \citep[see e.g.,][for a detailed discussion on the correspondence of correlation length and the foreground wedge]{Pal2022}.
\end{enumerate}

For each of these scenarios, we set the overall variance $\sigma_f^2 = 10^{12} \, {\rm mK}^2$, which corresponds to a power spectrum amplitude in the range $P(k_\parallel \approx 0) \sim 10^{14}\text{--}10^{16} \, {\rm mK}^2 \, {\rm Mpc}$, typical of foreground levels encountered in EoR experiments (see e.g., \citetalias{Elahi2025}). We generate an ensemble of 100 independent realizations for each configuration. We have used the \textsc{George} software package \citep{george} to generate these realizations.

Simulating foregrounds directly from their covariance models provides the flexibility to generate and test a wide variety of foreground scenarios. This framework also allows us to efficiently explore diverse spectral signatures, such as complex correlation structures arising from bandpass calibration errors \citep[see e.g.,][]{gayen2025}. We therefore avoid full interferometric simulations and their subsequent analysis in this work, which would provide no immediate benefit for our specific investigation into the effects of the missing frequency channels. We defer more detailed simulations and subsequent analyses on actual data to future work.

\subsection{Correlation structure of the simulated components}
\label{sec:sim_stat_props}

\begin{figure*}
    \centering
    \includegraphics[width=0.99\textwidth]{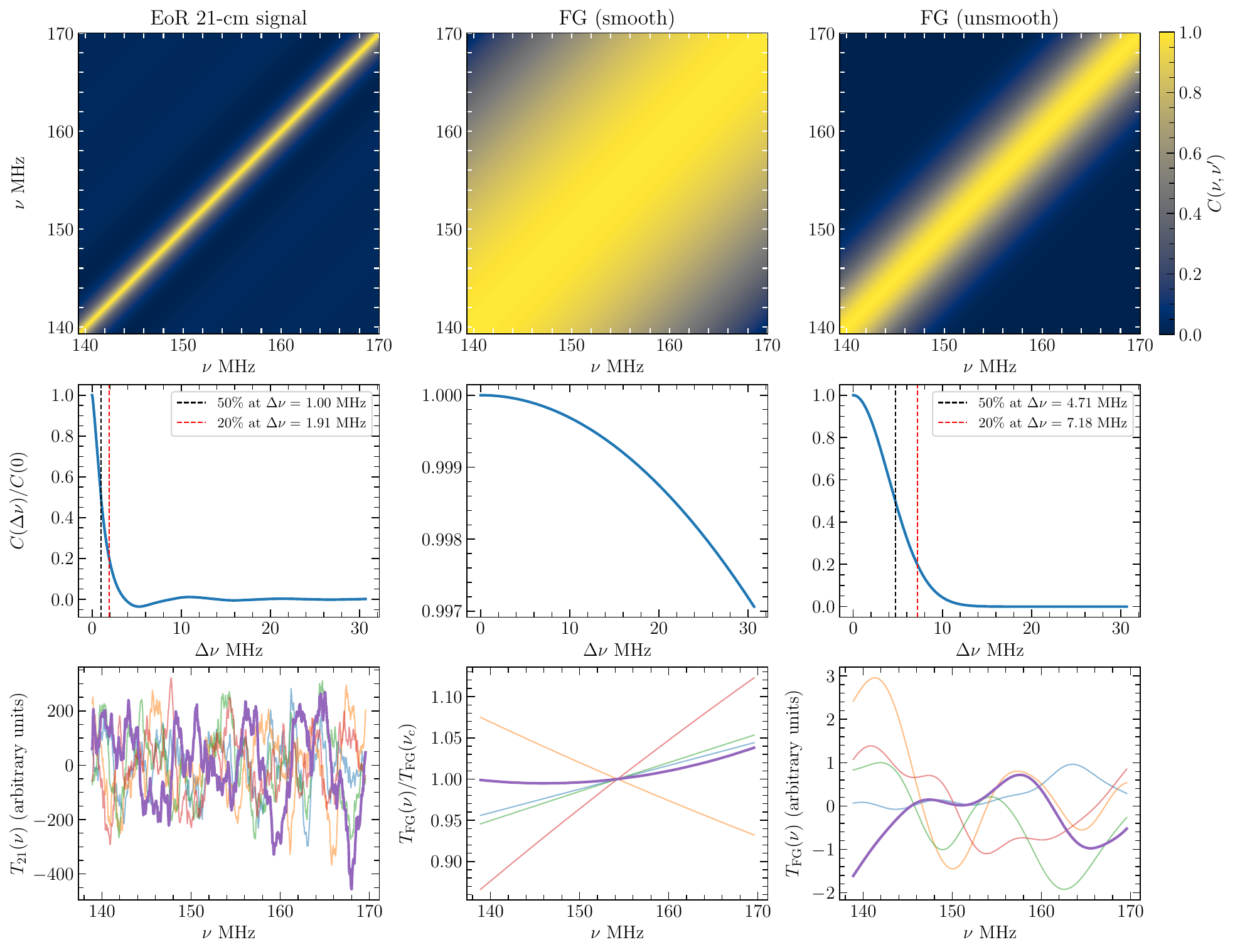}
    \caption{A comparison of the statistical properties of the three simulated components. The top row shows the frequency covariance matrices $\mathbf{C}$, normalised by their diagonal elements to highlight the correlation structure. The middle row plots the corresponding normalised correlation $C(\Delta\nu)/C(0)$. The bottom row displays a few representative signal realizations. The EoR signal (left) is characterised by short-scale fluctuations and rapid decorrelation. Smooth foregrounds (middle) exhibit correlated structure across the entire band. Unsmooth foregrounds (right) show intermediate correlation with significant off-diagonal structure, mimicking instrumental mode-mixing effects.}      
    \label{fig:fg_eor_cov_realizations}
\end{figure*}

The distinct components of the simulated data exhibit different correlation structures, which we quantify using the frequency covariance matrix $\mathbf{C}$. For the EoR 21-cm signal, we determine the covariance function $C(\Delta\nu)$ using,
\begin{align}
    C(\Delta\nu) =  \frac{1}{\pi}\int_{0}^{\infty} \mathrm{d}k_{\parallel}\, \cos\!\left(k_{\parallel} r^{\prime}\Delta\nu\right)\, P^{\rm M}(k_{\parallel}) \,,
\label{eq:cl_Pk_sinc}
\end{align}
which is the 1D LoS limit of the 3D multi-frequency angular power spectrum (MAPS) $C_\ell(\Delta\nu)$ formalism presented in \citet{Bharadwaj2005}. We assume statistical homogeneity along the LoS, which implies that the covariance depends only on the spectral separation.  In other words, the elements of the EoR covariance matrix are given by $\mathbf{C}_{ij} = C(|\nu_i - \nu_j|)$. Considering the foregrounds, the covariance function is already fully defined by the RBF kernel (equation~\ref{eq:rbf_kernel}), and in this case, the matrix elements are $\mathbf{C}_{ij} = K(\nu_i, \nu_j)$.

Figure \ref{fig:fg_eor_cov_realizations} compares the statistical behaviour of the simulated 21-cm signal and the foreground components. The top row displays the covariance matrices $\mathbf{C}$, which are normalized by their diagonals for a visual comparison of the structures independent of amplitude. The EoR signal is strongly diagonal, which indicates a short correlation length. In contrast, the smooth foregrounds exhibit near-constant covariance across the entire band. The unsmooth foregrounds show intermediate behaviour, with correlations extending significantly off-diagonal. 

To quantify the differences between the three types of covariances considered here, we use the stationary, normalised correlation coefficient, defined as the ratio $R \equiv C(\Delta\nu)/C(0)$. This is shown in the middle row. We define the characteristic correlation length $[\Delta\nu]_{\rm corr}$ as the separation at which this ratio drops to $R = 0.5$, and consider the correlation to be negligible when $R\leq0.2$. The EoR signal decorrelates rapidly, with a characteristic correlation length of $[\Delta\nu]_{\rm corr} = 1.00 \,{\rm MHz}$, and becomes negligible at $\Delta\nu \gtrsim 1.91\,{\rm MHz}$. Smooth foregrounds, at the other extreme, exhibit long-range spectral coherence ($[\Delta\nu]_{\rm corr} \gg B$), with the correlation coefficient remaining $>0.99$ (i.e., decorrelating by less than 1\%) across the full 30.72 MHz bandwidth. The unsmooth foregrounds have a reduced correlation length ($[\Delta\nu]_{\rm corr} = 4.71 \,{\rm MHz}$), yet significant correlation ($>0.2$) persists out to separations of $\Delta\nu = 7.18 \,{\rm MHz}$. 

The bottom row of Figure \ref{fig:fg_eor_cov_realizations} shows representative realizations of the three components of the simulated data. The EoR signal appears as rapid stochastic fluctuations, which reflect its short correlation length. The smooth foregrounds have a power-law-like, slowly varying trend. The unsmooth foregrounds exhibit additional oscillatory structures consistent with the spectral features often observed in interferometric data (see e.g., \citetalias{Elahi2025}, Figure~9).

\begin{figure*}
    \centering
    \includegraphics[width=\textwidth]{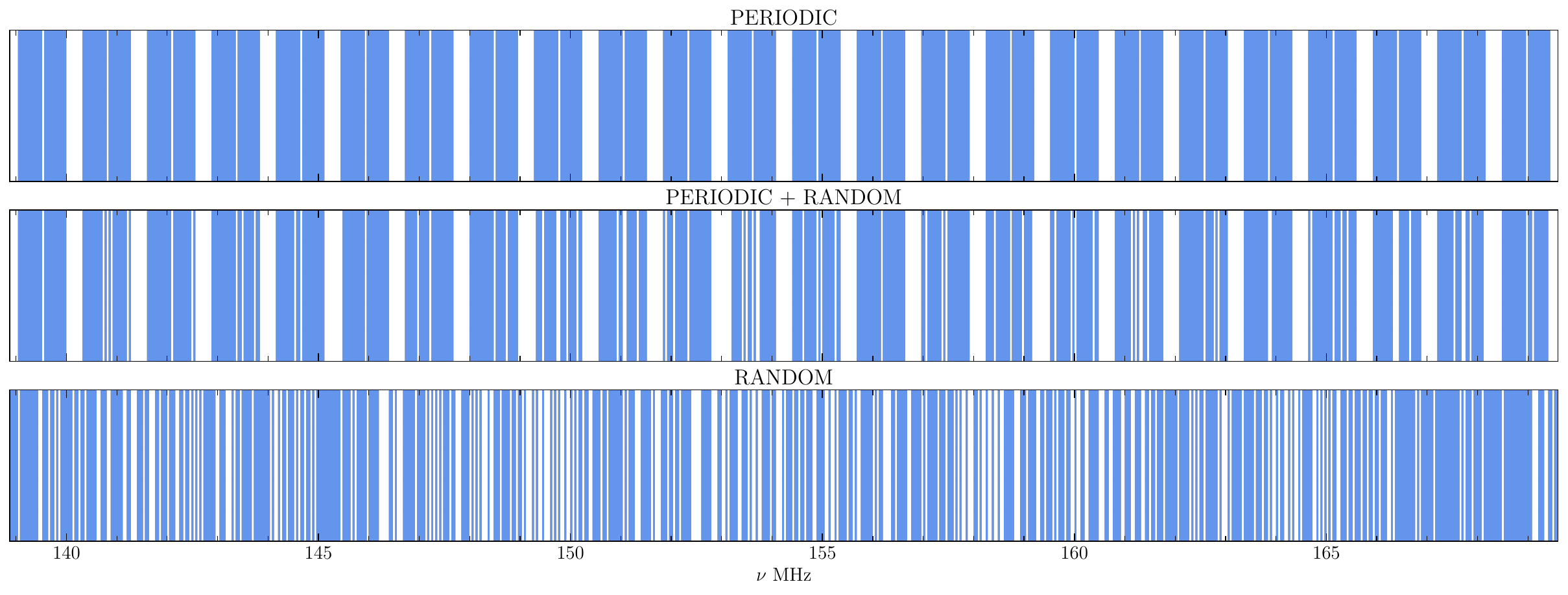}
    \caption{
    Representative patterns of missing frequency channels considered in this work.
    Unflagged channels are shown as blue-filled regions, while flagged channels appear as gaps.
    From top to bottom, the panels show:
    PERIODIC flagging, motivated by the coarse sub-band structure of the MWA;
    PERIODIC+RANDOM flagging, which combines periodic flagging with additional randomly flagged channels;
    and RANDOM flagging, in which a similar number of channels are randomly flagged without imposing any periodic structure.
    The NOFLAG case, in which no channels are flagged, is not shown.
    }
    \label{fig:flagging}
\end{figure*}

\section{Missing frequency channels}
\label{sec:flagging}

Flagging of frequency channels in visibility data is an unavoidable aspect of low-frequency radio interferometric observations. Such data loss arises from a variety of sources, including RFI, instrumental artefacts, and the necessary masking of bandpass edges.

MWA, in particular, employs a polyphase filter bank (PFB) that divides the observed band into equal-width coarse sub-bands of 1.28 MHz \citep{Prabu2015}. The spectral response of these coarse channels rolls off significantly at the edges, and the data there are often contaminated by aliasing effects. Consequently, these edge channels are routinely flagged during data processing, which creates a characteristic pattern of periodically missing data, as illustrated in Figure~\ref{fig:flagging}. These periodic gaps pose a serious challenge for 21-cm power spectrum estimation. As demonstrated in \citetalias{Elahi2025}, the resulting window function couples power across $k$-modes and distorts the frequency correlation structure. While correlation-based estimators, such as TGE, are generally more robust to gaps than direct Fourier transform-based methods, the interplay between missing channels and bright foregrounds can still introduce significant spectral artefacts if not properly mitigated.

To quantify the impact of missing frequency channels on the reconstruction of the 21-cm power spectrum and to evaluate our proposed mitigation strategies, we consider a set of representative flagging configurations, described below and shown schematically in Figure~\ref{fig:flagging}.

\begin{enumerate}
    \item[I.] NOFLAG: An idealised, optimistic scenario in which no frequency channels are flagged (not shown in the Figure). 

    \item[II.] PERIODIC: A deterministic flagging pattern motivated by the MWA coarse-channel structure. In this configuration, we flag the central DC channel and the four edge channels of each 1.28~MHz sub-band (totalling 9 flagged channels per sub-band). This removes approximately $28\%$ of the available bandwidth and introduces a strong periodic structure in the frequency domain.

    \item[III.] PERIODIC+RANDOM: A hybrid configuration combining the periodic flagging pattern described above with an additional stochastic component. We randomly flag $10\%$ additional channels on top of the periodic mask, yielding a total flagged fraction of $\simeq 35\%$. This scenario mimics realistic MWA observations in which RFI affects random channels, in addition to the systematic instrumental flagging.

    \item[IV.] RANDOM: A purely stochastic flagging configuration in which $35\%$ of the frequency channels are flagged randomly, with no imposed periodicity. This serves as a general case relevant for most interferometric observations.  
\end{enumerate}

In the following sections, we examine how each of these patterns of the missing frequency channels affects power spectrum estimation. We additionally test substantially higher RANDOM flagging fractions (up to $80\%$), as well as a single, wide contiguous gap that may arise from a broadband RFI event, later in this work.

\section{Power spectrum estimation}
\label{sec:pk_from_corr}

To estimate the 21-cm power spectrum, we first estimate the frequency covariance function $C(\Delta\nu)$ as
\begin{equation}
    C(\Delta\nu) = \langle T(\nu)\,T(\nu') \rangle ,
    \label{eq:corr_est}
\end{equation}
where we assume statistical homogeneity such that the covariance depends only on the separation $\Delta\nu = |\nu - \nu'|$. Here, the ensemble average is computed over the different realizations of the brightness temperature field. The corresponding 1D LoS power spectrum is then obtained via the cosine transform,
\begin{equation}
    P(k_\parallel) = 2\,r' \int_{0}^{\infty} \mathrm{d}(\Delta\nu)\, \cos\!\left(k_\parallel r'\Delta\nu\right)\, C(\Delta\nu)\,.
    \label{eq:dct}
\end{equation}
This approach has been developed for the 21-cm power spectrum in a series of papers (e.g., \citealt{Bharadwaj2001b, Bharadwaj2005, Datta2007, choudhuri2016b, Bharadwaj2018}), and we have adapted it here for our 1D LoS-only analysis. We refer to this estimator as $\mathcal{F}[C(\Delta\nu)]$, the Fourier transform of the correlation, or simply as the correlation-based estimator. 

\begin{figure}
    \centering
    \includegraphics[width=0.99\columnwidth]{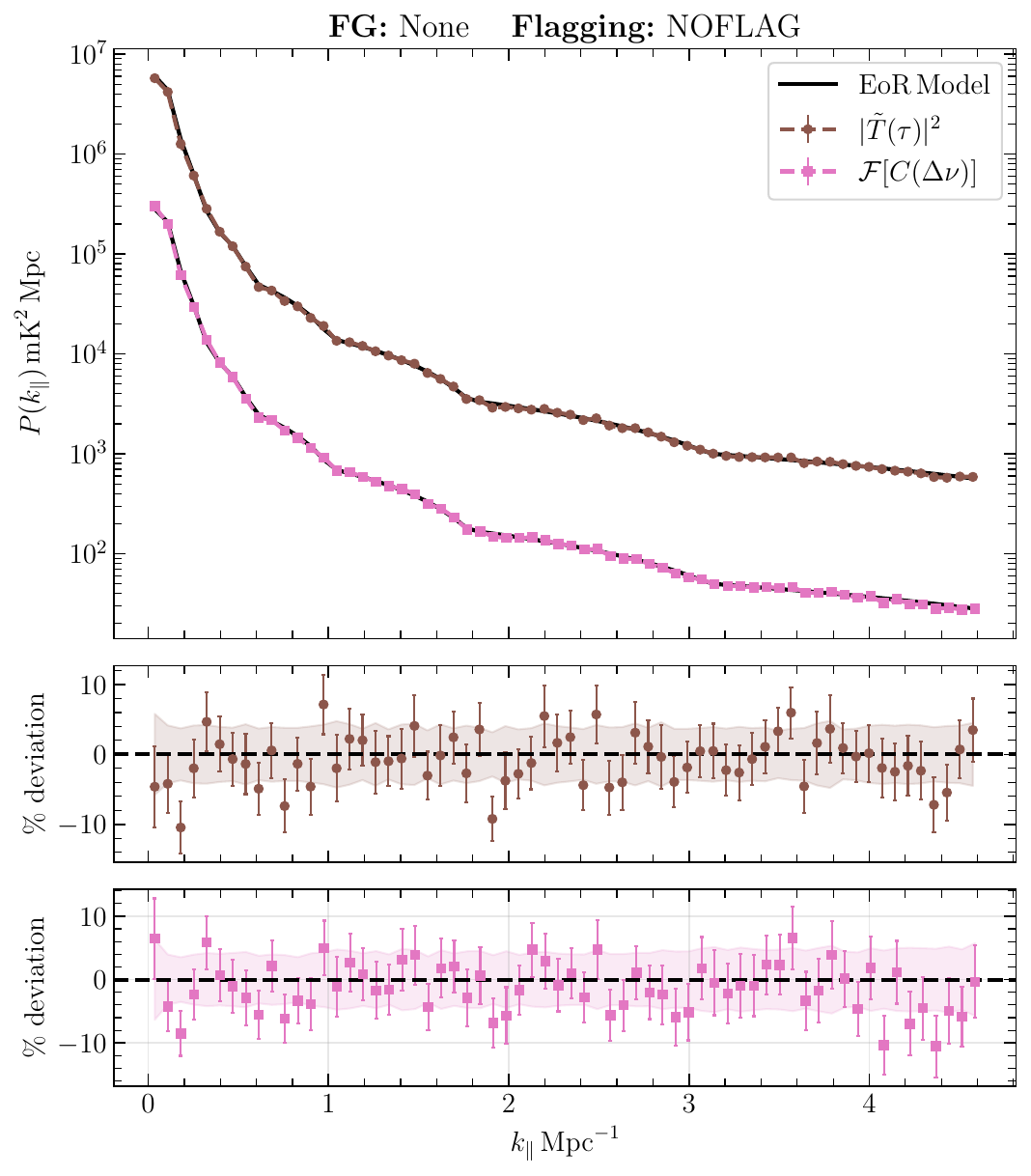}
    \caption{Comparison of the recovered 1D LoS power spectrum using the FFT-squared $|\tilde{T}(\tau)|^2$ (brown circles) and correlation-based $\mathcal{F}\left[ C{\Delta\nu} \right]$ (pink squares) estimators. The solid black line represents the theoretical input model EoR power spectrum, vertically offset for visual clarity.  
    The lower panel shows the percentage deviation relative to the model. All error bars are $1\sigma$ uncertainties. In the absence of flagging, the two estimators yield consistent results, in accordance with the Wiener–Khinchin theorem.}
    \label{fig:dctfft_Flag_NOFLAG}
\end{figure}

Another approach to obtain the 21-cm power spectrum is to perform a Fourier transform along the frequency axis to obtain the delay spectrum \citep{morales2005}, and then construct the power spectrum by squaring their amplitudes (equation~\ref{eq:delay_spectrum}). We refer to this approach as $|\tilde{T}(\tau)|^2$, the FFT-squared (or delay spectrum) estimator. These two approaches are mathematically equivalent when the underlying signal is wide-sense stationary, and the frequency sampling is complete (i.e., no missing channels). 
\Cref{fig:dctfft_Flag_NOFLAG} presents a numerical validation of the equivalence between these two approaches in the idealised, NOFLAG scenario. The upper panel displays the recovered  1D LoS power spectrum $P(k_\parallel)$ from the two estimators, compared with the input EoR power spectrum model $P^M(k_\parallel)$. Note that $P^M(k_\parallel)$ is vertically shifted in the plot for visual clarity. Across the entire $k_\parallel$ range, the two estimators yield consistent power spectra.  The lower panels show the fractional deviation of each estimator from the input model, which demonstrates consistency within the quoted $1\sigma$ uncertainties. This agreement verifies that, in the absence of data flagging or instrumental spectral distortions, the delay-spectrum and correlation-based estimators are mathematically equivalent, in agreement with the Wiener–Khinchin theorem. The mathematical equivalence of these approaches is presented in Appendix~\ref{app:wiener_khinchin}.

\begin{figure}
    \centering
    \includegraphics[width=0.99\columnwidth]{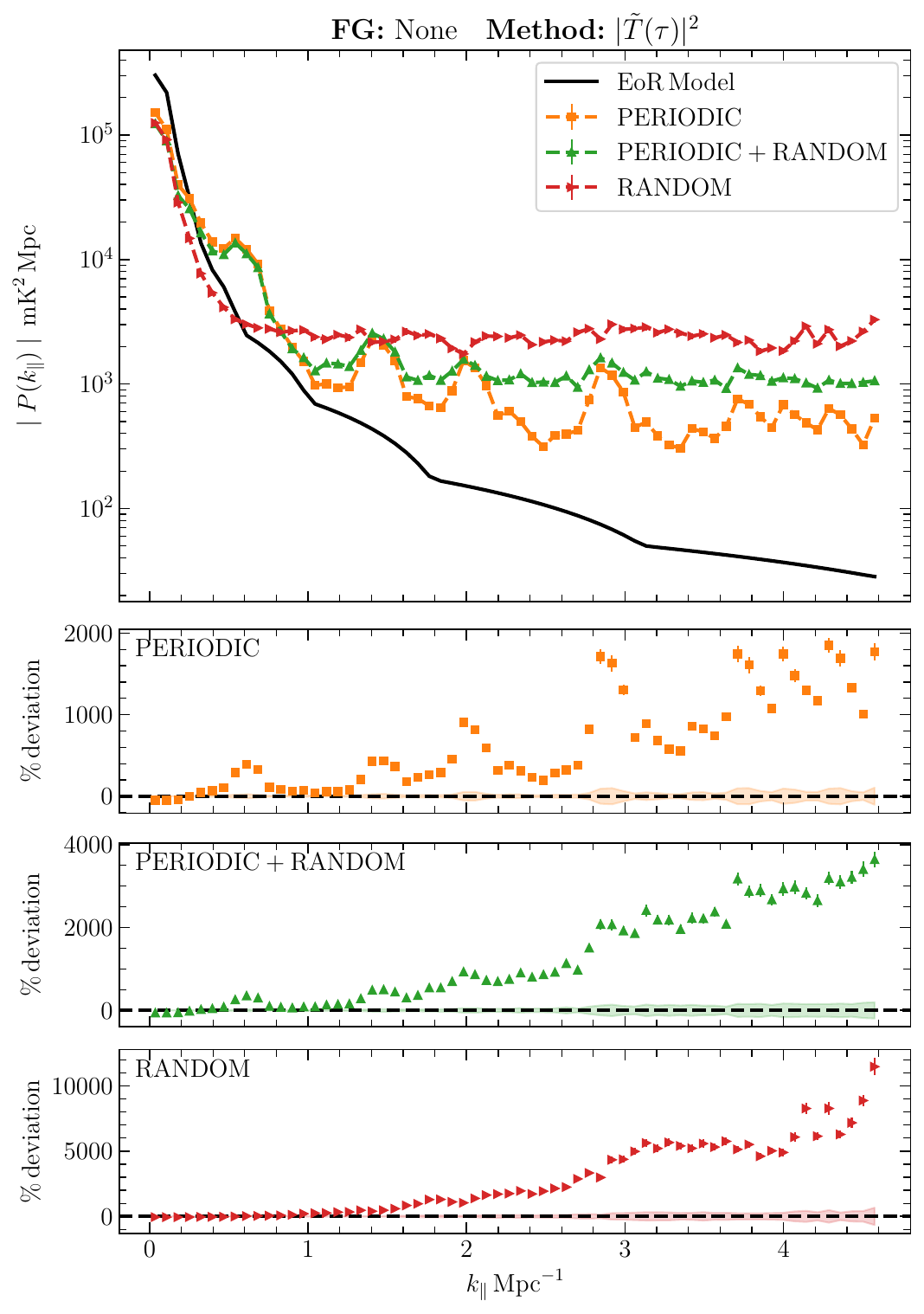}
    \caption{
    Recovered 1D LoS power spectrum using the FFT-squared $|\tilde{T}(\tau)|^2$ estimator for different flagging scenarios. The solid black line in the top panel, shows the input theoretical EoR power spectrum. The lower panel shows the percentage deviation of the recovered power spectrum from the model. Missing frequency channels introduce spectral discontinuities that lead to strong leakage across $k_\parallel$ modes.}
    \label{fig:fft_allflags}
\end{figure}

The equivalence breaks down in the presence of uneven frequency sampling due to flagged channels, which introduce non-stationary spectral structure. \Cref{fig:fft_allflags} presents the recovered power spectrum obtained with the FFT-squared estimator $|\tilde{T}(\tau)|^2$ after including different flagging patterns in the data. When frequency channels are missing, the FFT implicitly assigns zero values to the flagged samples, thereby introducing sharp discontinuities along the frequency axis. These discontinuities induce pronounced spectral leakage and redistribute power over a broad range of $k_\parallel$ modes. As a result, the FFT-squared estimator fails to recover the input EoR power spectrum. It exhibits significant biases and excess power across essentially the entire $k_\parallel$ domain. As discussed previously, several classes of methods can, in principle, correct this problem, including inpainting \citep{ewallwice2021}, deconvolution \citep{parsons12}, and Fourier transforms tailored to irregular sampling \citep{trott16}.

\begin{figure}
    \centering
    \includegraphics[width=0.99\columnwidth]{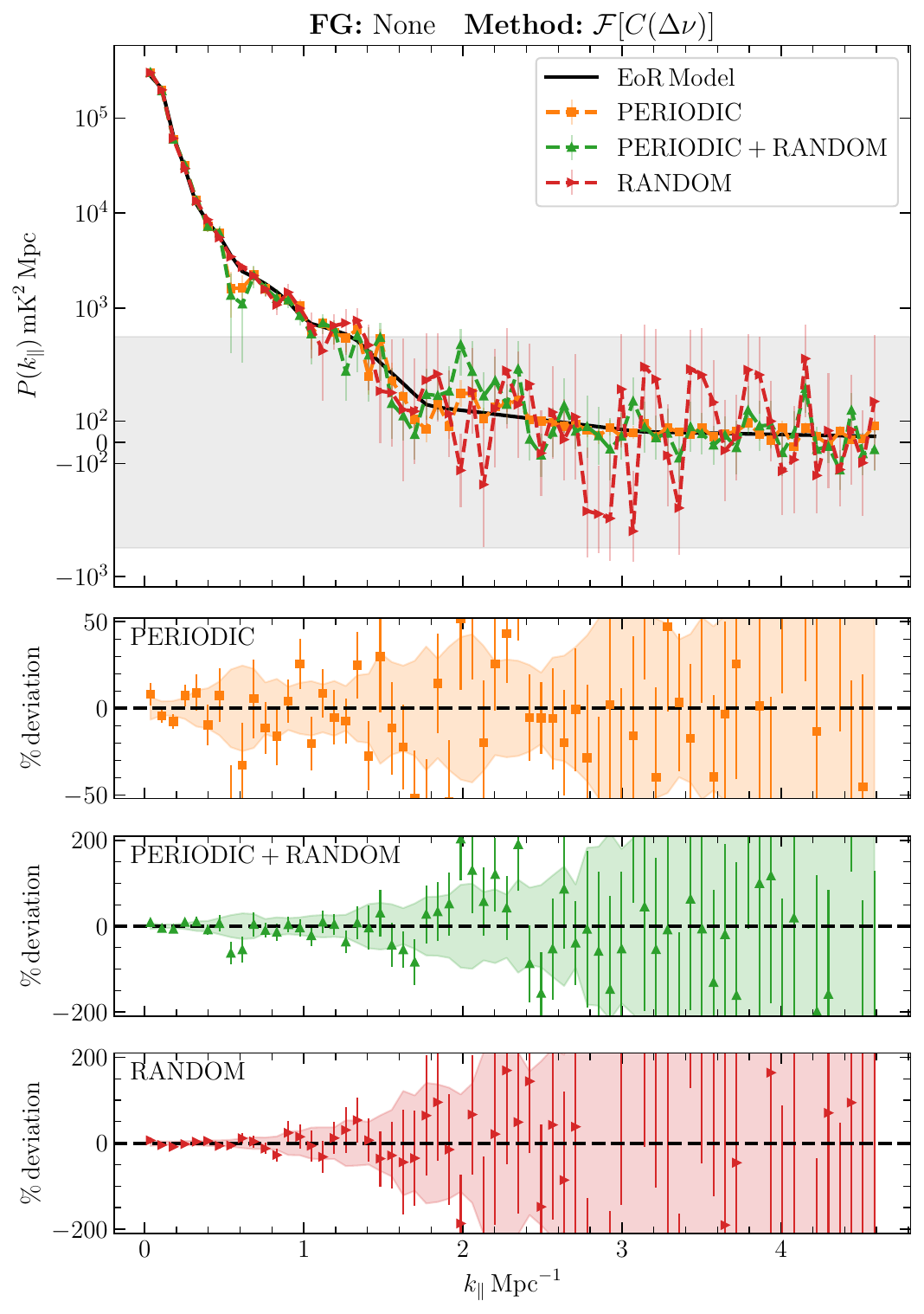}
    \caption{Recovered 1D LoS power spectrum using the correlation-based estimator $\mathcal{F}[C(\Delta\nu)]$ for different flagging scenarios. The solid black line shows the input theoretical EoR power spectrum. The y-axis uses a symmetric logarithmic (symlog) scale with the linear region $\mid P(k_\parallel) \mid < 5\times10^2\,{\rm mK^2\,Mpc}$ marked by the shaded grey band. The lower panel displays the percentage deviation relative to the model, with all error bars representing $1\sigma$ uncertainties. The correlation-based approach successfully recovers the input power spectrum over most of the $k_\parallel$ range.  The impact of missing channels appear primarily as increased uncertainties at large $k_\parallel$. A version of this figure zoomed in to $k_\parallel \leq 2\,{\rm Mpc}^{-1}$ is shown in \Cref{fig:21cm_SCF_None_allflags_zoom}.  }
    \label{fig:21cm_SCF_None_allflags}
\end{figure}

\begin{figure}
    \centering
    \includegraphics[width=0.99\columnwidth]{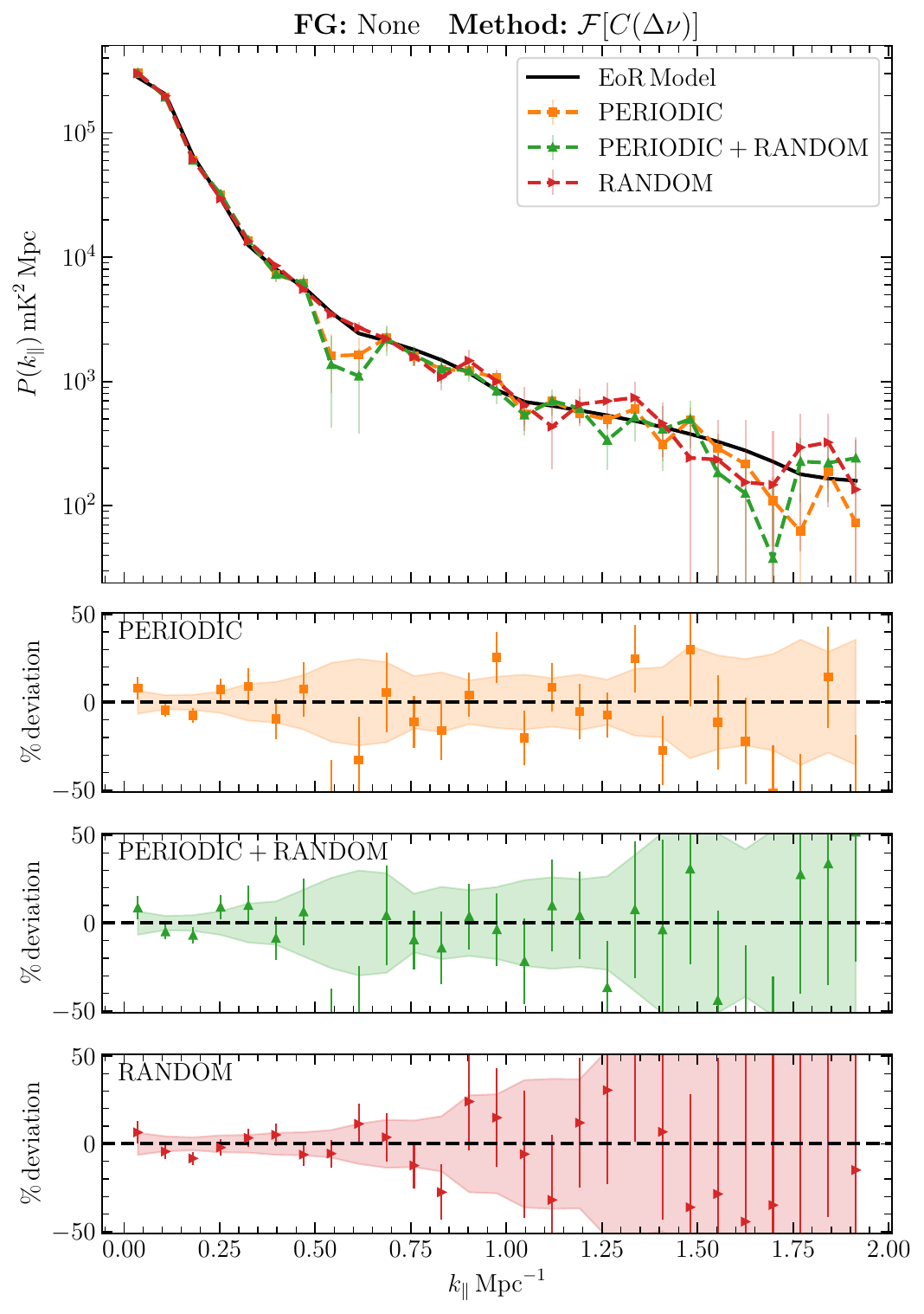}
    \caption{Same as \Cref{fig:21cm_SCF_None_allflags}, zoomed in to the region $k_\parallel \leq 2\,{\rm Mpc}^{-1}$. The $y$-axis of the top panel uses a logarithmic scale.}
    \label{fig:21cm_SCF_None_allflags_zoom}
\end{figure}

\Cref{fig:21cm_SCF_None_allflags} displays the estimated power spectrum obtained using the two-point correlation-based formalism (equations~\ref{eq:corr_est} and \ref{eq:dct}). The key here is that $C(\Delta\nu)$ is estimated only from the available frequency pairs. Here, the flagged channels reduce the number of contributing frequency pairs at a given frequency separation, but they do not introduce sharp spectral discontinuities in the frequency separation $(\Delta\nu)$ domain. 
As a consequence, the correlation-based estimator reliably recovers the input power spectrum over the bulk of the $k_\parallel$ range for all flagging configurations (top panel). The percentage deviation of the recovered power spectrum from the model (bottom panels) shows some scatter in the estimated values, which is mostly consistent with their associated uncertainties. 
We see that, particularly at large $k_\parallel$, the scatter is more prominent in the PERIODIC+RANDOM and RANDOM flagging scenarios, both of which have a higher total flagging percentage than the PERIODIC one. The primary impact of flagging manifests as an increase in the statistical uncertainties at large $k_\parallel$. We can therefore expect a noisier power spectrum at higher $k_\parallel$ with increased flagging.

\Cref{fig:21cm_SCF_None_allflags_zoom} zooms in to $k_\parallel \leq 2\,{\rm Mpc}^{-1}$, the large scale regime that is of primary scientific interest in current 21-cm experiments, to illustrate how the different flagging patterns affect these scales. We notice that PERIODIC flagging increases the uncertainties at relatively small $k_\parallel$ regimes, whereas RANDOM flagging increases the uncertainties at larger $k_\parallel$. This can be understood as a consequence of the spectral scale of each flagging pattern as presented in \Cref{fig:flagging}. PERIODIC flagging is a comparatively large-scale feature in frequency, while RANDOM flagging introduces small-scale features in the data, and the resulting uncertainties propagate to the corresponding LoS scales accordingly. We discuss this point further later in the text.

These results demonstrate that, although the FFT-squared and correlation-based estimators are equivalent in the idealised limit of complete, unflagged data, the presence of missing channels explicitly breaks this equivalence in practical applications. Under realistic 21-cm observational conditions, the correlation-based estimator is therefore intrinsically more robust to flagging. In this approach, it is neither necessary to inpaint or reconstruct missing data, nor to explicitly deconvolve the sampling function. This approach has therefore been widely adopted in analyses of the 21-cm power spectrum \citep[e.g.,][]{Pal2020, Elahi2024, Chatterjee2024}. In the subsequent sections of this paper, we adopt the correlation-based approach (equations~\ref{eq:corr_est} and \ref{eq:dct}) to estimate the power spectrum.

\begin{figure}
    \centering
    \includegraphics[width=0.99\columnwidth]{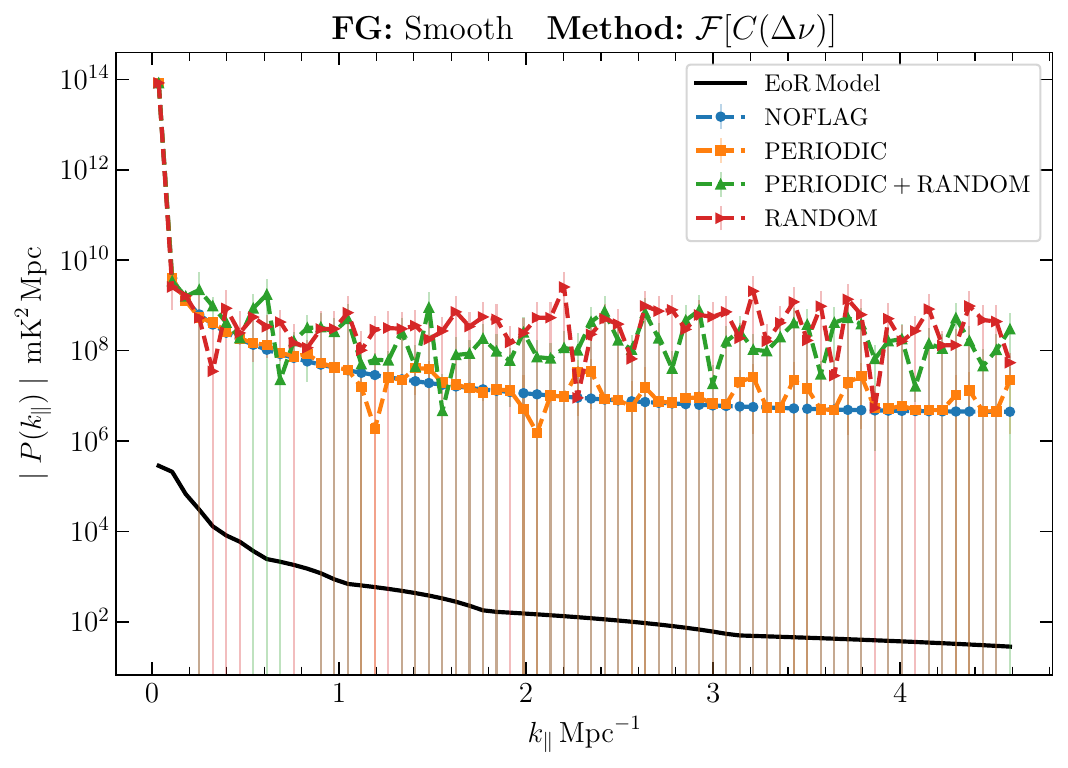}
    \caption{Recovered 1D LoS power spectrum using the correlation-based estimator $\mathcal{F}[C(\Delta\nu)]$ for different flagging scenarios. The input data is a combination of smooth FG and the EoR 21-cm signal. The solid black line denotes the input theoretical EoR power spectrum. Without SCF, the correlation-based estimator fails to recover the EoR signal in the presence of foregrounds.}
    \label{fig:FG_SCF_None_allflags}
\end{figure}

\citetalias{Elahi2025} demonstrated that in the case of foreground-dominated data with large-scale structured flagging, such as the periodic MWA flagging pattern, the correlation-based estimator can still suffer from residual foreground contamination. To show this using our simulated data, we add smooth foregrounds to the EoR 21-cm signal and then use the correlation-based estimator to estimate the power spectrum. As we see in \Cref{fig:FG_SCF_None_allflags}, when the correlation-based estimator is applied directly to data containing spectrally smooth foregrounds, the recovered power spectrum is dominated by foreground power across the entire $k_\parallel$ range. The estimated spectrum exhibits several orders of magnitude higher bias relative to the input EoR model. The correlation-based estimator, therefore, cannot reliably recover the underlying 21-cm power spectrum in the presence of foregrounds. To address this, \citetalias{Elahi2025} introduced the Smooth Component Filtering (SCF) technique, which removes the spectrally smooth component of the data before estimating the power spectrum.

\section{Smooth Component Filtering (SCF)}
\label{sec:scf}

We model the total brightness temperature vector $\mathbf{T}$ along each line of sight as a superposition of a smooth component ($\mathbf{T}_{\rm S}$), a rough component ($\mathbf{T}_{\rm R}$) and noise ($\textbf{n}$):
\begin{equation} 
    \mathbf{T} = \mathbf{T}_{\rm S} + \mathbf{T}_{\rm R} + \mathbf{n}\,. 
\label{eq:model} 
\end{equation}
In SCF, we estimate the  spectrally smooth, foreground-dominated component $\mathbf{T}_{\rm S}$, and subtract it from the total data $\mathbf{T}$ to obtain the (smooth component) filtered data   
\begin{equation}
    \mathbf{T}_{\rm F} = \mathbf{T} - \mathbf{T}_{\rm S} \, , 
\end{equation}
which we use in the correlation-based estimator (equations~\ref{eq:corr_est} and \ref{eq:dct}) to obtain the power spectrum. Below, we describe two implementations for estimating the smooth component $\mathbf{T}_{\rm S}$. Note that in the actual implementation on interferometric data, SCF independently operates on real and imaginary parts of the complex visibilities as a function of frequency. Here, in our 1D LoS-only analysis, we apply SCF directly to $\mathbf{T}$, the real brightness temperature field. 

\subsection{SCF using a Hann window (Hann-SCF)}
\label{subsec:scf_hann}

In this approach, the smooth component is estimated by convolving the data with a Hann window along the frequency axis,
\begin{equation}
    T_{\rm S}(\nu_n) = \sum_m T(\nu_m)\,H(n-m),
\end{equation}
where the window function is defined as
\begin{equation}
    H(n) = \mathcal{A} \left[1 + \cos\!\left(2\pi\frac{n}{2N_W}\right)\right], \quad -N_W \le n \le N_W \,.
    \label{eq:methods_hann}
\end{equation}
Here, $N_W$ determines the half-width of the window, and $\mathcal{A}$ is a normalisation constant ensuring that $\sum H(n) = 1$. The performance of several other window functions, such as the Kaiser and Blackman functions, has been found to be similar when applied to actual MWA data (\citetalias{Elahi2025}).   

The width of the Hann window determines the smoothing scale. Following \citetalias{Elahi2025}, we adopt $N_W=50$, which corresponds to a full window width of $\simeq 4$~MHz. When applied to simulated data, this choice suppresses spectrally smooth foreground power at $k_\parallel \gtrsim 0.135\,\mathrm{Mpc}^{-1}$, which corresponds to a frequency scale of $\simeq 2.7$~MHz \citep{Sarkar2026}.

\begin{figure}
    \centering
    \includegraphics[width=0.99\columnwidth]{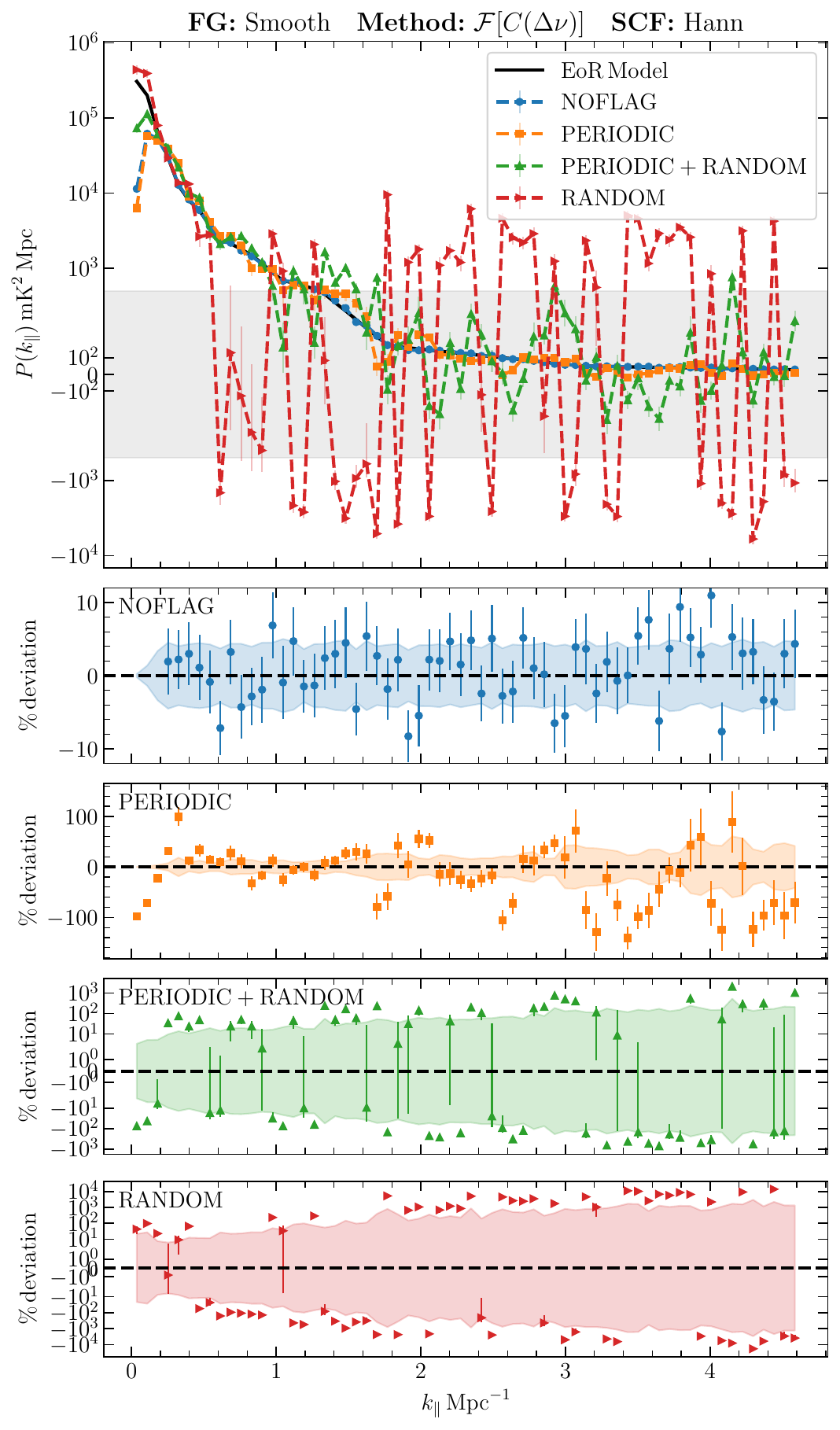}
    \caption{Recovered 1D LoS power spectrum using the Hann-SCF followed by the correlation-based estimator $\mathcal{F}[C(\Delta\nu)]$ for different flagging scenarios. The input data is the same as that used in \Cref{fig:FG_SCF_None_allflags}, i.e., a combination of smooth FG and the EoR 21-cm signal. The solid black line denotes the input theoretical EoR power spectrum. The y-axis uses a symlog scale with the linear region $\mid P(k_\parallel) \mid < 5\times10^2\,{\rm mK^2\,Mpc}$ marked by the shaded grey band. The lower panel shows the fractional deviation from the model; all error bars correspond to $1\sigma$ uncertainties. For NOFLAG, when the Hann-SCF is applied, the correlation-based estimator successfully recovers the EoR signal in the presence of foregrounds Its performance is reasonable for PERIODIC flagging. However, it fails to remove the foregrounds adequately for PERIODIC+RANDOM and RANDOM flagging, both shown in symlog scale.  A version of this figure zoomed in to $k_\parallel \leq 2\,{\rm Mpc}^{-1}$ is shown in \Cref{fig:FG_Hann_allflags_zoom}. }
    \label{fig:FG_Hann_allflags}
\end{figure}

\begin{figure}
    \centering
    \includegraphics[width=0.99\columnwidth]{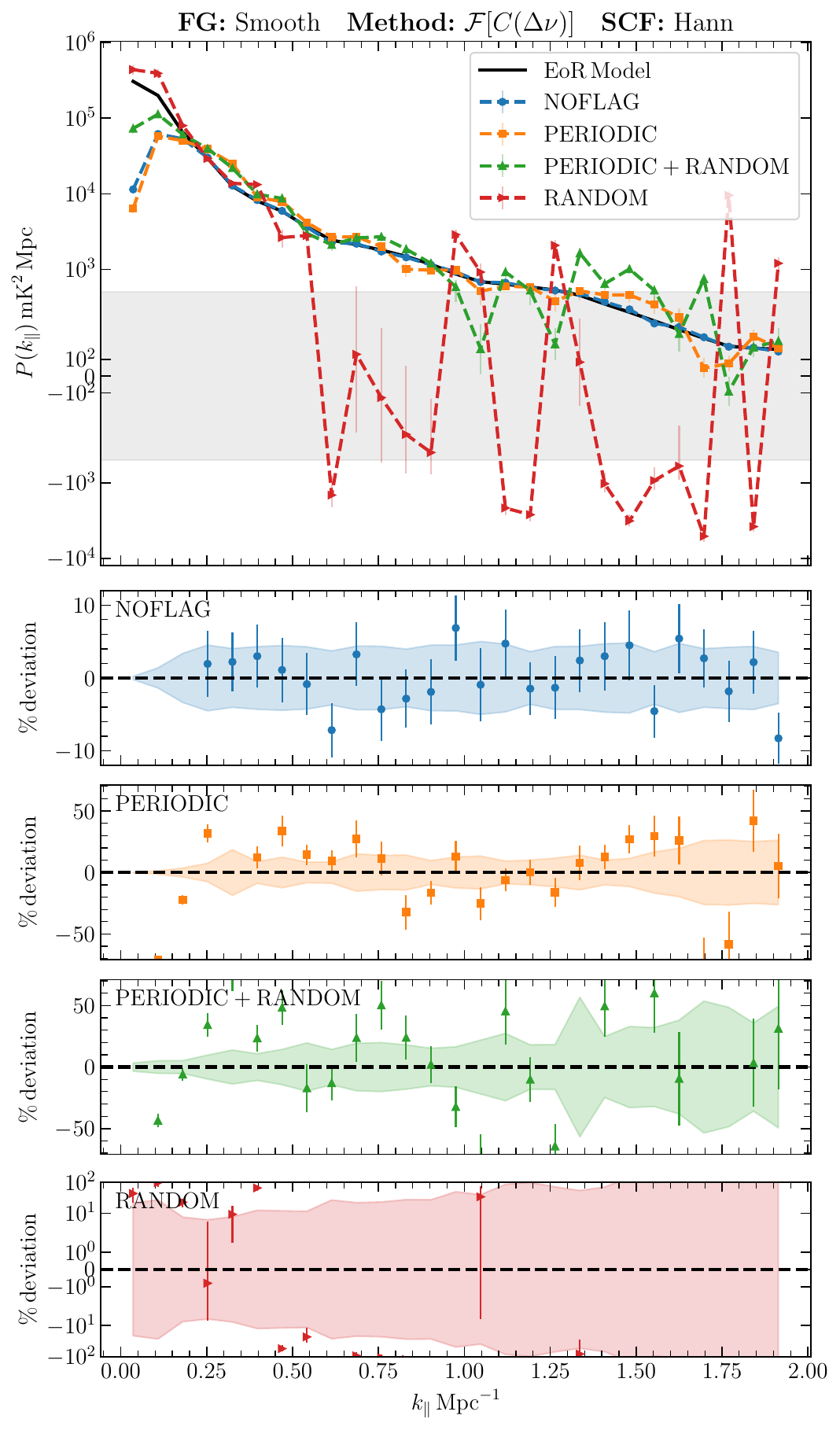}
    \caption{Same as \Cref{fig:FG_Hann_allflags}, zoomed in to the region $k_\parallel \leq 2\,{\rm Mpc}^{-1}$. In the bottom panel, the RANDOM flagging case is shown in symlog scale.}
    \label{fig:FG_Hann_allflags_zoom}
\end{figure}

\Cref{fig:FG_Hann_allflags}, and its zoomed in counterpart \Cref{fig:FG_Hann_allflags_zoom}, demonstrate that, once the Hann-based SCF (Hann-SCF, hereafter) is employed to suppress the smooth foreground component, the correlation-based estimator is able to recover an unbiased estimate of the EoR 21-cm power spectrum for the NOFLAG and PERIODIC flagging configurations. The signal loss at the lowest  modes $(k_\parallel<0.2\,{\rm Mpc}^{-1})$ is consistent with the expected loss of large-scale, spectrally smooth modes induced by the filter. Over the range of $k_\parallel$ modes that are not affected by the filtering, the recovered power spectrum is consistent with the input theoretical model within the $1\sigma$ error bars for these two flagging configurations. For PERIODIC+RANDOM and RANDOM flagging, however, the recovered power spectrum is mostly inconsistent with the model within the quoted $1\sigma$ error bars across the full $k_\parallel$ range. The performance is particularly  worse for RANDOM flagging, which indicates that the Hann-SCF fails to adequately filter the smooth component in the presence of RANDOM flagging, even though the foreground itself is spectrally smooth.

This finding is in line with what has been found considering actual MWA data with PERIODIC flagging in \citetalias{Elahi2025} and more recently in \cite{Sarkar2026}. There, the Hann-SCF has shown promising results at large angular scales $k_\perp \leq 0.05\,{\rm Mpc}^{-1}$, at which foregrounds exhibit broad spectral coherence that resembles the smooth foregrounds we have simulated here. While effective at removing strictly smooth foreground components, the Hann-SCF has several inherent limitations. Simulations in \citetalias{Elahi2025} show that when foregrounds acquire additional spectral structure due to baseline migration, e.g., at $k_\perp > 0.05\,{\rm Mpc}^{-1}$, the Hann filter fails to adequately model the smooth component. Even if narrower windows are used to model those spectral structures,  some foreground residuals can persist within the wedge and subsequently leak into higher $k_\parallel$ modes. Recently, \cite{Sarkar2026} used relatively more sensitive data than \citetalias{Elahi2025} by incoherently combining the measured power spectrum from different observations and found that there are some very faint streaks around specific $k_\parallel$ values that survive the Hann-SCF. The authors masked out those $k_\parallel$ ranges before calculating the spherically averaged power spectrum $P(k)$. Both the simulations and actual data, therefore, indicate a limitation in the Hann-SCF.

\begin{figure}
    \centering
    \includegraphics[width=0.99\columnwidth]{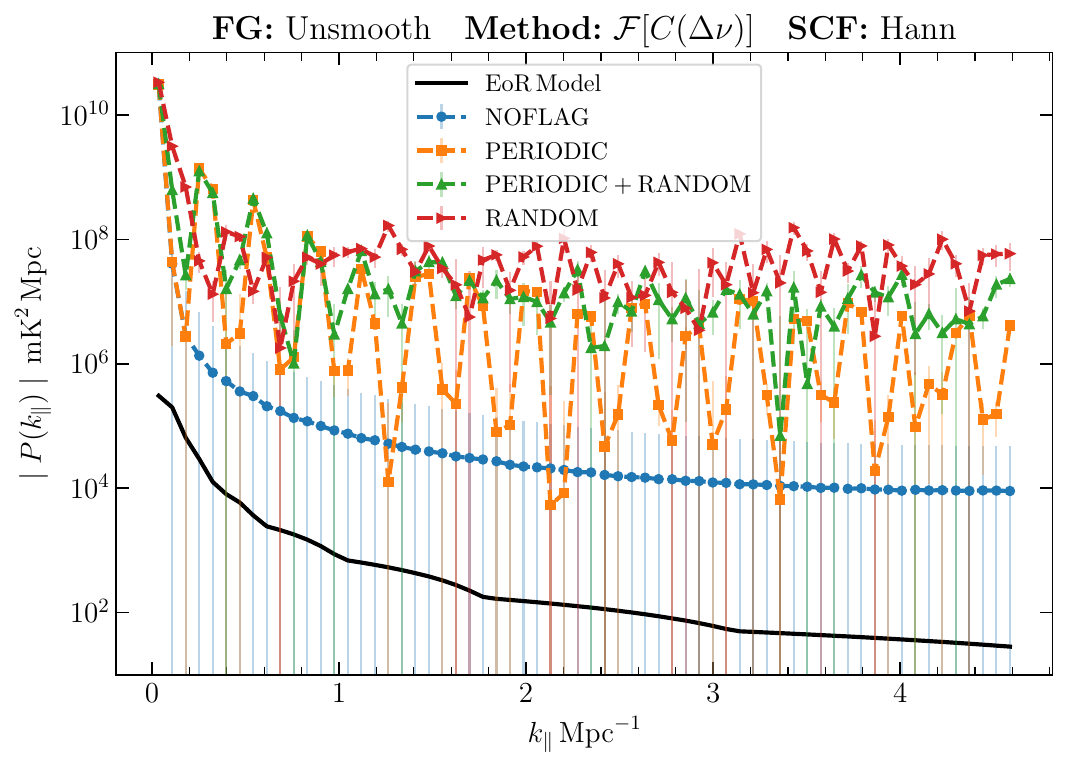}
    \caption{Recovered 1D LoS power spectrum obtained with the Hann-SCF followed by the correlation-based estimator $\mathcal{F}[C(\Delta\nu)]$ for different flagging configurations. The input data is a combination of unsmooth FG and the EoR 21-cm signal. The solid black curve represents the theoretical input EoR power spectrum. In the presence of unsmooth foregrounds, the correlation-based estimator, even when combined with the Hann-SCF, is unable to  recover the input EoR power spectrum.}
    \label{fig:FGMM_Hann_allflags}
\end{figure}

We now test the performance of the Hann-SCF in a more challenging and realistic scenario in which the data contain strong spectral structure, as introduced in \Cref{sec:gp_foregrounds} as `unsmooth' foregrounds. \Cref{fig:FGMM_Hann_allflags} shows that the Hann-SCF is no longer sufficient to fully remove the foreground contribution once strong mode-mixing is present. Although the filter suppresses the spectrally smooth component from the data, the residual foregrounds are dominant enough to obscure the 21-cm signal. The recovered power spectra exhibit a clear residual bias relative to the input EoR model. The bias also worsens in the presence of flagging. Foreground components affected by mode-mixing, equivalently, those characterized by short spectral correlation lengths, thus expose a fundamental limitation of the Hann-SCF approach. While this framework is reasonably effective at mitigating smoothly varying spectral foregrounds, it lacks the flexibility required to accurately model and subtract the more complex spectral structures generated by instrumental chromaticity.

We note that a Hann filter with a smaller width, $N<50$, should improve results, but it comes at the cost of losing large-scale features. Furthermore, as shown in the Appendix~\ref{app:appendix_limitations_of_hann}, a fixed window filter is fundamentally limited to fully capture the strong spectral components. We also note that, because the Hann window has finite support, the convolution is ill-defined near the band edges, which requires $N_W$ channels to be discarded at each end of the band, thereby reducing the effective bandwidth and sensitivity. 

\subsection{Bayesian SCF (Bayes-SCF)}
\label{subsec:scf_gp}

We present a Bayesian formulation of the SCF framework based on Gaussian Process (GP), which overcomes the inherent limitations of the Hann-SCF approach. We have used the \textsc{George} software package \citep{george} to perform this analysis.

A GP is defined as a collection of random variables such that any finite subset follows a joint multivariate Gaussian distribution \citep{bishop}. It is fully specified by a mean function $m(\nu)$ and a covariance function (kernel) $K(\nu,\nu')$:
\begin{equation}
    f(\nu) \sim \mathcal{GP}\!\left[m(\nu), K(\nu,\nu')\right].
    \label{eq:GP}
\end{equation}
In this framework, each component of the total brightness temperature vector $\mathbf{T}$ (equation~\ref{eq:model}) is described by a kernel. To model the dominant smooth component, we employ an RBF kernel
\begin{equation}
    k_{\rm S}(\nu, \nu') = A_{\rm S}^2 \exp\left[ - \frac{(\nu - \nu')^2}{2 (N_{\rm S} \Delta\nu_c)^2} \right] \, ,
    \label{eq:kernel_smooth}
\end{equation}
which is infinitely differentiable and is well-suited for modelling the smooth component of the data. Here, the amplitude $A_{\rm S}$ and the correlation length $N_{\rm S}$ are the two hyperparameters of the smooth kernel $\mathbf{K}_{\rm S}$. 
To model the rough component, we choose a Matérn-3/2 kernel \citep{Rasmussen2006}
\begin{equation}
    k_{\rm R}(\nu, \nu') = A_{\rm R}^2 \left[ 1 + \frac{\sqrt{3} \, |\nu - \nu'|}{(N_{\rm R}\Delta\nu_c)} \right] \exp\left[ - \frac{\sqrt{3} \, |\nu - \nu'|}{(N_{\rm R}\Delta\nu_c)} \right] \, ,
    \label{eq:kernel_rough}
\end{equation}
which is only once differentiable, making it suitable for modelling rapid fluctuations. Here, the amplitude $A_{\rm R}$ and the correlation length $N_{\rm R}$ are the two hyperparameters of the rough kernel $\mathbf{K}_{\rm R}$.  The total kernel is given by an addition 
\begin{equation}
    \mathbf{K} = \mathbf{K}_{\rm S} + \mathbf{K}_{\rm R} + \sigma_n^2 \mathbf{I}\, ,
    \label{eq:kernel_tot}
\end{equation}
where $\sigma_n^2$ is the variance of the white noise kernel. 

The novelty of Bayes-SCF lies in how we set the prior distribution over the kernels' hyperparameters. Considering the smooth kernel, we keep the correlation length scale of the kernel to a \textit{fixed} value $(N_{\rm S} = N_{\rm GP})$. The fixed value $N_{\rm GP}$ ensures that the kernel $\mathbf{K}_{\rm S}$ captures only a fixed, large-scale spectral trend and cannot absorb the fluctuations finer than that set by $N_{\rm GP}$. On the other hand, we allow the correlation length scale for the rough component $N_{\rm R}$ to be optimized during the fit -- subject to a strict upper bound $N_{\rm R} < N_{\rm GP}$. This constraint ensures that the kernel $\mathbf{K}_{\rm R}$ can adapt to rapid variations in the data, while also preventing it from degenerating into a smooth function that could covary with the foreground model. The hyperparameters $A_{\rm S}, A_{\rm R}$ and $N_{\rm R}$  are estimated by maximising the marginal log-likelihood of the observed data on the available (non-flagged) channels. Considering noise, we set $\sigma_n$ to a fixed, relatively small value $(10^{-5})$ and do not optimize it. Since our simulated data does not contain noise, the purpose of including a relatively small amount of white noise is solely to maintain numerical stability in the code. Table~\ref{tab:gp_hyperparameters} provides a summary of the hyperparameters. 

We reiterate that $N_{\rm S}$ is not optimized and is kept fixed to the value $N_{\rm GP}$ to ensure controlled smoothing.  We note that the choice of $N_{\rm GP}$ can either be data driven, or motivated by the foreground wedge boundary using the relation 
\begin{equation}
    N_{\rm GP}   = \frac{2\pi} {r^\prime \Delta\nu_c [k_\parallel]_{\rm H} } = \frac{2\pi \nu_c} { rk_\perp\Delta\nu_c},
    \label{eq:N_GP}
\end{equation}
where $[k_\parallel]_{\rm H} = (r/r^\prime \nu_c) \, k_\perp$ is the `horizon limit' or the theoretically predicted boundary of the foreground wedge. For subsequent analysis, we set $N_{\rm GP} = 96$, which corresponds to a correlation length of $ [N_{\rm GP}~\Delta\nu_c] = 3.84$~MHz. We discuss the rationale for this choice in Appendix~\ref{app:smoothing_scale}.

\begin{table}
	\centering
	\caption{Summary of the GP kernel hyperparameters and their priors used in the Bayes-SCF. }
	\label{tab:gp_hyperparameters}
	\begin{tabular}{lll} 
		\hline
		\textbf{Kernel} & Hyperparameters & Prior / Value \\
		\hline
		\rule{0pt}{3ex} 
		\multirow{2}{*}{Smooth (RBF)} & Amplitude ($A_{\rm S}$) & $\mathcal{U}(0, \infty)$ \\
		    & Length Scale ($N_{\rm S}$) & Fixed ($N_{\rm GP}$) \\
		\hline
		\rule{0pt}{3ex}%
		\multirow{2}{*}{Rough (Matérn-3/2)} & Amplitude ($A_{\rm R}$) & $\mathcal{U}(0, \infty)$ \\
		    & Length Scale ($N_{\rm R}$) & $\mathcal{U}(0, N_{\rm GP})$ \\
		\hline
		\rule{0pt}{3ex}%
		Noise (White Noise) & RMS ($\sigma_n$) & Fixed ($10^{-5}$) \\
		
		\hline
	\end{tabular}
\end{table}

\begin{figure}
    \centering
    \includegraphics[width=0.99\columnwidth]{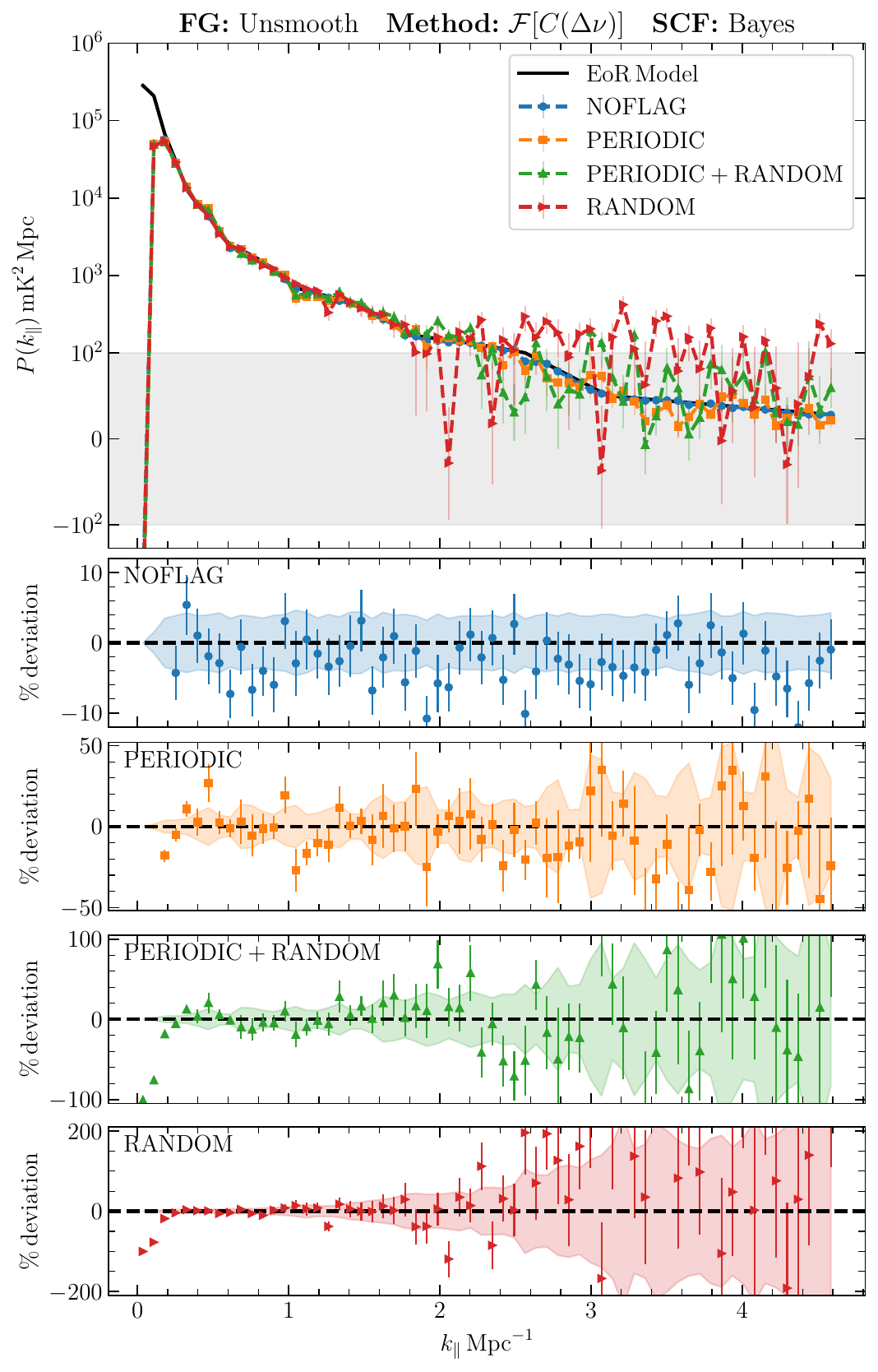}
    \caption{Recovered 1D LoS power spectrum obtained with the Bayes-SCF followed by the correlation-based estimator $\mathcal{F}[C(\Delta\nu)]$, for different flagging scenarios. The input data is the same as that used in \Cref{fig:FGMM_Hann_allflags}, i.e., a combination of unsmooth FG and the EoR 21-cm signal. The solid black curve represents the theoretical input EoR power spectrum. The y-axis uses a symlog scale with the linear region $\mid P(k_\parallel) \mid < 10^2\,{\rm mK^2\,Mpc}$ marked by the shaded grey band. The lower panels show the fractional deviation with respect to the theoretical model for each flagging scenario, with $1\sigma$ error bars. The correlation-based estimator, when coupled to the Bayes-SCF, successfully recovers the input model even in the presence of strong, unsmooth foregrounds. A version of this figure zoomed in to $k_\parallel \leq 2\,{\rm Mpc}^{-1}$ is shown in \Cref{fig:FGMM_GP_allflags_zoom}.}
    \label{fig:FGMM_GP_allflags}
\end{figure}
 
\begin{figure}
    \centering
    \includegraphics[width=0.99\columnwidth]{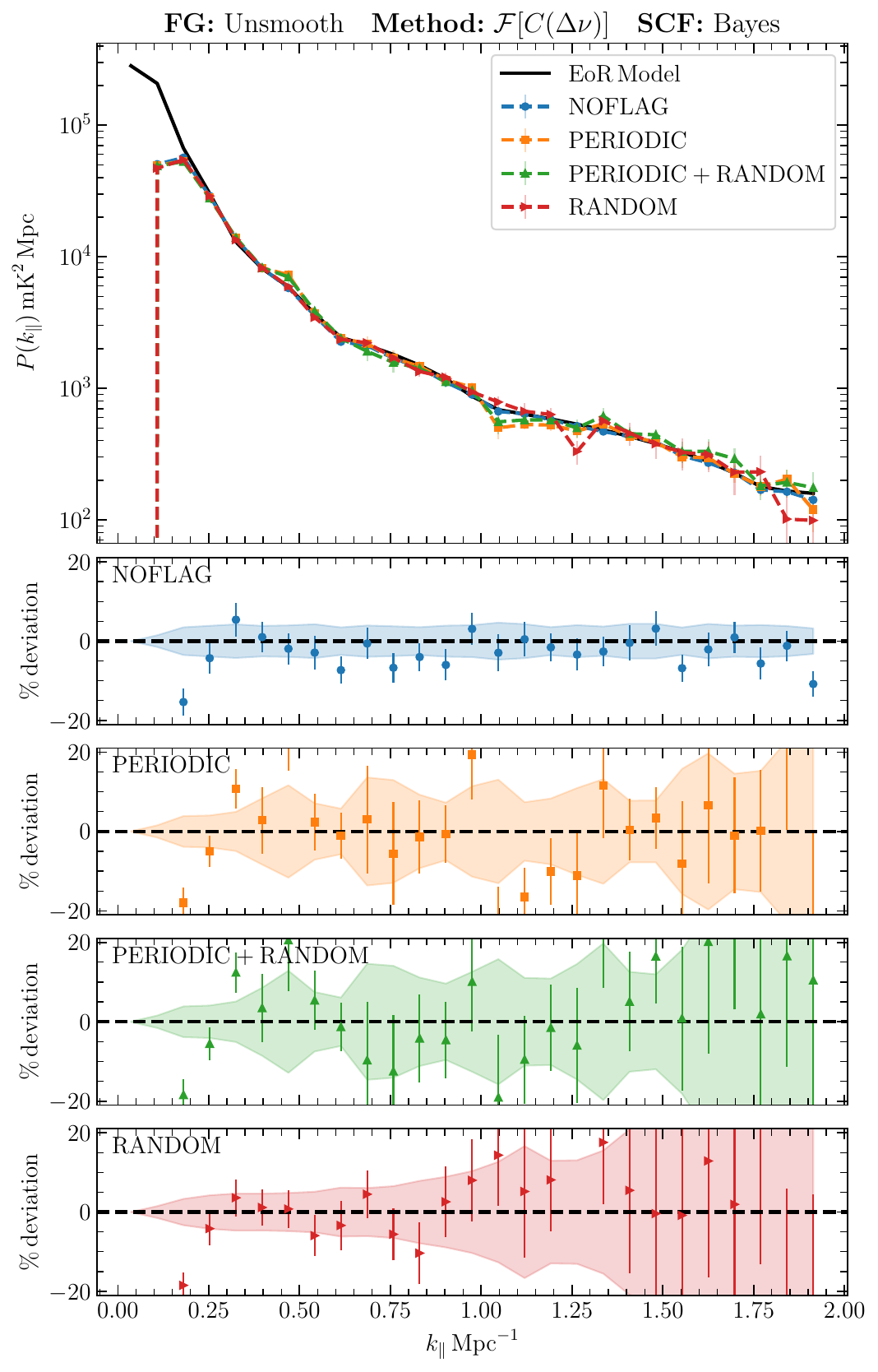}
    \caption{Same as \Cref{fig:FGMM_GP_allflags}, zoomed in to the region $k_\parallel \leq 2\,{\rm Mpc}^{-1}$. The $y$-axis of the top panel uses a logarithmic scale.}
    \label{fig:FGMM_GP_allflags_zoom}
\end{figure}

Once the parameters are optimized, we extract the smooth component using the linear properties of Gaussian distributions. The posterior mean of the smooth component is given by \citep{Rasmussen2006}:
\begin{equation}
    \hat{\mathbf{T}}_{\rm S} = \mathbf{K}_{\rm S} \mathbf{K}^{-1} \mathbf{T} \,.
    \label{eq:gp_prediction}
\end{equation}
Finally, we obtain the filtered component: 
\begin{equation}
    \mathbf{T}_{\rm F} = \mathbf{T} - \hat{\mathbf{T}}_{\rm S} \, .
    \label{eq:gp_filtered}
\end{equation}

This formulation involves an important distinction from the GPR framework presented in \citet{mertens18} and in several subsequent works, in which the covariance kernels are constructed to model the foregrounds and the 21-cm signal, thereby separating them. In contrast, our objective here is to isolate the dominant (smooth) component from the data, regardless of whether it is the foregrounds or the cosmological signal. We do not aim to separate the 21-cm signal from the foregrounds at this stage. The motivation is to filter out the dominant non-stationary component of the data and pass the residual component to the correlation-based estimator, which is sensitive to the non-stationary component, but otherwise robust against the missing frequency channels.

We apply the Bayes-SCF to the same dataset (unsmooth FG + 21-cm signal) where the Hann-SCF failed (Figure~\ref{fig:FGMM_Hann_allflags}). \Cref{fig:FGMM_GP_allflags} shows that once the Bayes-SCF is applied to the data, the correlation-based estimator yields a power spectrum that is consistent with the input EoR model across the accessible range of $k_\parallel$. The residuals in the lower panels scatter around zero and remain within the $1\sigma$ error bars for all flagging configurations. 
For RANDOM flagging, we observe slight deviations at the largest $k_\parallel$,which, although mostly within the error bars, can partly be attributed to the leakage of unfiltered components in the data due to the missing channels. 
A natural question that arises at this point is how the Bayes-SCF performs with increased flagging. To test this, we apply the Bayes-SCF to the same dataset with a substantially higher fraction of randomly flagged channels, and the details are provided in Appendix~\ref{app:higher_flag_fraction}. We have also tested the algorithm in the presence of a single, wide contiguous gap in the frequency band, which is presented in Appendix~\ref{app:broadband_rfi}.

\Cref{fig:FGMM_GP_allflags_zoom} further zooms in to the large scale range $k_\parallel \leq 2 \, {\rm Mpc}^{-1}$. We see that the Bayes-SCF recovers the input EoR model quite well in the $0.2 \lesssim k_\parallel \leq 2 \, {\rm Mpc}^{-1}$ region. The uncertainties in this range are much smaller than large $k_\parallel$. Note here that the range $k_\parallel \lesssim 0.2 \, {\rm Mpc}^{-1}$ is filtered by SCF, and we do not recover the EoR signal at those scales. The filtered length scale is related to the choice of $N_{\rm GP}$, and further details with different choices of $N_{\rm GP}$ are provided in Appendix~\ref{app:smoothing_scale}.

We note that the uncertainties from the different flagging patterns behave in a very similar way to what we described earlier in \Cref{fig:21cm_SCF_None_allflags_zoom} for EoR-only simulations without SCF. Here, the PERIODIC flagging increases the uncertainties at small $k_\parallel$, whereas RANDOM flagging increases the uncertainties at a larger $k_\parallel$. It is interesting to note that after Bayes-SCF, the uncertainties have decreased from the EoR-only simulations where SCF is not applied. We interpret this reduction in sample variance as being due to the fact that Bayes-SCF reduces the power at small $k_\parallel$, which results in reduced leakage of power from small to large $k_\parallel$. However, note that these uncertainties are due to sample variance only; to quantify the total error in these estimates, it is necessary to evaluate the error covariance that includes the uncertainties in the estimated smooth components.

\begin{figure*}
	\centering
	\includegraphics[width=0.99\textwidth]{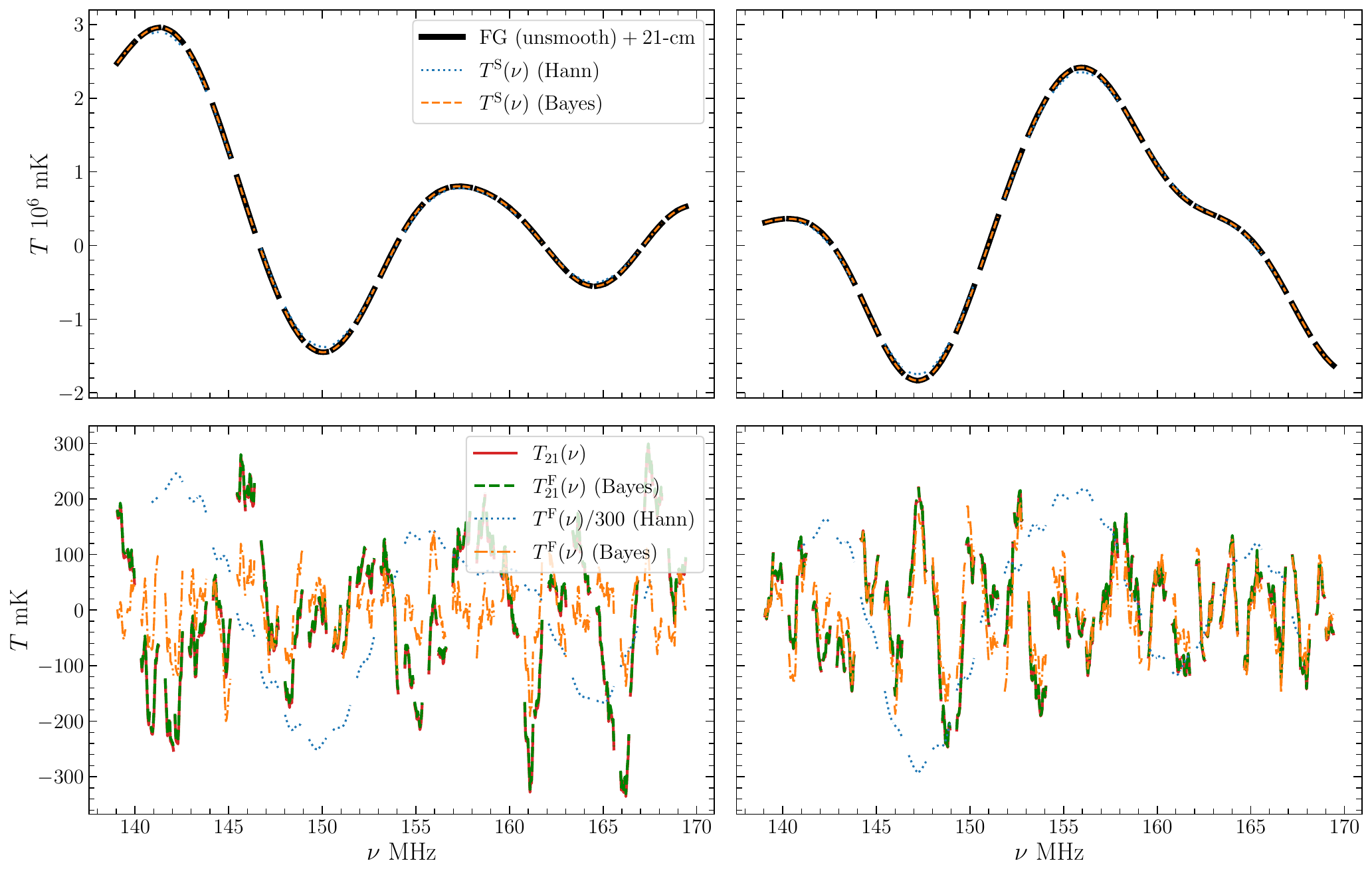}
	\caption{Demonstration of SCF on two independent realizations (left and right columns) of the simulated data. The top panels show the total input brightness temperature (black solid line), due to unsmooth foregrounds  and the EoR 21-cm signal. The estimated smooth components $T_{\rm S}(\nu)$ recovered by the Hann- (blue dotted) and Bayes-SCF (orange dashed) are also shown. The gaps indicate PERIODIC missing channels.  
	Considering the bottom panels, the red solid and green dashed lines are reference lines, which nearly overlap, respectively show the unfiltered and filtered (with Bayes-SCF) EoR-signal-only data. 
    When Bayes-SCF is applied on the total input brightness temperature, the residuals (orange dashed-dot) accurately recover the small-scale features of the 21-cm signal across the full band. In contrast, the  residuals from Hann-SCF (blue dotted) exhibit the presence of significant systematic residual foregrounds and are scaled down by a factor of 300 for clarity.  It demonstrates that the Bayes-SCF outperforms the Hann-SCF at extracting the EoR 21-cm signal, in the presence of unsmooth foregrounds.}
	\label{fig:scf_comparison}
\end{figure*}

To understand the physical origin of the residual bias observed in the Hann-SCF (\Cref{fig:FGMM_Hann_allflags}) and the subsequent success of Bayes-SCF, we examine the performance of both filters in the frequency domain for individual realizations.
For this, we consider two independent realizations of the unsmooth foregrounds and the PERIODIC flagging scenario. 
The top panels of \Cref{fig:scf_comparison} show the total input signal, which consists of the bright, unsmooth foregrounds and the EoR signal. The smooth components derived from both the Hann- (blue dotted) and Bayes-SCF (orange dashed) appear to trace the large-scale trend of the foregrounds reasonably well. However, a closer inspection shows that Hann-SCF fails when the data has sharp features, e.g., near 140, 150 and 165~MHz in the left panel, and near 145 and 155~MHz in the right panel. This is clearly seen in the bottom panels, which show the filtered components. 
Here, the red solid and green dashed lines show the unfiltered and Bayes-SCF filtered EoR-signal-only data, respectively.  Both of these lines nearly overlap, indicating that there is hardly any signal loss when Bayes-SCF is applied to EoR-signal-only data. This reference line is similar for the Hann-SCF, and it is not shown here.
Once applied to the unsmooth foreground dominated data, we find that the residuals from the Hann-SCF (blue dotted) exhibit significant systematic foregrounds and are scaled down by a factor of 300 for clarity. On the other hand, the residuals from Bayes-SCF (orange dashed) accurately recover the small-scale features of the 21-cm signal across the full band, with signal loss mostly appearing at large scales. It is therefore clear that Bayes-SCF is better than the Hann-SCF at extracting the EoR 21-cm signal in the presence of unsmooth foregrounds. Note that a similar comparison in the delay space is presented in Appendix~\ref{app:scfindelay}.

\section{Conclusions}
\label{sec:conclusions}

Estimation of 21-cm power spectrum (as well as higher-order statistics such as bispectrum) from radio interferometric visibility data is a challenging task due to the many orders of magnitude brighter foregrounds and instrumental systematics. The missing frequency channels in the data particularly create artefacts in the estimated power spectrum. This problem can be circumvented by first estimating the two-point correlation, and then performing a Fourier transform. However, this method works only when the underlying data is statistically homogeneous (ergodic), which is not the case in the presence of foregrounds, which have a strong spectral structure. 

Smooth Component filtering (SCF; \citealt{Elahi2025}, henceforth \citetalias{Elahi2025}) solves this problem by first identifying the dominant, spectrally smooth component, and then subtracting it from the measured visibilities. \citetalias{Elahi2025} implemented the SCF by convolving the visibility data with a Hann window to find the smooth component. In this work, we use simulated data to demonstrate that the Hann-SCF performs adequately when the foregrounds are smooth in frequency, but its efficacy is severely limited when the foregrounds have pronounced spectral structure, such as those that give rise to the wedge in the 21-cm power spectrum. Its capability degrades further with increased flagging. We also find that the limitation of window-based SCF is of a fundamental nature, and it cannot be overcome by using a different size of the filter (Appendix~\ref{app:appendix_limitations_of_hann}).

We present a Bayesian framework for SCF that uses a Gaussian Process (GP) to model the smooth component from the data. In Bayes-SCF, we model the data as a GP that is described by a kernel consisting of a smooth and a rough component. We use a \textit{fixed} prior on the correlation length scale of the smooth kernel to control the degree of smoothness. We allow the other hyperparameters to vary to obtain the best-fit hyperparameters. Once the hyperparameters are optimized, we use the smooth kernel, total kernel, and the observed data to predict the smooth component of the data. Our analysis shows that Bayes-SCF outperforms Hann-SCF, particularly in scenarios where the foregrounds exhibit pronounced spectral structure (i.e., spectrally unsmooth foregrounds).  

In addition, Bayes-SCF offers two further advantages over the Hann-SCF: (1) The convolution operation in the Hann-SCF is ill-defined at the boundaries of the frequency band, which necessitates the removal of a half-window’s worth of frequency channels from both edges of the band. By contrast, Bayes-SCF remains well-defined across the full band and thus preserves the entire available bandwidth. (2) In the presence of flagging, the Hann-SCF requires missing data to be replaced with zeros, followed by an application of the SCF to the sampling function and a division of the data by this result to correct for the sampling pattern (\citetalias{Elahi2025}). In contrast, the GP formulation underlying Bayes-SCF is defined only on the unflagged channels and operates directly on the available data, without the need for zero-filling or separate corrections for the sampling function. We find that Hann-SCF fails to remove even smooth foregrounds in the presence of merely 35\% flagging. In contrast, Bayes-SCF can tolerate up to 80\% of flagging, even in the presence of unsmooth foregrounds (Appendix~\ref{app:random_flagging}).

A key outcome of this analysis is that the influence of missing frequency channels is not restricted to the MWA’s periodic flagging pattern; purely random flagging can also produce significant foreground leakage. Accordingly, we anticipate that SCF will provide more stringent constraints for other experiments, including post-reionization wideband intensity-mapping (e.g., \citealt{Elahi2024, Carucci2025, Deng2026}). We further find that missing channels increase the uncertainties at large $k_\parallel$, with the magnitude of this effect scaling approximately linearly with the flagging fraction. As a result, we expect weaker constraints at large $k_\parallel$, particularly for $k_\parallel > 2~{\rm Mpc}^{-1}$. In all cases considered, the application of SCF leads to a reduction in these uncertainties by reducing foreground leakage.

Our preliminary analysis (Appendix~\ref{app:scfindelay}) further demonstrates that SCF can effectively mitigate artefacts arising from missing channels even when the delay-spectrum formalism is employed. Hence, integrating Bayes-SCF into existing 21-cm analysis pipelines, such as TGE \citep{choudhuri2016b}, CHIPS \citep{trott16} or $\mathcal{E}$ppsilon \citep{Barry2019eppsilon}, is straightforward, as it requires only an additional, independent processing stage applied to the gridded (or per-baseline) visibility data. 

Future research will focus on applying this methodology to real observational datasets from the MWA, particularly targeting those parts of the sky where strong sources such as Fornax~A cause severe foreground leakage \citep{Jong2025, Sarkar2026}. In general, at low $k_\perp$, we anticipate that Bayes-SCF will further suppress any residual foreground contamination. At high $k_\perp$, which are not adequately probed by the Hann-SCF approach, we expect Bayes-SCF to recover the corresponding cosmological information. We also expect access to higher $k_\perp$ modes to increase the number of available triangle configurations in the 21-cm bispectrum, thereby enabling more stringent constraints \citep{Gill_2025_mwa1, Gill2026}. Nonetheless, a principal current limitation of Bayes-SCF is its substantial computational cost. As a result, applying this methodology to real visibility data will require significant optimization of the implementation, which we intend to pursue in future work.

%%%%%%%%%%%%%%%%%%%%%%%%%%%%%%%%%%%%%%%%%%%%%%%%%%

% \balance
\section*{Acknowledgements}

The author thanks the anonymous reviewer for carefully reading the manuscript and for the constructive comments that helped improve the work. The author acknowledges support from the Centres of Excellence (CoE) research grant at IIT Madras. The author is grateful to Dr. Samir Choudhuri for his encouragement and for supporting this research by providing the time and freedom to pursue it independently.

%%%%%%%%%%%%%%%%%%%%%%%%%%%%%%%%%%%%%%%%%%%%%%%%%%

\section*{Data Availability}

All source code and simulated datasets used in this study are publicly available at \url{https://github.com/aecosmo/Bayes-SCF}.

\bibliographystyle{mnras}
\bibliography{references}

\begin{thebibliography}{}
\makeatletter
\relax
\def\mn@urlcharsother{\let\do\@makeother \do\$\do\&\do\#\do\^\do\_\do\%\do\~}
\def\mn@doi{\begingroup\mn@urlcharsother \@ifnextchar [ {\mn@doi@} {\mn@doi@[]}}
\def\mn@doi@[#1]#2{\def\@tempa{#1}\ifx\@tempa\@empty \href {http://dx.doi.org/#2} {doi:#2}\else \href {http://dx.doi.org/#2} {#1}\fi \endgroup}
\def\mn@eprint#1#2{\mn@eprint@#1:#2::\@nil}
\def\mn@eprint@arXiv#1{\href {http://arxiv.org/abs/#1} {{\tt arXiv:#1}}}
\def\mn@eprint@dblp#1{\href {http://dblp.uni-trier.de/rec/bibtex/#1.xml} {dblp:#1}}
\def\mn@eprint@#1:#2:#3:#4\@nil{\def\@tempa {#1}\def\@tempb {#2}\def\@tempc {#3}\ifx \@tempc \@empty \let \@tempc \@tempb \let \@tempb \@tempa \fi \ifx \@tempb \@empty \def\@tempb {arXiv}\fi \@ifundefined {mn@eprint@\@tempb}{\@tempb:\@tempc}{\expandafter \expandafter \csname mn@eprint@\@tempb\endcsname \expandafter{\@tempc}}}

\bibitem[\protect\citeauthoryear{{Ali}, {Bharadwaj}  \& {Chengalur}}{{Ali} et~al.}{2008}]{Ali2008}
{Ali} S.~S.,  {Bharadwaj} S.,   {Chengalur} J.~N.,  2008, \mn@doi [\mnras] {10.1111/j.1365-2966.2008.12984.x}, \href {http://adsabs.harvard.edu/abs/2008MNRAS.385.2166A} {385, 2166}

\bibitem[\protect\citeauthoryear{{Ambikasaran}, {Foreman-Mackey}, {Greengard}, {Hogg}  \& {O'Neil}}{{Ambikasaran} et~al.}{2015}]{george}
{Ambikasaran} S.,  {Foreman-Mackey} D.,  {Greengard} L.,  {Hogg} D.~W.,   {O'Neil} M.,  2015, \mn@doi [IEEE Transactions on Pattern Analysis and Machine Intelligence] {10.1109/TPAMI.2015.2448083}, \href {https://ui.adsabs.harvard.edu/abs/2015ITPAM..38..252A} {38, 252}

\bibitem[\protect\citeauthoryear{{Amiri} et~al.,}{{Amiri} et~al.}{2024}]{chimelya}
{Amiri} M.,  et~al., 2024, \mn@doi [\apj] {10.3847/1538-4357/ad0f1d}, \href {https://ui.adsabs.harvard.edu/abs/2024ApJ...963...23A} {963, 23}

\bibitem[\protect\citeauthoryear{{Barry}, {Beardsley}, {Byrne}, {Hazelton}, {Morales}, {Pober}  \& {Sullivan}}{{Barry} et~al.}{2019}]{Barry2019eppsilon}
{Barry} N.,  {Beardsley} A.~P.,  {Byrne} R.,  {Hazelton} B.,  {Morales} M.~F.,  {Pober} J.~C.,   {Sullivan} I.,  2019, \mn@doi [\pasa] {10.1017/pasa.2019.21}, \href {https://ui.adsabs.harvard.edu/abs/2019PASA...36...26B} {36, e026}

\bibitem[\protect\citeauthoryear{{Bernardi} et~al.,}{{Bernardi} et~al.}{2009}]{Bernardi2009}
{Bernardi} G.,  et~al., 2009, \mn@doi [\aap] {10.1051/0004-6361/200911627}, \href {http://adsabs.harvard.edu/abs/2009A%26A...500..965B} {500, 965}

\bibitem[\protect\citeauthoryear{{Bharadwaj} \& {Ali}}{{Bharadwaj} \& {Ali}}{2004}]{Bharadwaj2004a}
{Bharadwaj} S.,  {Ali} S.~S.,  2004, \mn@doi [\mnras] {10.1111/j.1365-2966.2004.07907.x}, \href {http://adsabs.harvard.edu/abs/2004MNRAS.352..142B} {352, 142}

\bibitem[\protect\citeauthoryear{{Bharadwaj} \& {Ali}}{{Bharadwaj} \& {Ali}}{2005}]{Bharadwaj2005}
{Bharadwaj} S.,  {Ali} S.~S.,  2005, \mn@doi [\mnras] {10.1111/j.1365-2966.2004.08604.x}, \href {http://adsabs.harvard.edu/abs/2005MNRAS.356.1519B} {356, 1519}

\bibitem[\protect\citeauthoryear{{Bharadwaj} \& {Sethi}}{{Bharadwaj} \& {Sethi}}{2001}]{Bharadwaj2001b}
{Bharadwaj} S.,  {Sethi} S.~K.,  2001, \mn@doi [J. Astrophys. Astron.] {10.1007/BF02702273}, \href {http://adsabs.harvard.edu/abs/2001JApA...22..293B} {22, 293}

\bibitem[\protect\citeauthoryear{{Bharadwaj}, {Nath}  \& {Sethi}}{{Bharadwaj} et~al.}{2001}]{Bharadwaj2001a}
{Bharadwaj} S.,  {Nath} B.~B.,   {Sethi} S.~K.,  2001, \mn@doi [J. Astrophys. Astron.] {10.1007/BF02933588}, \href {http://adsabs.harvard.edu/abs/2001JApA...22...21B} {22, 21}

\bibitem[\protect\citeauthoryear{Bharadwaj, Pal, Choudhuri  \& Dutta}{Bharadwaj et~al.}{2018}]{Bharadwaj2018}
Bharadwaj S.,  Pal S.,  Choudhuri S.,   Dutta P.,  2018, \mn@doi [\mnras] {10.1093/mnras/sty3501}, 483, 5694

\bibitem[\protect\citeauthoryear{Bishop}{Bishop}{2006}]{bishop}
Bishop C.~M.,  [2006], Pattern recognition and machine learning.
New York : Springer, [2006] {\textcopyright}2006, \url {https://link.springer.com/book/9780387310732}

\bibitem[\protect\citeauthoryear{{Carucci} et~al.,}{{Carucci} et~al.}{2025}]{Carucci2025}
{Carucci} I.~P.,  et~al., 2025, \mn@doi [\aap] {10.1051/0004-6361/202453461}, \href {https://ui.adsabs.harvard.edu/abs/2025A&A...703A.222C} {703, A222}

\bibitem[\protect\citeauthoryear{{Chatterjee}, {Elahi}, {Bharadwaj}, {Sarkar}, {Choudhuri}, {Sethi}  \& {Patwa}}{{Chatterjee} et~al.}{2024}]{Chatterjee2024}
{Chatterjee} S.,  {Elahi} K. M.~A.,  {Bharadwaj} S.,  {Sarkar} S.,  {Choudhuri} S.,  {Sethi} S.~K.,   {Patwa} A.~K.,  2024, \mn@doi [\pasa] {10.1017/pasa.2024.45}, \href {https://ui.adsabs.harvard.edu/abs/2024PASA...41...77C} {41, e077}

\bibitem[\protect\citeauthoryear{{Chen} et~al.,}{{Chen} et~al.}{2025}]{Chen2025}
{Chen} K.-F.,  et~al., 2025, \mn@doi [\apj] {10.3847/1538-4357/ad9b91}, \href {https://ui.adsabs.harvard.edu/abs/2025ApJ...979..191C} {979, 191}

\bibitem[\protect\citeauthoryear{Choudhuri, Bharadwaj, Chatterjee, Ali, Roy  \& Ghosh}{Choudhuri et~al.}{2016}]{choudhuri2016b}
Choudhuri S.,  Bharadwaj S.,  Chatterjee S.,  Ali S.~S.,  Roy N.,   Ghosh A.,  2016, \mn@doi [\mnras] {10.1093/mnras/stw2254}, 463, 4093

\bibitem[\protect\citeauthoryear{{Datta}, {Choudhury}  \& {Bharadwaj}}{{Datta} et~al.}{2007}]{Datta2007}
{Datta} K.~K.,  {Choudhury} T.~R.,   {Bharadwaj} S.,  2007, \mn@doi [\mnras] {10.1111/j.1365-2966.2007.11747.x}, \href {http://adsabs.harvard.edu/abs/2007MNRAS.378..119D} {378, 119}

\bibitem[\protect\citeauthoryear{{Datta}, {Bowman}  \& {Carilli}}{{Datta} et~al.}{2010}]{datta2010}
{Datta} A.,  {Bowman} J.~D.,   {Carilli} C.~L.,  2010, \mn@doi [\apj] {10.1088/0004-637X/724/1/526}, \href {http://adsabs.harvard.edu/abs/2010ApJ...724..526D} {724, 526}

\bibitem[\protect\citeauthoryear{DeBoer et~al.,}{DeBoer et~al.}{2017}]{Deboer2017}
DeBoer D.~R.,  et~al., 2017, Publications of the Astronomical Society of the Pacific, 129, 045001

\bibitem[\protect\citeauthoryear{{Deng}, {Zuo}, {Li}, {Wang}  \& {Chen}}{{Deng} et~al.}{2026}]{Deng2026}
{Deng} Y.,  {Zuo} S.,  {Li} J.,  {Wang} Y.,   {Chen} X.,  2026, \mn@doi [Research in Astronomy and Astrophysics] {10.1088/1674-4527/ae5e3f}, \href {https://ui.adsabs.harvard.edu/abs/2026RAA....26i5005D} {26, 095005}

\bibitem[\protect\citeauthoryear{{Elahi} et~al.,}{{Elahi} et~al.}{2023a}]{Elahi2023}
{Elahi} K. M.~A.,  et~al., 2023a, \mn@doi [\mnras] {10.1093/mnras/stad191}, \href {https://ui.adsabs.harvard.edu/abs/2023MNRAS.520.2094E} {520, 2094}

\bibitem[\protect\citeauthoryear{{Elahi} et~al.,}{{Elahi} et~al.}{2023b}]{Elahi2023b}
{Elahi} K. M.~A.,  et~al., 2023b, \mn@doi [\mnras] {10.1093/mnras/stad2495}, \href {https://ui.adsabs.harvard.edu/abs/2023MNRAS.525.3439E} {525, 3439}

\bibitem[\protect\citeauthoryear{{Elahi} et~al.,}{{Elahi} et~al.}{2024}]{Elahi2024}
{Elahi} K. M.~A.,  et~al., 2024, \mn@doi [\mnras] {10.1093/mnras/stae740}, \href {https://ui.adsabs.harvard.edu/abs/2024MNRAS.529.3372E} {529, 3372}

\bibitem[\protect\citeauthoryear{{Elahi}, {Bharadwaj}, {Chatterjee}, {Sarkar}, {Choudhuri}, {Sethi}  \& {Patwa}}{{Elahi} et~al.}{2025}]{Elahi2025}
{Elahi} K. M.~A.,  {Bharadwaj} S.,  {Chatterjee} S.,  {Sarkar} S.,  {Choudhuri} S.,  {Sethi} S.,   {Patwa} A.~K.,  2025, \mn@doi [\mnras] {10.1093/mnras/staf896}, \href {https://ui.adsabs.harvard.edu/abs/2025MNRAS.540.2745E} {540, 2745}

\bibitem[\protect\citeauthoryear{{Ewall-Wice} et~al.,}{{Ewall-Wice} et~al.}{2021}]{ewallwice2021}
{Ewall-Wice} A.,  et~al., 2021, \mn@doi [\mnras] {10.1093/mnras/staa3293}, \href {https://ui.adsabs.harvard.edu/abs/2021MNRAS.500.5195E} {500, 5195}

\bibitem[\protect\citeauthoryear{{Gayen}, {Kumar}, {Dutta}, {Elahi}, {Choudhuri}  \& {Roy}}{{Gayen} et~al.}{2025}]{gayen2025}
{Gayen} S.,  {Kumar} J.,  {Dutta} P.,  {Elahi} K. M.~A.,  {Choudhuri} S.,   {Roy} N.,  2025, \mn@doi [\jcap] {10.1088/1475-7516/2025/07/024}, \href {https://ui.adsabs.harvard.edu/abs/2025JCAP...07..024G} {2025, 024}

\bibitem[\protect\citeauthoryear{{Ghosh}, {Prasad}, {Bharadwaj}, {Ali}  \& {Chengalur}}{{Ghosh} et~al.}{2012}]{Ghosh2012}
{Ghosh} A.,  {Prasad} J.,  {Bharadwaj} S.,  {Ali} S.~S.,   {Chengalur} J.~N.,  2012, \mn@doi [\mnras] {10.1111/j.1365-2966.2012.21889.x}, \href {http://adsabs.harvard.edu/abs/2012MNRAS.426.3295G} {426, 3295}

\bibitem[\protect\citeauthoryear{{Gill}, {Bharadwaj}, {Elahi}, {Sethi}  \& {Patwa}}{{Gill} et~al.}{2025}]{Gill_2025_mwa1}
{Gill} S.~S.,  {Bharadwaj} S.,  {Elahi} K. M.~A.,  {Sethi} S.~K.,   {Patwa} A.~K.,  2025, \mn@doi [\apj] {10.3847/1538-4357/ae0463}, \href {https://ui.adsabs.harvard.edu/abs/2025ApJ...993...56G} {993, 56}

\bibitem[\protect\citeauthoryear{{Gill}, {Elahi}, {Bharadwaj}, {Sethi}  \& {Patwa}}{{Gill} et~al.}{2026}]{Gill2026}
{Gill} S.~S.,  {Elahi} K. M.~A.,  {Bharadwaj} S.,  {Sethi} S.~K.,   {Patwa} A.~K.,  2026, \mn@doi [\apj] {10.3847/1538-4357/ae66f4}, \href {https://ui.adsabs.harvard.edu/abs/2026ApJ..1003..189G} {1003, 189}

\bibitem[\protect\citeauthoryear{{Jong}, {Trott}, {Nunhokee}  \& {Zheng}}{{Jong} et~al.}{2025}]{Jong2025}
{Jong} E.,  {Trott} C.,  {Nunhokee} C.~D.,   {Zheng} Q.,  2025, \mn@doi [\pasa] {10.1017/pasa.2025.10100}, \href {https://ui.adsabs.harvard.edu/abs/2025PASA...42..135J} {42, e135}

\bibitem[\protect\citeauthoryear{{Kennedy}, {Bull}, {Wilensky}, {Burba}  \& {Choudhuri}}{{Kennedy} et~al.}{2023}]{Kennedy2023}
{Kennedy} F.,  {Bull} P.,  {Wilensky} M.~J.,  {Burba} J.,   {Choudhuri} S.,  2023, \mn@doi [\apjs] {10.3847/1538-4365/acc324}, \href {https://ui.adsabs.harvard.edu/abs/2023ApJS..266...23K} {266, 23}

\bibitem[\protect\citeauthoryear{{Kern} \& {Liu}}{{Kern} \& {Liu}}{2021}]{Kern2021}
{Kern} N.~S.,  {Liu} A.,  2021, \mn@doi [\mnras] {10.1093/mnras/staa3736}, \href {https://ui.adsabs.harvard.edu/abs/2021MNRAS.501.1463K} {501, 1463}

\bibitem[\protect\citeauthoryear{Liu, Parsons  \& Trott}{Liu et~al.}{2014}]{liu14a}
Liu A.,  Parsons A.~R.,   Trott C.~M.,  2014, \mn@doi [Phys. Rev. D] {10.1103/PhysRevD.90.023018}, 90, 023018

\bibitem[\protect\citeauthoryear{{Madau}, {Meiksin}  \& {Rees}}{{Madau} et~al.}{1997}]{Madau1997}
{Madau} P.,  {Meiksin} A.,   {Rees} M.~J.,  1997, \apj, \href {http://adsabs.harvard.edu/abs/1997ApJ...475..429M} {475, 429}

\bibitem[\protect\citeauthoryear{{Mertens}, {Ghosh}  \& {Koopmans}}{{Mertens} et~al.}{2018}]{mertens18}
{Mertens} F.~G.,  {Ghosh} A.,   {Koopmans} L.~V.~E.,  2018, \mn@doi [\mnras] {10.1093/mnras/sty1207}, \href {https://ui.adsabs.harvard.edu/abs/2018MNRAS.478.3640M} {478, 3640}

\bibitem[\protect\citeauthoryear{Mertens et~al.,}{Mertens et~al.}{2020}]{Mertens2020}
Mertens F.~G.,  et~al., 2020, \mn@doi [MNRAS] {10.1093/mnras/staa327}, 493, 1662

\bibitem[\protect\citeauthoryear{{Mondal}, {Bharadwaj}  \& {Majumdar}}{{Mondal} et~al.}{2017}]{Mondal2017}
{Mondal} R.,  {Bharadwaj} S.,   {Majumdar} S.,  2017, \mn@doi [\mnras] {10.1093/mnras/stw2599}, \href {https://ui.adsabs.harvard.edu/abs/2017MNRAS.464.2992M} {464, 2992}

\bibitem[\protect\citeauthoryear{{Morales}}{{Morales}}{2005}]{morales2005}
{Morales} M.~F.,  2005, \mn@doi [\apj] {10.1086/426730}, \href {http://adsabs.harvard.edu/abs/2005ApJ...619..678M} {619, 678}

\bibitem[\protect\citeauthoryear{{Morales}, {Hazelton}, {Sullivan}  \& {Beardsley}}{{Morales} et~al.}{2012}]{Morales2012}
{Morales} M.~F.,  {Hazelton} B.,  {Sullivan} I.,   {Beardsley} A.,  2012, \mn@doi [\apj] {10.1088/0004-637X/752/2/137}, \href {http://adsabs.harvard.edu/abs/2012ApJ...752..137M} {752, 137}

\bibitem[\protect\citeauthoryear{{Neal}}{{Neal}}{1997}]{Neal1997}
{Neal} R.~M.,  1997, \mn@doi [arXiv e-prints] {10.48550/arXiv.physics/9701026}, \href {https://ui.adsabs.harvard.edu/abs/1997physics...1026N} {p. physics/9701026}

\bibitem[\protect\citeauthoryear{{Nunhokee} et~al.,}{{Nunhokee} et~al.}{2025}]{Nunhokee2025}
{Nunhokee} C.~D.,  et~al., 2025, \mn@doi [\apj] {10.3847/1538-4357/adda45}, \href {https://ui.adsabs.harvard.edu/abs/2025ApJ...989...57N} {989, 57}

\bibitem[\protect\citeauthoryear{{Pal}, {Bharadwaj}, {Ghosh}  \& {Choudhuri}}{{Pal} et~al.}{2021}]{Pal2020}
{Pal} S.,  {Bharadwaj} S.,  {Ghosh} A.,   {Choudhuri} S.,  2021, \mn@doi [\mnras] {10.1093/mnras/staa3831}, \href {https://ui.adsabs.harvard.edu/abs/2021MNRAS.501.3378P} {501, 3378}

\bibitem[\protect\citeauthoryear{Pal et~al.,}{Pal et~al.}{2022}]{Pal2022}
Pal S.,  et~al., 2022, \mn@doi [\mnras] {10.1093/mnras/stac2419}, 516, 2851

\bibitem[\protect\citeauthoryear{{Parsons} \& {Backer}}{{Parsons} \& {Backer}}{2009}]{Parsons2009}
{Parsons} A.~R.,  {Backer} D.~C.,  2009, \mn@doi [\aj] {10.1088/0004-6256/138/1/219}, \href {https://ui.adsabs.harvard.edu/abs/2009AJ....138..219P} {138, 219}

\bibitem[\protect\citeauthoryear{{Parsons}, {Pober}, {Aguirre}, {Carilli}, {Jacobs}  \& {Moore}}{{Parsons} et~al.}{2012}]{parsons12}
{Parsons} A.~R.,  {Pober} J.~C.,  {Aguirre} J.~E.,  {Carilli} C.~L.,  {Jacobs} D.~C.,   {Moore} D.~F.,  2012, \mn@doi [\apj] {10.1088/0004-637X/756/2/165}, \href {https://ui.adsabs.harvard.edu/abs/2012ApJ...756..165P} {756, 165}

\bibitem[\protect\citeauthoryear{Patil et~al.,}{Patil et~al.}{2017}]{Patil2017}
Patil A.~H.,  et~al., 2017, ApJ, 838, 65

\bibitem[\protect\citeauthoryear{{Planck Collaboration} et~al.,}{{Planck Collaboration} et~al.}{2020}]{Planck2020f}
{Planck Collaboration} et~al., 2020, \mn@doi [\aap] {10.1051/0004-6361/201833910}, \href {https://ui.adsabs.harvard.edu/abs/2020A&A...641A...6P} {641, A6}

\bibitem[\protect\citeauthoryear{{Prabu} et~al.,}{{Prabu} et~al.}{2015}]{Prabu2015}
{Prabu} T.,  et~al., 2015, \mn@doi [Experimental Astronomy] {10.1007/s10686-015-9444-3}, \href {https://ui.adsabs.harvard.edu/abs/2015ExA....39...73P} {39, 73}

\bibitem[\protect\citeauthoryear{Rasmussen \& Williams}{Rasmussen \& Williams}{2006}]{Rasmussen2006}
Rasmussen C.,  Williams C.,  2006, Gaussian Processes for Machine Learning.
Adaptive Computation and Machine Learning, MIT Press, Cambridge, MA, USA

\bibitem[\protect\citeauthoryear{{Sarkar}, {Elahi}, {Choudhuri}, {Bharadwaj}, {Chatterjee}, {Bhattacharyya}, {Sethi}  \& {Patwa}}{{Sarkar} et~al.}{2026}]{Sarkar2026}
{Sarkar} S.,  {Elahi} K. M.~A.,  {Choudhuri} S.,  {Bharadwaj} S.,  {Chatterjee} S.,  {Bhattacharyya} B.,  {Sethi} S.,   {Patwa} A.~K.,  2026, \mn@doi [\mnras] {10.1093/mnras/stag801}, \href {https://ui.adsabs.harvard.edu/abs/2026MNRAS.549ag801S} {549, stag801}

\bibitem[\protect\citeauthoryear{{Tingay} et~al.,}{{Tingay} et~al.}{2013}]{Tingay2013}
{Tingay} S.~J.,  et~al., 2013, \mn@doi [\pasa] {10.1017/pasa.2012.007}, \href {http://adsabs.harvard.edu/abs/2013PASA...30....7T} {30, e007}

\bibitem[\protect\citeauthoryear{{Trott}, {Wayth}  \& {Tingay}}{{Trott} et~al.}{2012}]{trott1}
{Trott} C.~M.,  {Wayth} R.~B.,   {Tingay} S.~J.,  2012, \mn@doi [\apj] {10.1088/0004-637X/757/1/101}, \href {https://ui.adsabs.harvard.edu/abs/2012ApJ...757..101T} {757, 101}

\bibitem[\protect\citeauthoryear{{Trott} et~al.,}{{Trott} et~al.}{2016}]{trott16}
{Trott} C.~M.,  et~al., 2016, \mn@doi [\apj] {10.3847/0004-637X/818/2/139}, \href {https://ui.adsabs.harvard.edu/abs/2016ApJ...818..139T} {818, 139}

\bibitem[\protect\citeauthoryear{Trott et~al.,}{Trott et~al.}{2020}]{Trott2020}
Trott C.~M.,  et~al., 2020, \mn@doi [\mnras] {10.1093/mnras/staa414}, 493, 4711

\bibitem[\protect\citeauthoryear{{Vedantham}, {Udaya Shankar}  \& {Subrahmanyan}}{{Vedantham} et~al.}{2012}]{vedantham12}
{Vedantham} H.,  {Udaya Shankar} N.,   {Subrahmanyan} R.,  2012, \mn@doi [\apj] {10.1088/0004-637X/745/2/176}, \href {https://ui.adsabs.harvard.edu/abs/2012ApJ...745..176V} {745, 176}

\bibitem[\protect\citeauthoryear{Williams}{Williams}{1998}]{williams1998}
Williams C. K.~I.,  1998, Prediction with Gaussian Processes: From Linear Regression to Linear Prediction and Beyond.
Springer Netherlands, Dordrecht, pp 599--621, \url {https://doi.org/10.1007/978-94-011-5014-9_23}

\bibitem[\protect\citeauthoryear{Williams \& Rasmussen}{Williams \& Rasmussen}{1996}]{WR1996}
Williams C.,  Rasmussen C.,  1996, Gaussian processes for regression.
MIT

\bibitem[\protect\citeauthoryear{{Zaldarriaga}, {Furlanetto}  \& {Hernquist}}{{Zaldarriaga} et~al.}{2004}]{Zaldarriaga2004}
{Zaldarriaga} M.,  {Furlanetto} S.~R.,   {Hernquist} L.,  2004, \mn@doi [\apj] {10.1086/386327}, \href {http://adsabs.harvard.edu/abs/2004ApJ...608..622Z} {608, 622}

\bibitem[\protect\citeauthoryear{{van Haarlem} et~al.,}{{van Haarlem} et~al.}{2013}]{vanharlem2013}
{van Haarlem} M.~P.,  et~al., 2013, \mn@doi [\aap] {10.1051/0004-6361/201220873}, \href {http://adsabs.harvard.edu/abs/2013A%26A...556A...2V} {556, A2}

\makeatother
\end{thebibliography}

\appendix

\section{Equivalence between delay-spectrum and correlation-based estimators}
\label{app:wiener_khinchin}

In this appendix, we explicitly demonstrate the equivalence between the delay-spectrum estimator and the correlation-based estimator for the 1D LoS power spectrum, following the Wiener-Khinchin theorem.

We consider a statistically homogeneous, zero-mean 21-cm brightness temperature fluctuation field $T(\nu)$ defined along the frequency axis. The Fourier transform of the field is defined as
\begin{equation}
    \tilde{T}(\tau)\equiv\int_{-\infty}^{\infty} \mathrm{d}\nu \,T(\nu)\, e^{-i \tau \nu},
    \label{eq:app_delay_transform}
\end{equation}
where $\tau$ is the Fourier conjugate of the observing frequency $\nu$.

The two-point correlation function along the frequency axis is given by
\begin{equation}
    C(\nu,\nu')\equiv\left\langle T(\nu)\, T^*(\nu') \right\rangle.
    \end{equation}
Assuming statistical homogeneity (ergodicity) along the LoS, the correlation depends only on the frequency separation $\Delta\nu = \nu-\nu'$, such that
\begin{equation}
    C(\nu,\nu') = C(\Delta\nu).
    \label{eq:app_stationary}
\end{equation}

The covariance of the delay-domain modes can be written as
\begin{align}
    \left\langle \tilde{T}(\tau)\, \tilde{T}^*(\tau') \right\rangle
    &= \int \mathrm{d}\nu \int \mathrm{d}\nu'\, e^{-i \tau \nu} e^{+i \tau' \nu'}\, \left\langle T(\nu)\, T^*(\nu') \right\rangle \nonumber \\
    &= \int \mathrm{d}\nu \int \mathrm{d}\nu'\, e^{-i \tau \nu} e^{+i \tau' \nu'}\, C(\nu-\nu').
\end{align}

Introducing the change of variables $\Delta\nu = \nu-\nu'$ and $\bar{\nu} = \nu'$, the above expression becomes
\begin{equation}
    \left\langle \tilde{T}(\tau)\, \tilde{T}^*(\tau') \right\rangle = \int \mathrm{d}(\Delta\nu)\,e^{-i \tau \Delta\nu} C(\Delta\nu)\int \mathrm{d}\bar{\nu}\, e^{-i (\tau-\tau') \bar{\nu}}.
\end{equation}
The integral over $\bar{\nu}$ yields a Dirac delta function,
\begin{equation}
    \int \mathrm{d}\bar{\nu}\,e^{-i (\tau-\tau') \bar{\nu}} = 2\pi\, \delta(\tau-\tau').
\end{equation}

The covariance of the delay-domain modes therefore reduces to
\begin{equation}
    \left\langle \tilde{T}(\tau)\, \tilde{T}^*(\tau') \right\rangle = 2\pi\, \delta(\tau-\tau') \int_{-\infty}^{\infty} \mathrm{d}(\Delta\nu)\,e^{-i \tau \Delta\nu}\,C(\Delta\nu).
    \label{eq:app_wk}
\end{equation}

Comparing equation~(\ref{eq:app_wk}) with the definition of the 1D power spectrum,
\begin{equation}
    \left\langle \tilde{T}(\tau)\, \tilde{T}^*(\tau') \right\rangle = 2\pi\, \delta(\tau-\tau')\, P_\tau(\tau),
\end{equation}
we identify
\begin{equation}
    P_\tau(\tau) = \int_{-\infty}^{\infty} \mathrm{d}(\Delta\nu)\,e^{-i \tau \Delta\nu}\,C(\Delta\nu).
    \label{eq:app_ps_eta}
\end{equation}

Using the mapping between frequency separation and comoving LoS distance, $\Delta r_\parallel = r^\prime\,\Delta\nu$ with $r^\prime=\mathrm{d}r/\mathrm{d}\nu$, the Fourier conjugate variables are related by $\tau = k_\parallel r^\prime$. Since $P_\tau(\tau)\,\mathrm{d}\tau = P(k_\parallel)\,\mathrm{d}k_\parallel$, the Jacobian $\mathrm{d}\tau/\mathrm{d}k_\parallel = r^\prime$ leads to
\begin{equation}
    P(k_\parallel) = r^\prime\int_{-\infty}^{\infty} \mathrm{d}(\Delta\nu)\,e^{-i k_\parallel r^\prime \Delta\nu}\,C(\Delta\nu),
\end{equation}

Since the brightness temperature field is real-valued, the correlation function is an even function of the frequency separation, $C(\Delta\nu)=C(-\Delta\nu)$. As a result, the imaginary (sine) component of the Fourier kernel vanishes, and the above expression reduces to a Fourier cosine transform,
\begin{equation}
    P(k_\parallel) = 2\, r^\prime \int_{0}^{\infty} \mathrm{d}(\Delta\nu)\, \cos\!\left(k_\parallel r^\prime \Delta\nu\right)\, C(\Delta\nu),
\end{equation}
which corresponds to equation~(\ref{eq:dct}) of the main text.

\section{Limitations of Linear Window-Based Filtering}
\label{app:appendix_limitations_of_hann}

In this appendix, we demonstrate why fixed linear filters, such as the Hann window, are fundamentally incapable of perfectly separating spectrally smooth foregrounds from the 21-cm signal, regardless of the chosen window width.

We assume that the observed signal $T(\nu)$ is processed by a convolution filter defined by a normalized window $W(\nu)$. The smooth component can be written as:
\begin{equation}
   T_{\rm S}(\nu) = T(\nu) * W(\nu) \, ,
\end{equation}
The filtered component is
\begin{equation}
    T_{\rm F}(\nu) = T(\nu) - T_{\rm S}(\nu). 
    \label{eq:filtered_component}
\end{equation}
In the Fourier (delay) domain equation~(\ref{eq:filtered_component}) becomes
\begin{equation}
    \tilde{T}_{\rm F}(\tau) = \tilde{T}(\tau) \left[ 1 - \tilde{W}(\tau) \right]\, ,
\end{equation}
where, $\tilde{T}(\tau)$ and $\tilde{W}(\tau)$ denote the Fourier transforms of $T(\nu)$ and $W(\nu)$, respectively. $\tilde{W}(\tau)$ is referred to as the \textit{Transfer Function} of the filter. 
For perfect filtering, we require $\tilde{W}(\tau) = 1$ at all delays where foregrounds possess significant power. Conversely, to perfectly preserve the 21-cm signal, we require $\tilde{W}(\tau) = 0$ at delays dominated by the cosmological signal.

% Now discuss the broad  and the narrow filters in \Cref{fig:transfer_combined}  
\begin{figure}
    \centering
    \includegraphics[width=0.99\columnwidth]{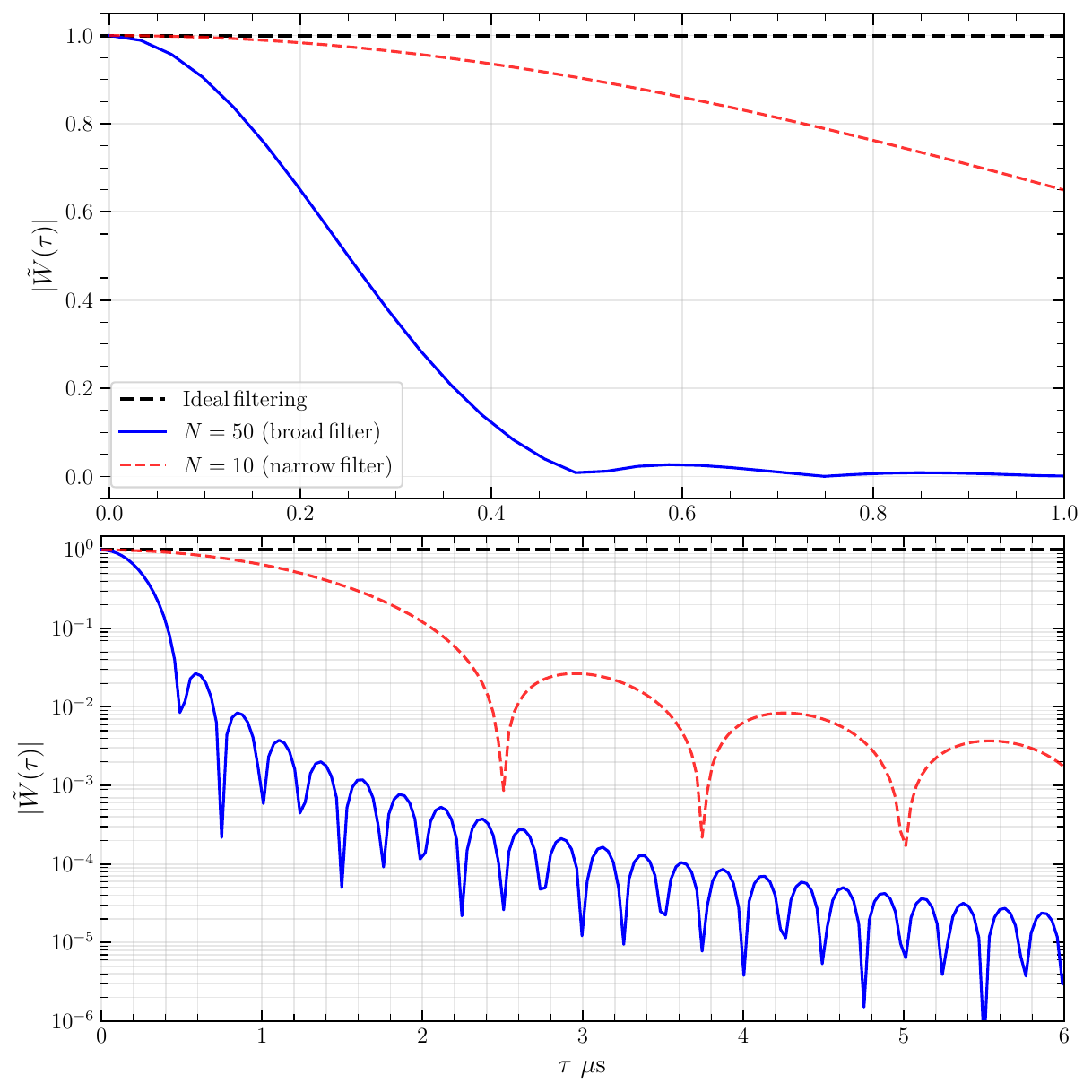}
    \caption{$|\tilde{W}(\tau)|$ spectral response (transfer function) of the Hann window filter for two different widths of the window: a broad filter ($N=50$, blue solid) and a narrow filter ($N=10$, red dashed). The top and bottom panels show the y-axis on a linear and log scale, respectively.}
    \label{fig:transfer_combined}
\end{figure}

\Cref{fig:transfer_combined} shows the transfer function of a Hann window for two different window widths: a broad filter ($N=50$, blue solid) and a narrow filter ($N=10$, red dashed). The broad filter has a sharp transfer function that drops to $\sim10^{-2}$ within a delay of $\sim0.5\,\mu \rm{s}$. The narrow window, on the other hand, has a slowly varying transfer function that extends to larger delays. 

In the ideal case where the smooth foreground component is strictly localized to the zero-delay mode ($\tau=0$), a perfect SCF  is trivial: any low-pass filter with a transfer function satisfying $\tilde{W}(0)=1$ will completely capture and subtract the smooth component without affecting the 21-cm signal at $\tau > 0$. Real foregrounds, however, are not delta functions at $\tau=0$. They exhibit a `spectral width' in delay space. The success of a linear filter depends entirely on whether its transfer function $\tilde{W}(\tau)$ remains effectively unity ($\tilde{W} \approx 1$) over the entire delay range occupied by the foregrounds.

The spectrally smooth foregrounds decay rapidly in the delay space. For this, the foreground power fall significantly within the main lobe of the transfer function of an $N=50$ Hann window, even though it is narrow in delay space. Therefore a broad filter is sufficient for smooth foregrounds. In contrast,  the spectrally unsmooth foregrounds decay gradually in the delay space, and extends to large delays (e.g., $\tau \approx 0.5\,\mu$s). The transfer function of the $N=50$ filter, as shown in \Cref{fig:transfer_combined} (blue dashed line), begins to roll off at very low delays, and therefore it can not effectively filter the foregrounds. 
Even if the window width $N$ is tuned to be `narrow' ($N=50$), the transfer function $\tilde{W}(\tau)$ is not a flat top; it curves downwards. At the delay $\tau = 0.5\,\mu$s, the response is strictly less than unity ($\tilde{W}(\tau_{\rm mix}) < 1$). 
In other words, if the foregrounds have significant power at $\tau = 0.5\,\mu$s, but the filter response has dropped to $\tilde{W}(0.5) \approx 0.99$, the filter captures only $99\%$ of the foreground power. The remaining $1\%$  leaks  into the residuals. 
Due to the extreme dynamic range of the foregrounds ($T_{\rm fg} \sim 10^4 - 10^5\,$K), even a fractional deviation of $1\%$ results in leakage of order $100\,$K, which is orders of magnitude larger than the 21-cm signal. Thus, it is extremely difficult to sufficiently remove the unsmooth foregrounds by tuning the width of the Hann-like filter.  
We also note that the transfer function of the broad filter decays quickly to zero, and therefore it preserves the 21-cm signal at the high-$\tau$ modes. The transfer function of the narrow filter on the other hand has strong sidelobes that can distort the 21-cm signal at high delays.

\section{Choice of smoothing scale in Bayes-SCF}
\label{app:smoothing_scale}

\begin{figure}
    \centering
    \includegraphics[width=0.99\columnwidth]{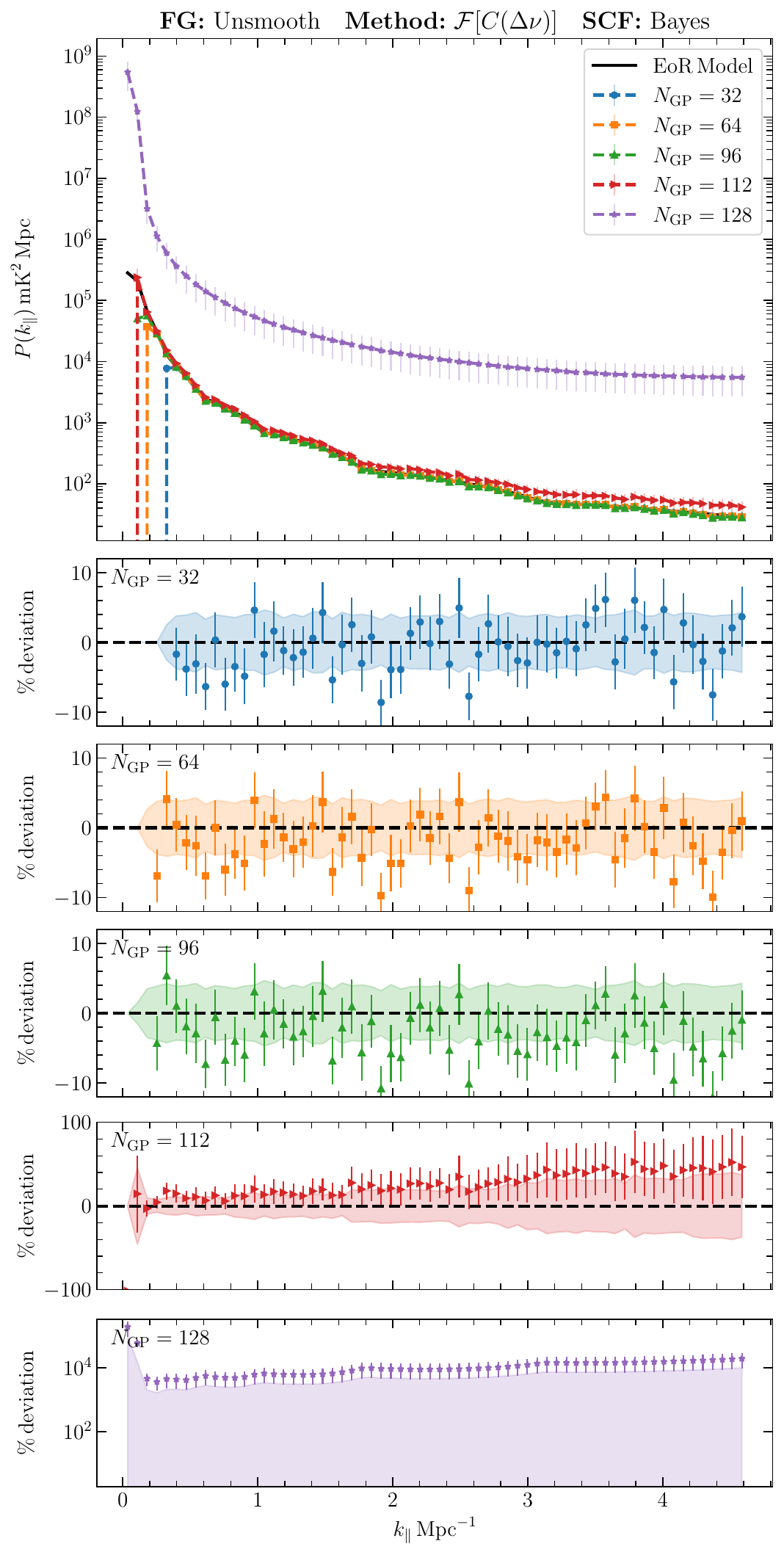}
    \caption{Recovered 1D LoS power spectrum obtained with the Bayes-SCF followed by the correlation-based estimator $\mathcal{F}[C(\Delta\nu)]$, for different values of the fixed smoothing scale $N_{\rm GP}$. The input data is a combination of unsmooth FG and the EoR 21-cm signal under PERIODIC flagging. The solid black curve represents the input theoretical EoR power spectrum. The lower panels show the percentage deviation relative to the model, with $1\sigma$ error bars, for each value of $N_{\rm GP}$. The recovered spectrum is consistent with the input model for $N_{\rm GP}\leq112$, but becomes severely biased for $N_{\rm GP}=128$, for which the smooth kernel can no longer adequately track the spectral structure of the foreground.}
    \label{fig:FGMM_GP_NN}
\end{figure}

In Bayes-SCF the degree of smoothing is controlled by $N_{\rm S}$ the fixed correlation length scale of the smooth kernel. In \Cref{sec:scf}, we fixed it to be $N_{\rm S} = N_{\rm GP}=96$ for all the results presented in the main text. The role of $N_{\rm GP}$ is to set the smoothing scale below which we filter information, regardless of whether that information originates from the foregrounds or the 21-cm signal. In this appendix, we examine the robustness of our results against this choice and quantify the trade-off involved in selecting a particular value of $N_{\rm GP}$.

We repeat the analysis presented in \Cref{sec:scf} for the unsmooth FG + EoR 21-cm signal dataset under the NOFLAG scenario, varying $N_{\rm GP}$ in the range 32 --128, which correspond to smoothing length scales $[N_{\rm GP}\,\Delta\nu_c]$ in the range $1.28-5.12\,$MHz. \Cref{fig:FGMM_GP_NN} shows the recovered power spectrum for each value of $N_{\rm GP}$. For $N_{\rm GP} \leq 112$, the Bayes-SCF successfully recovers the input EoR model within the $1\sigma$ uncertainties over the bulk of the $k_\parallel$ range, with only the low $k_\parallel$ modes excluded due to filtering. However, at $N_{\rm GP}=128$, the recovered power spectrum is systematically biased by several orders of magnitude higher values across the entire range. This is because when the correlation length the smooth kernel is set too large, it can no longer adequately model the more complex spectral structure present in our `unsmooth' foreground simulation. The GP regression effectively under-fits the foreground, which leaves substantial residual foregrounds that subsequently leaks into the power spectrum. This demonstrates that the framework is robust against the choice of $N_{\rm GP}$ as long as the correlation length remains comparable to or shorter than the correlation length of the spectral structure present in the data. 

\begin{figure}
    \centering
    \includegraphics[width=0.99\columnwidth]{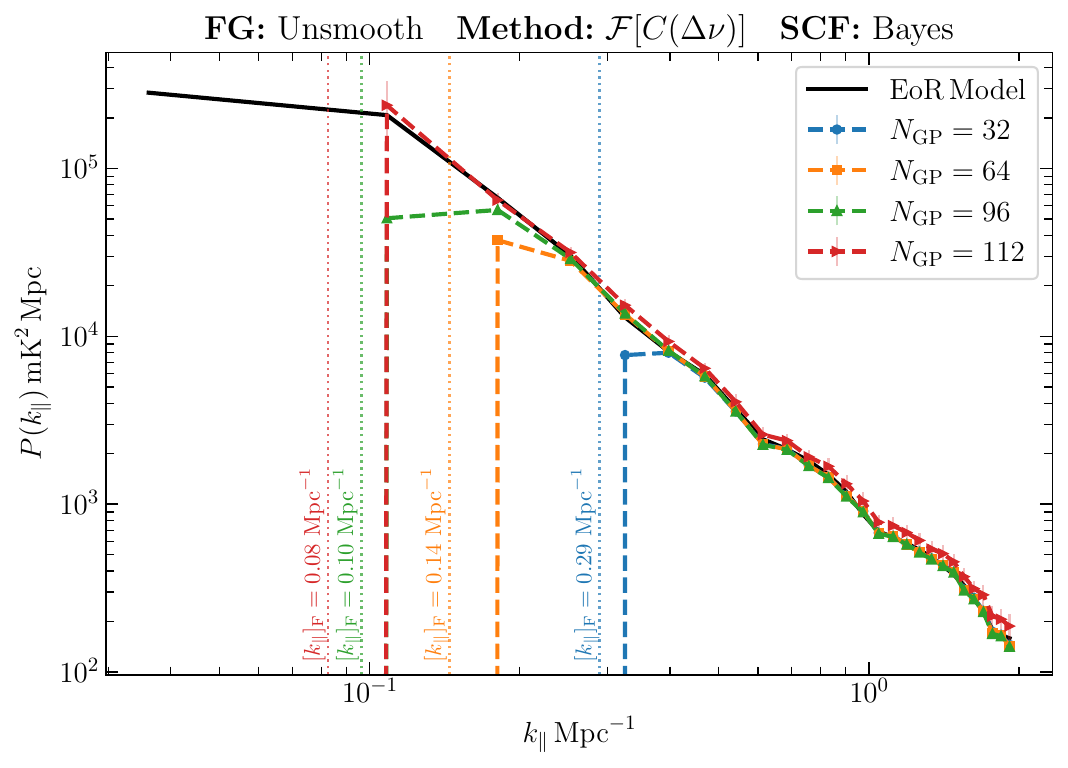}
    \caption{Same as \Cref{fig:FGMM_GP_NN} for $N_{\rm GP}=32,64,96$ and $112$, zoomed in to the low-$k_\parallel$ region. The vertical dotted lines mark the theoretical filtered scale $[k_\parallel]_{\rm F}$ (equation~\ref{eq:k_par_F}) associated with each $N_{\rm GP}$. A larger $N_{\rm GP}$ moves $[k_\parallel]_{\rm F}$ to smaller $k_\parallel$, preserving more large-scale modes at the cost of weaker foreground suppression.}
    \label{fig:FGMM_GP_NN_filter}
\end{figure}

The trade-off in choosing $N_{\rm GP}$ is illustrated more directly in \Cref{fig:FGMM_GP_NN_filter}, which shows the same recovered power spectra zoomed in to the low-$k_\parallel$ region. The vertical dotted lines mark the theoretical filtered scale
\begin{equation}
    [k_{\parallel}]_{\rm F} = \frac{2 \pi}{r^{'} \, N_{\rm GP} \, \Delta\nu_c} \, ,
    \label{eq:k_par_F}
\end{equation}
associated with each $N_{\rm GP}$. A larger $N_{\rm GP}$, which corresponds to a longer smoothing length, pushes $[k_\parallel]_{\rm F}$ to smaller $k_\parallel$, which means fewer large-scale modes are discarded by the filter. Conversely, a smaller $N_{\rm GP}$ filters out more low-$k_\parallel$ modes, at the benefit of more robustly removing complex foreground structure. Therefore, a larger $N_{\rm GP}$ preserves more large-scale cosmological information but risks under-fitting spectrally structured foregrounds, whereas a smaller $N_{\rm GP}$ reduces foreground leakage at the cost of discarding a larger range of low-$k_\parallel$ modes. In practice, $N_{\rm GP}$ should therefore be chosen based on the expected spectral complexity of the foregrounds for a given baseline length. We can set $[k_\parallel]_{\rm F} = [k_\parallel]_{\rm H}$, and filter the foregrounds along the theoretically predicted forgeound wedge boundary. Specifically, it is ideal to choose smaller $N_{\rm GP}$ for long baselines, where mode-mixing produces a more rapidly varying foreground, and larger $N_{\rm GP}$ for short baselines, where the foregrounds are expected to remain smooth over a wider frequency range (equation~\ref{eq:N_GP}).

\begin{figure*}
    \centering
    \includegraphics[width=0.95\textwidth]{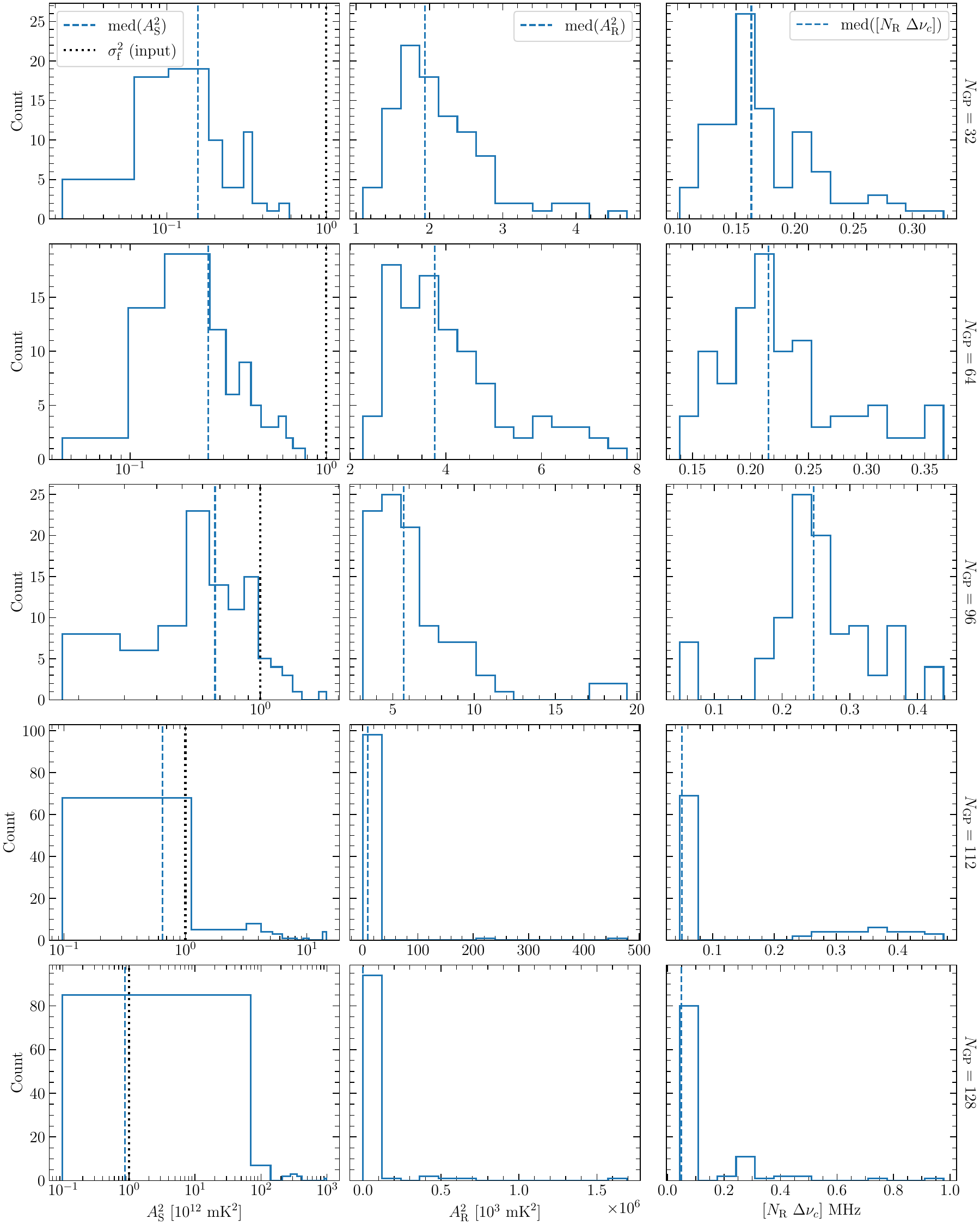}
    \caption{Distributions of the best-fit GP hyperparameters $A_{\rm S}^2$ (left column), $A_{\rm R}^2$ (middle column) and $[N_{\rm R}\,\Delta\nu_c]$ (right column), obtained by maximising the marginal log-likelihood independently for each of the 100 realizations of the unsmooth FG + 21-cm signal dataset, for $N_{\rm GP}=32,64,96,112$ and $128$ (top to bottom row). The blue dashed vertical lines mark the median of the recovered amplitudes, $\mathrm{med}(A_{\rm S}^2)$, $\mathrm{med}(A_{\rm R}^2)$ and $\mathrm{med}([N_{\rm R}\,\Delta\nu_c])$. In the left column, the black dotted vertical line marks the input foreground variance, $\sigma_{\rm f}^2=10^{12}\,{\rm mK}^2$ (\Cref{sec:gp_foregrounds}), used to generate the simulated unsmooth FG. For $N_{\rm GP}\geq112$, the hyperparameter fits become increasingly unstable, with long tails extending to large amplitudes.}
    \label{fig:optimised_hyperparameters}
\end{figure*}

We next examine the stability of the optimized hyperparameters $A_{\rm S}^2$, $A_{\rm R}^2$ and $[N_{\rm R}\,\Delta\nu_c]$ that we obtain for each of the 100 realizations of the dataset. \Cref{fig:optimised_hyperparameters} shows the resulting distributions of these hyperparameters across the realizations, for each $N_{\rm GP}$. The dashed vertical lines mark the median of the recovered amplitudes, $\mathrm{med}(A_{\rm S}^2)$, $\mathrm{med}(A_{\rm R}^2)$ and $\mathrm{med}([N_{\rm R}\,\Delta\nu_c])$. In the left column, the dotted vertical line marks the input foreground variance $\sigma_{\rm f}^2=10^{12}\,{\rm mK}^2$ used to generate the simulated unsmooth FG (\Cref{sec:gp_foregrounds}). For $N_{\rm GP}=32,64$ and $96$, the recovered $\mathrm{med}(A_{\rm S}^2)$ is broadly consistent with $\sigma_{\rm f}^2$, indicating that the smooth kernel correctly captures the dominant input foreground amplitude at these smoothing scales. For $N_{\rm GP}=112$ and $N_{\rm GP}=128$, the $A_{\rm S}^2$ and $A_{\rm R}^2$ distributions develop long tails extending to much larger values, suggesting that the fit is not good for several realizations of the input signal.
 
The recovered median rough correlation length $[N_{\rm R}\,\Delta\nu_c]$, also reported in \Cref{tab:gp_hyperparams_NGP}, remains well below $[N_{\rm GP}\,\Delta\nu_c]$ throughout, confirming that the prescribed bound $N_{\rm R} < N_{\rm GP}$ (\Cref{tab:gp_hyperparameters}) is followed. These diagnostics indicate that $N_{\rm GP}=96$, the value that we adopted in the main text, lies safely within the stable regime for our simulated unsmooth foregrounds, while also confirming that the qualitative conclusions of \Cref{sec:scf} are not sensitive to the precise choice of $N_{\rm GP}$ as long as it remains comparable to, or smaller than, the spectral correlation length of the foregrounds being filtered. 

\begin{table}
\centering
\renewcommand{\arraystretch}{1.4}
\setlength{\tabcolsep}{8pt}
\caption{A summary of the optimized hyperparameters of the Gaussian Process (GP) kernels for different choices of $N_{\rm GP}$. The reported values correspond to the median, with uncertainty ranges given by the 16th – 84th percentile intervals.}
\label{tab:gp_hyperparams_NGP}
\begin{tabular}{c c c c}
\hline
$N_{\rm GP}$ & $A_S^2\ [10^{11}\ \mathrm{mK}^2]$ & $A_R^2\ [10^{3}\ \mathrm{mK}^2]$ & $ [N_{\rm R}~\Delta\nu_c]\ (\mathrm{MHz})$ \\
\hline
32 & $1.57_{-0.62}^{+1.64}$ & $1.94_{-0.37}^{+0.74}$ & $0.16_{-0.03}^{+0.05}$ \\
64 & $2.51_{-0.99}^{+1.80}$ & $3.76_{-0.86}^{+1.58}$ & $0.22_{-0.04}^{+0.08}$ \\
96 & $6.70_{-2.36}^{+2.93}$ & $5.68_{-1.67}^{+2.81}$ & $0.25_{-0.04}^{+0.08}$ \\
112 & $6.48_{-2.57}^{+32.38}$ & $9.66_{-4.42}^{+6.94}$ & $0.05_{-0.00}^{+0.30}$ \\
128 & $8.70_{-4.33}^{+586.31}$ & $42.55_{-34.77}^{+6718.75}$ & $0.05_{-0.00}^{+0.21}$ \\
\hline
\end{tabular}
\end{table}

\section{Further test on Bayes-SCF with RANDOM flagging}
\label{app:random_flagging}
 
In the main text, we considered flagging fractions of up to $\simeq 35\%$ (\Cref{sec:flagging}). A natural question is at what point the framework breaks down as the fraction of missing channels increases further, and how this depends on the nature of the patterns in the missing channels. In this appendix, we explore two such limits: a substantially higher fraction of randomly flagged channels (Appendix~\ref{app:higher_flag_fraction}), and a single, wide contiguous gap that qualitatively resembles a broadband RFI event (Appendix~\ref{app:broadband_rfi}).
 
\subsection{Higher flag fraction}
\label{app:higher_flag_fraction}
 
We extend the RANDOM flagging scenario introduced in \Cref{sec:flagging} to substantially higher flagging fractions ($ff$) to up to $80\%$. As before, we use the unsmooth FG + 21-cm signal dataset, and apply the Bayes-SCF with $N_{\rm GP}=96$ followed by the correlation-based estimator.

\begin{figure}
    \centering
    \includegraphics[width=0.99\columnwidth]{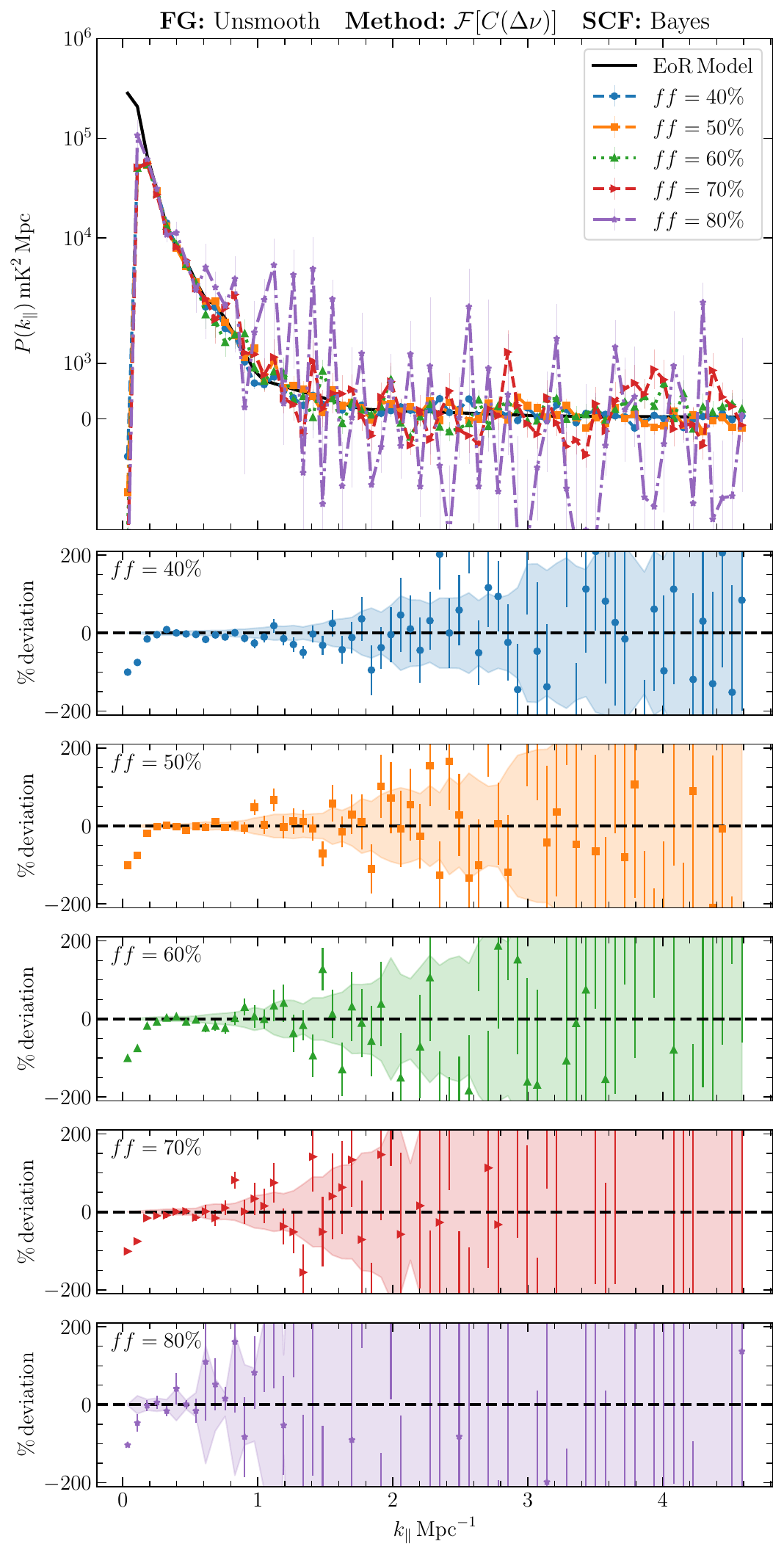}
    \caption{Recovered 1D LoS power spectrum obtained with the Bayes-SCF followed by the correlation-based estimator $\mathcal{F}[C(\Delta\nu)]$, for RANDOM flagging fractions $ff=40,50,60,70$ and $80\%$. The input data is a combination of unsmooth FG and the EoR 21-cm signal. The solid black curve represents the input theoretical EoR power spectrum. The lower panels show the percentage deviation relative to the model, with $1\sigma$ error bars, for each value of $ff$. The recovered spectrum remains consistent with the input model for all flagging fractions, with the scatter and uncertainty growing systematically with $ff$.}
    \label{fig:FGMM_GP_RANDOM_flag}
\end{figure}
 
\begin{figure}
    \centering
    \includegraphics[width=0.99\columnwidth]{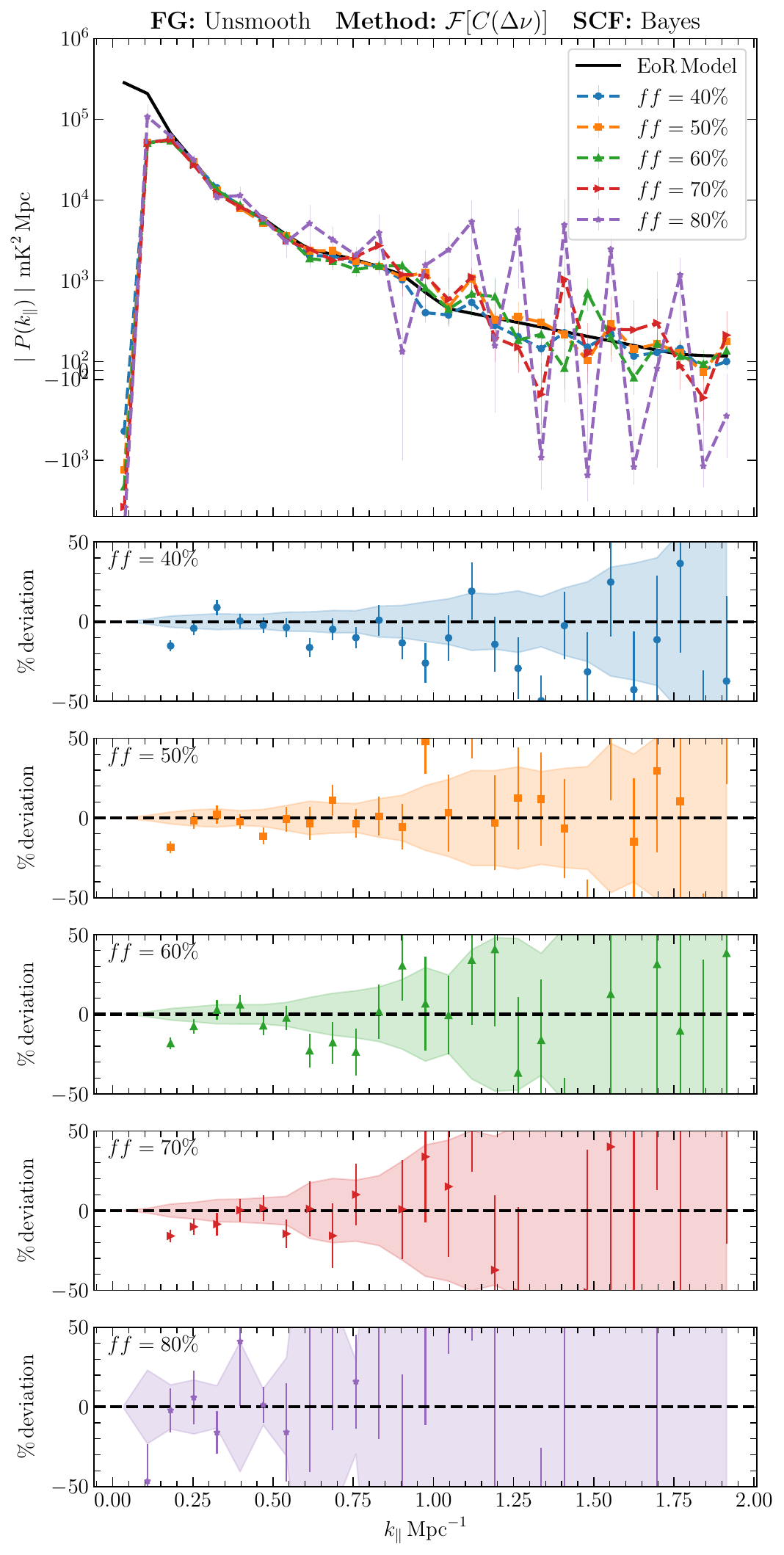}
    \caption{Same as \Cref{fig:FGMM_GP_RANDOM_flag}, zoomed in to $k_\parallel \leq 2\,{\rm Mpc^{-1}}$. The percentage deviations remain confined to a narrower range and stay consistent with the input model within the $1\sigma$ uncertainties for all flagging fractions considered.}
    \label{fig:FGMM_Bayes_RANDOM_flag_zoom}
\end{figure}

\Cref{fig:FGMM_GP_RANDOM_flag} shows the recovered power spectrum across the full $k_\parallel$ range for each flagging fraction. The recovered spectra remain consistent with the input EoR model for all values of $ff$ considered, including the most aggressive case of $ff=80\%$. As expected, the scatter and the width of the $1\sigma$ uncertainty band grow systematically with $ff$, since the sample variance in the estimated $C(\Delta\nu)$ increases as more channels are removed. \Cref{fig:FGMM_Bayes_RANDOM_flag_zoom} shows the same comparison zoomed in to $k_\parallel \leq 2\,{\rm Mpc^{-1}}$, where the percentage deviations remain confined to a narrower range ($\lesssim 50\%$) and stay consistent with the input model within the $1\sigma$ uncertainties for all flagging fractions.

\begin{figure}
    \centering
    \includegraphics[width=0.99\columnwidth]{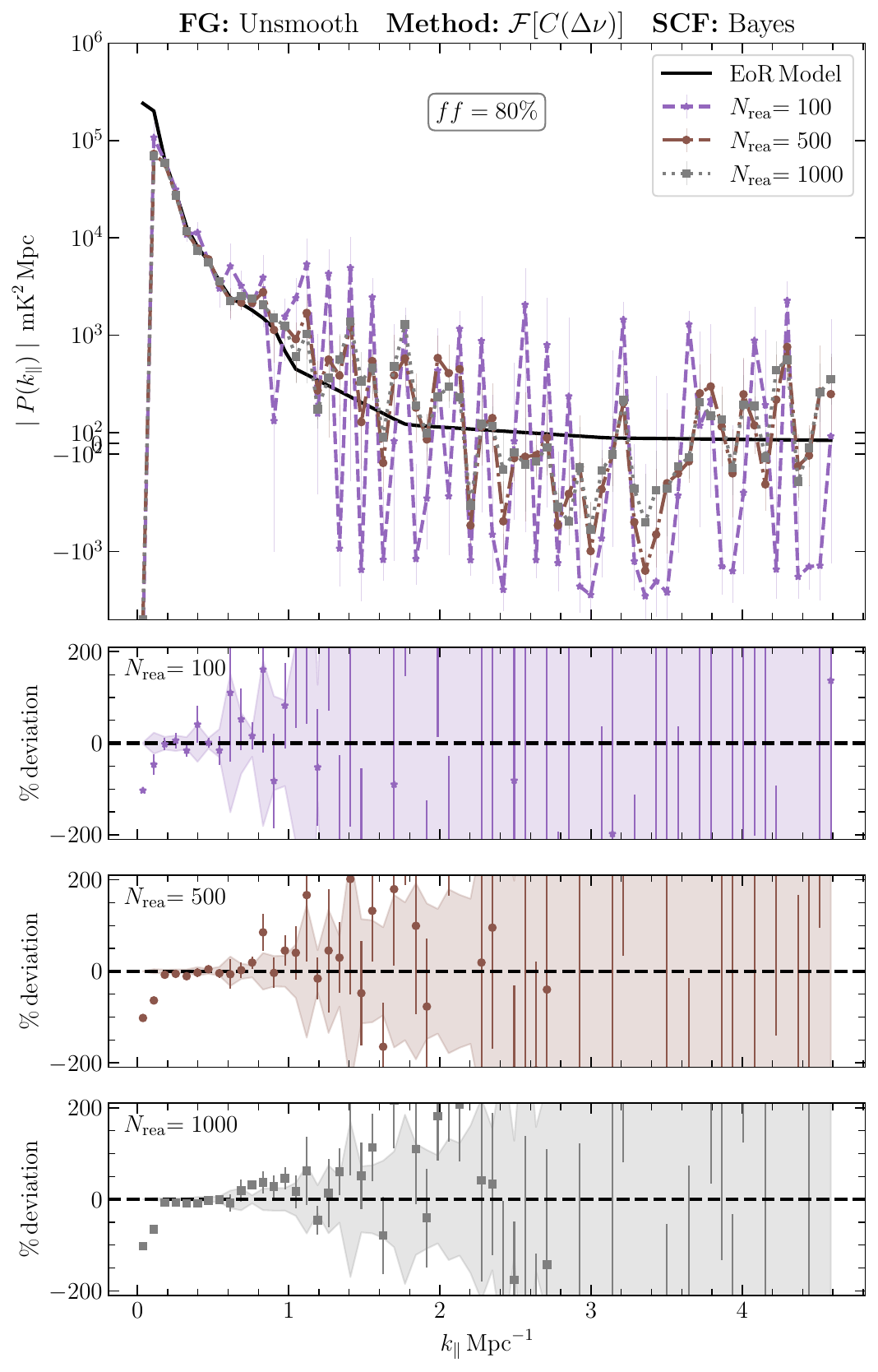}
    \caption{Recovered 1D LoS power spectrum at a fixed flagging fraction $ff=80\%$, obtained with the Bayes-SCF followed by the correlation-based estimator, for $N_{\rm rea}=100,500$ and $1000$ independent realizations of the unsmooth FG + 21-cm signal dataset. The solid black curve represents the input theoretical EoR power spectrum. The lower panels show the percentage deviation relative to the model, with $1\sigma$ error bars, for each $N_{\rm rea}$. The recovered power spectrum and its scatter converge as $N_{\rm rea}$ increases.}
    \label{fig:FGMM_GP_RANDOM_flag80}
\end{figure}
 
\begin{figure}
    \centering
    \includegraphics[width=0.99\columnwidth]{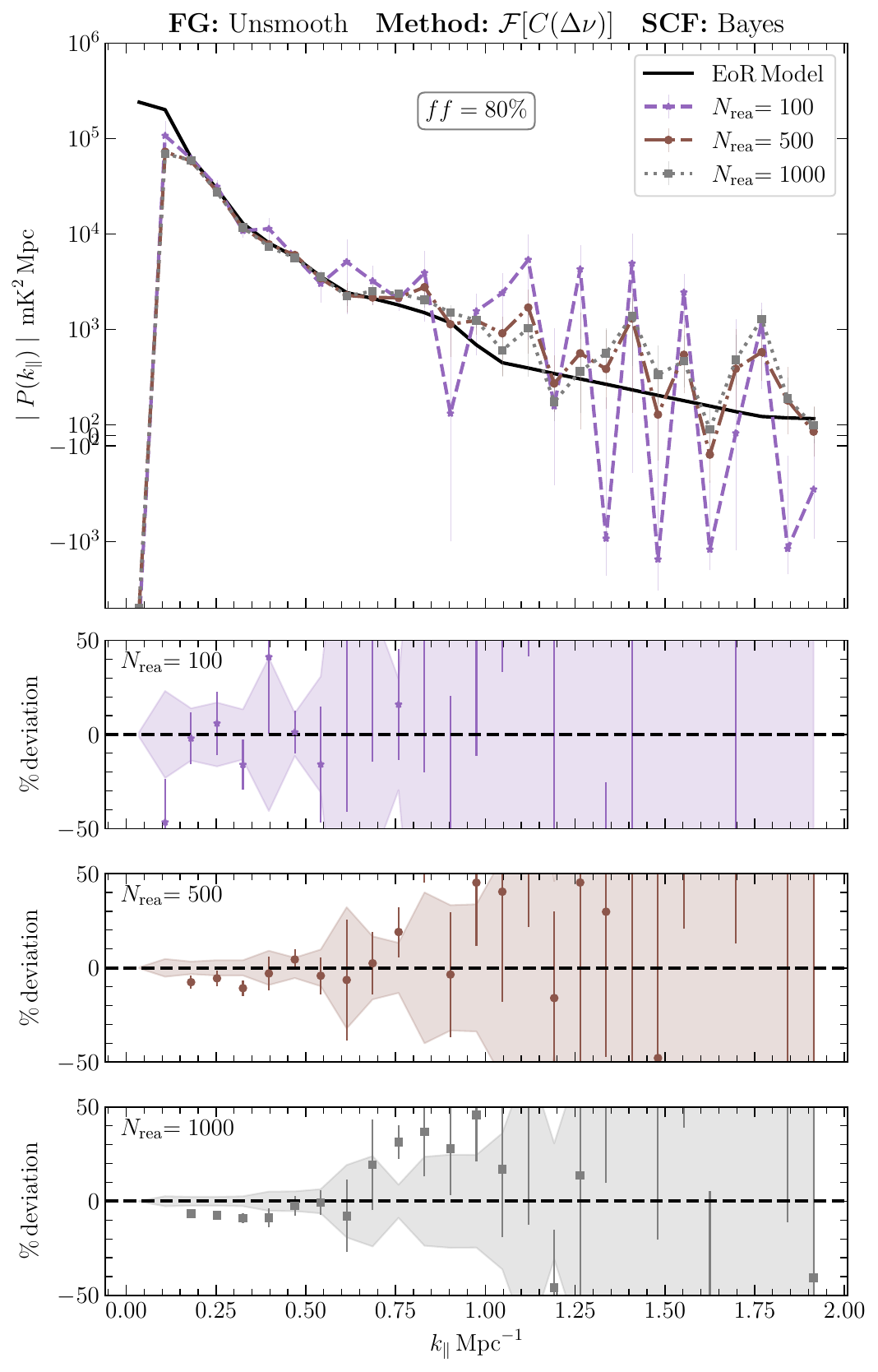}
    \caption{Same as \Cref{fig:FGMM_GP_RANDOM_flag80}, zoomed in to $k_\parallel \leq 2\,{\rm Mpc^{-1}}$.}
    \label{fig:FGMM_Bayes_RANDOM_flag80_zoom}
\end{figure}

To assess the robustness of this conclusion against the limited number of available realizations, we repeat the $ff=80\%$ case with 100, 500, and 1000 independent realizations of the unsmooth FG + 21-cm signal dataset. \Cref{fig:FGMM_GP_RANDOM_flag80} and its zoomed-in counterpart, \Cref{fig:FGMM_Bayes_RANDOM_flag80_zoom}, show that the recovered power spectrum and its associated scatter converge as the number of realization $(N_{\rm rea})$ increases, with the $N_{\rm rea}=500$ and $N_{\rm rea}=1000$ results closely tracking one another. This confirms that the apparent scatter at $ff=80\%$ in \Cref{fig:FGMM_GP_RANDOM_flag} is consistent with sampling variance rather than a systematic failure of the method, and that $N_{\rm rea}=100$ realizations, as used throughout the main text, is sufficient to characterise the typical behaviour of the estimator, albeit with a noisier estimate of the uncertainty itself.

\begin{figure}
    \centering
    \includegraphics[width=0.99\columnwidth]{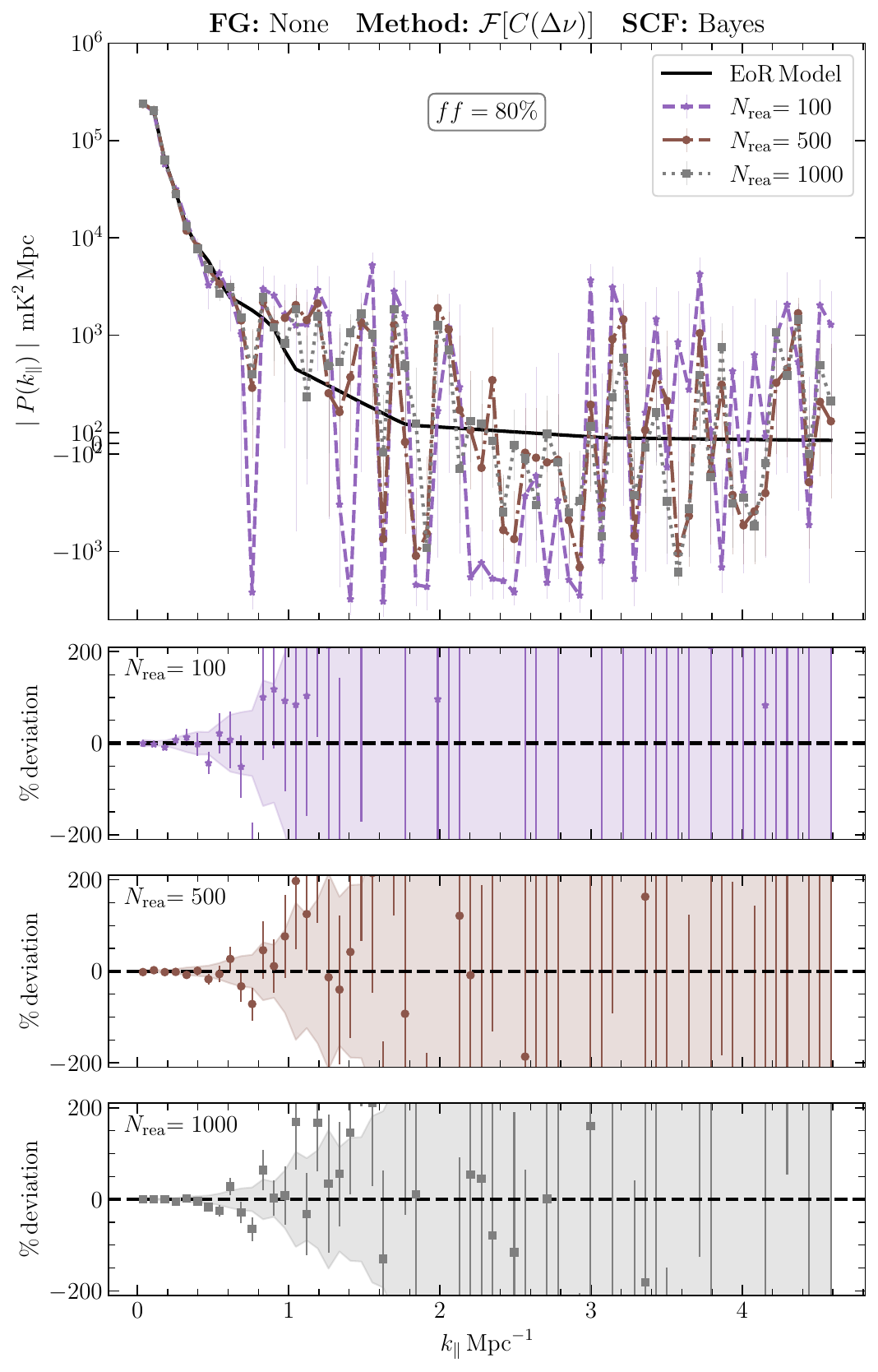}
    \caption{Same as \Cref{fig:FGMM_GP_RANDOM_flag80}, but for an EoR-signal-only dataset (no foreground component), isolating the effect of the $ff=80\%$ RANDOM flagging pattern alone. The recovered power spectrum is consistent with the input EoR model for all $N_{\rm rea}$, with scatter comparable to that seen in \Cref{fig:FGMM_GP_RANDOM_flag80}.}
    \label{fig:21cm_GP_RANDOM_flag80}
\end{figure}
 
\begin{figure}
    \centering
    \includegraphics[width=0.99\columnwidth]{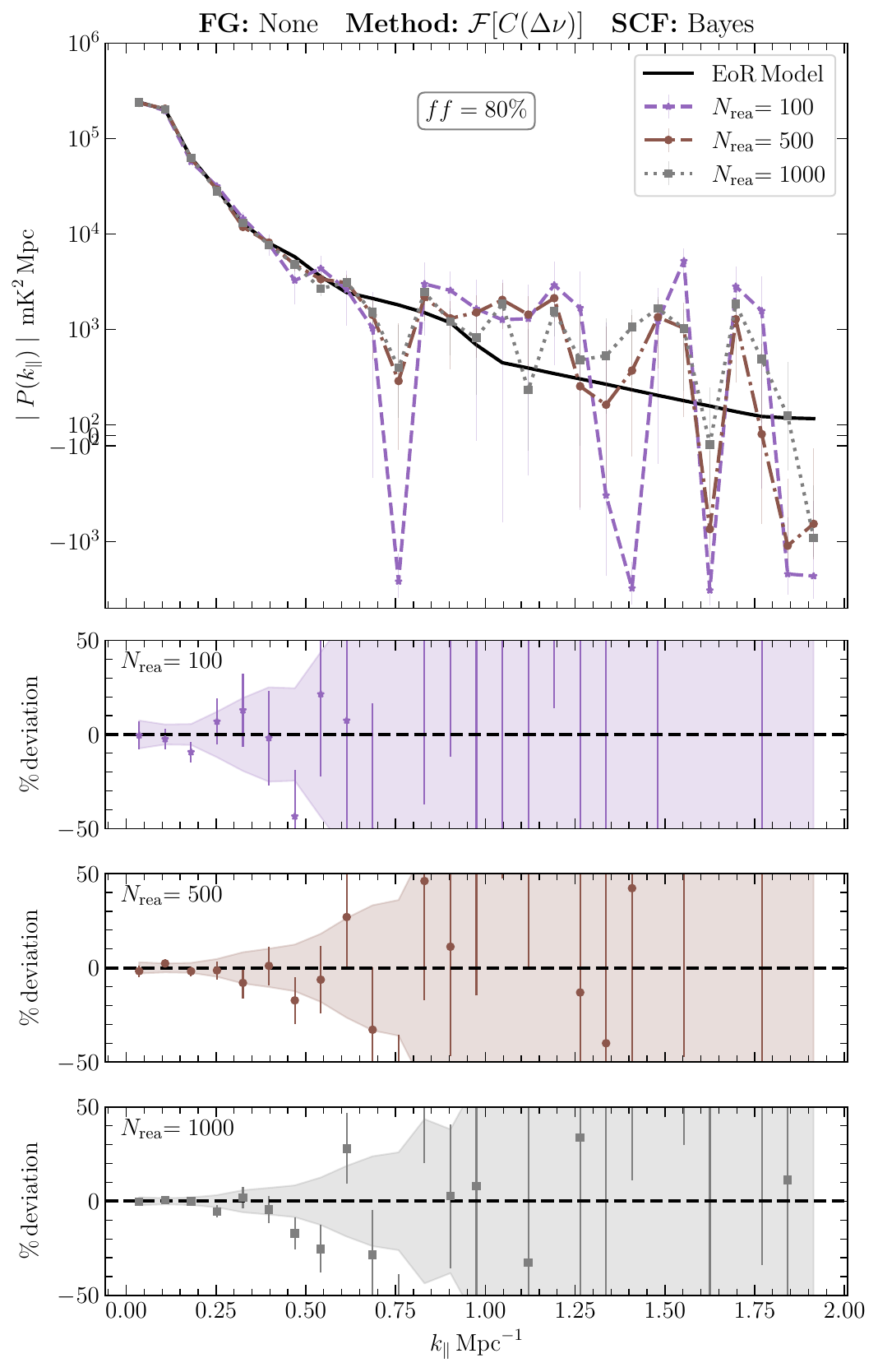}
    \caption{Same as \Cref{fig:21cm_GP_RANDOM_flag80}, zoomed in to $k_\parallel \leq 2\,{\rm Mpc^{-1}}$.}
    \label{fig:21cm_Bayes_RANDOM_flag80_zoom}
\end{figure} 

We repeat the same $N_{\rm rea}$ convergence test for an EoR-signal-only dataset (i.e., without any foreground component) at $ff=80\%$, isolating the effect of missing channels alone. \Cref{fig:21cm_GP_RANDOM_flag80} and \Cref{fig:21cm_Bayes_RANDOM_flag80_zoom} show that the recovered power spectrum is consistent with the input EoR model across the full $k_\parallel$ range for all $N_{\rm rea}$, with the scatter again decreasing as $N_{\rm rea}$ increases. The close similarity between the EoR-only results presented here and the unsmooth FG + 21-cm signal results in \Cref{fig:FGMM_GP_RANDOM_flag80} confirms that, once the foreground component has been correctly filtered by the Bayes-SCF, the residual scatter at high flagging fractions is dominated by the reduced number of available frequency pairs rather than by any residual foreground contamination. We therefore conclude that the correlation-based estimator, combined with the Bayes-SCF, remains robust to RANDOM flagging fractions of up to at least $80\%$, broadly consistent with the conclusions of \citet{Bharadwaj2018} who tested the correlation-based approach without foregrounds. For practical purposes, given the rapid growth of statistical uncertainties illustrated in \Cref{fig:FGMM_GP_RANDOM_flag}, we recommend discarding datasets in which the flagging fraction exceeds $\simeq 80\%$.

\subsection{Mimicking broadband RFIs}
\label{app:broadband_rfi}

\begin{figure}
    \centering
    \includegraphics[width=0.99\columnwidth]{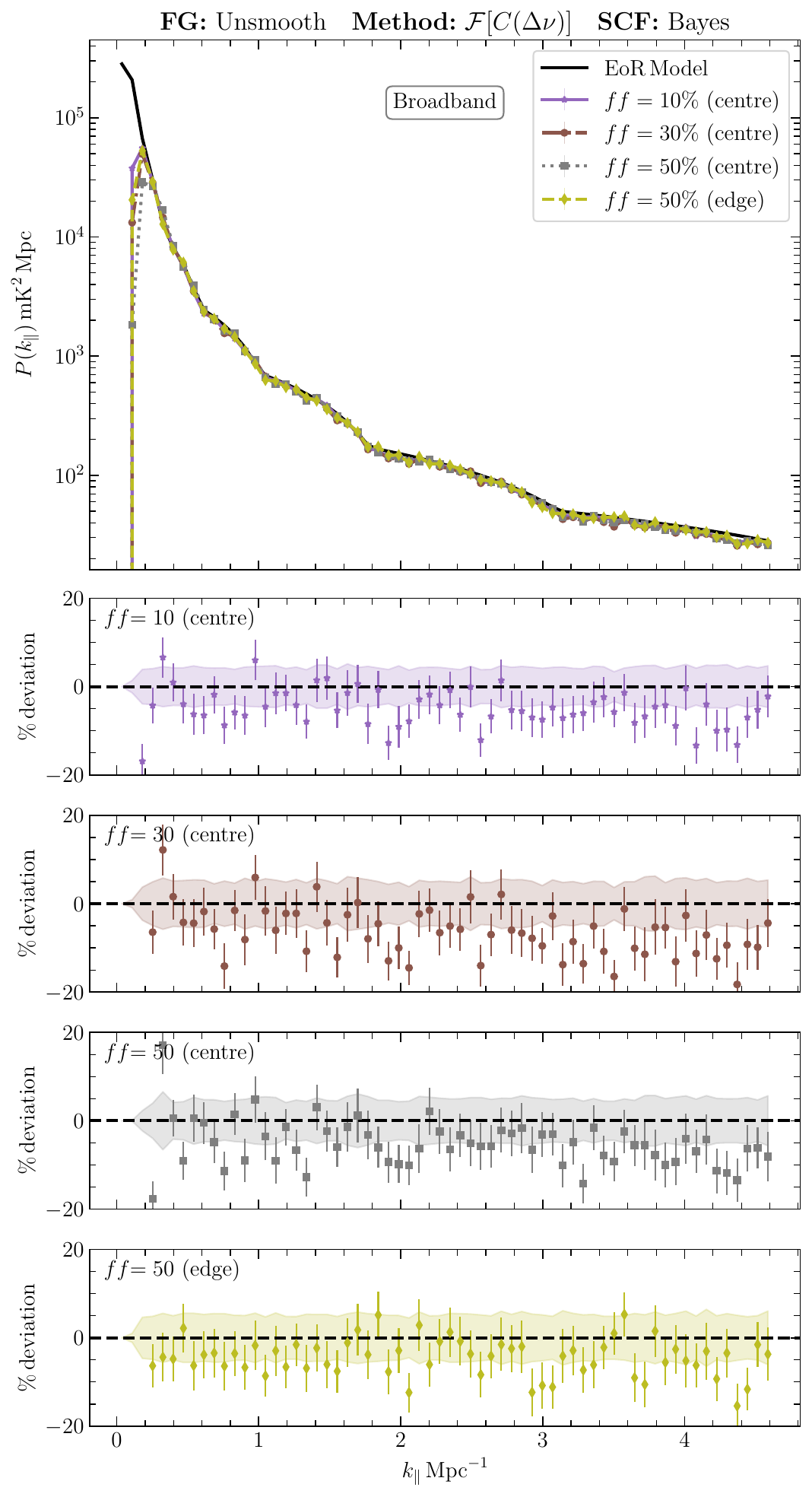}
    \caption{Recovered 1D LoS power spectrum obtained with the Bayes-SCF followed by the correlation-based estimator $\mathcal{F}[C(\Delta\nu)]$, for a single contiguous gap covering $10,30$ and $50\%$ of the bandwidth, centred on the band, and for a further $50\%$ gap placed at one edge of the band. The input data is a combination of unsmooth FG and the EoR 21-cm signal. The solid black curve represents the input theoretical EoR power spectrum. The lower panels show the percentage deviation relative to the model, with $1\sigma$ error bars, for each case. The recovered spectrum remains consistent with the input model across the full $k_\parallel$ range for all four cases considered.}
    \label{fig:FGMM_GP_broadband}
\end{figure}

Unlike the RANDOM pattern considered in Appendix~\ref{app:higher_flag_fraction}, broadband RFI typically removes a single, wide, contiguous block of frequency channels rather than a scattered set of individual channels. To test the impact of such a contiguous gap, we flag a single block of channels of width $ff=10,30$ and $50\%$ of the bandwidth, centred on the band, and repeat our analysis for the unsmooth FG + 21-cm signal dataset using the Bayes-SCF ($N_{\rm GP}=96$) followed by the correlation-based estimator. We additionally test a $ff=50\%$ gap placed at one edge of the band, rather than at the centre.
 
A contiguous gap affects the correlation-based estimator differently from a randomly distributed one. RANDOM flagging removes individual channels scattered across the band, so the uneven sampling of frequency pairs is spread  over all separations $\Delta\nu$, which is the origin of the increased uncertainty at all $k_\parallel$ seen in Appendix~\ref{app:higher_flag_fraction}. On the other hand, for a contiguous gap centred on the band, the two unflagged segments remain on either side of the band, whereas a gap placed at one edge leaves only a single contiguous unflagged segment. We therefore expect the position of the gap, in addition to its width, to influence which $\Delta\nu$ separations are most affected.
 
\Cref{fig:FGMM_GP_broadband} shows that our framework continues to recover the input EoR power spectrum within the $1\sigma$ uncertainties across the full $k_\parallel$ range, for all three centred gap widths considered, including $ff=50\%$. The reason is that a broadband flagging is a large scale feature, and it affects only the small $k_\parallel$ modes which are already filtered out by SCF. It should also be noted that even though the contiguous gap depletes the pair count severely at separations comparable to its own width, sufficient pairs remain at neighbouring separations, and the correlation-based estimator therefore still return an unbiased estimate at the affected scales.
 
\Cref{fig:FGMM_GP_broadband} shows that the recovered power spectrum for the edge-positioned gap (bottom panel) is comparable to the centred $ff=50\%$ case across the full $k_\parallel$ range, with no excess deviation or uncertainty at high $k_\parallel$. For the edge gap, the single surviving band spans only half of the original bandwidth $B$, so frequency pairs are only available up to $\Delta\nu \lesssim B/2$ rather than the full bandwidth. We simply discard those $\Delta\nu$ bins and restrict the Fourier cosine transform (equation~\ref{eq:dct}) to the available range. The effective bandwidth used to recover $P(k_\parallel)$ is therefore halved for the edge-positioned gap. As a consequence the resolution along $k_\parallel$ also becomes 2 times coarser.   

% \newpage
\section{SCF for delay spectrum}
\label{app:scfindelay}

We have discussed how missing channels impact the estimated power spectrum in a delay spectrum approach (Section~\ref{sec:pk_from_corr}). Here, we present a preliminary assessment of how SCF can help recover the EoR power spectrum in the presence of missing channels even in the delay spectrum approach.  For this we choose one realization of the EoR 21-cm signal, and add one realization of smooth and unsmooth foregrounds each, and apply both Hann- and Bayes-SCF on the data. Only the PERIODIC flagging pattern is used here for demonstration. 

\begin{figure*}
    \centering
    \includegraphics[width=0.99\textwidth]{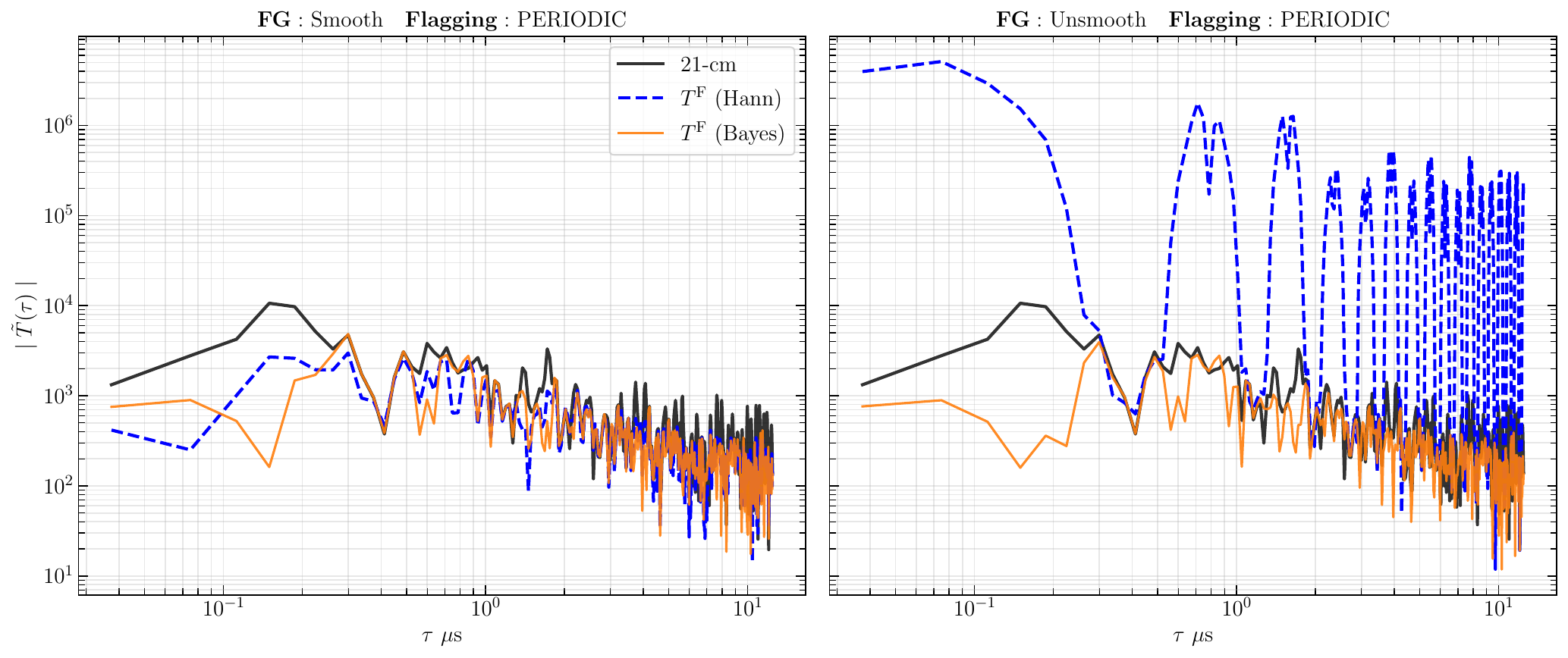}
    \caption{Comparison of the Hann- and Bayes-SCF in delay space for the PERIODIC flagging scenario. The input data has foregrounds and the EoR 21-cm signal. The left panel depicts the case of spectrally smooth foreground emission, for which both smoothing strategies yield satisfactory performance. The right panel illustrates the failure of window-based filtering in the presence of unsmooth foregrounds. In this regime, the Hann-SCF (blue dashed) fails to adequately remove the foreground contribution, and the residual emission produces pronounced periodic ringing in delay space that fully obscures the 21-cm signal. In contrast, the Bayes-SCF (orange solid) effectively models and subtracts the smooth component, resulting in a clean delay spectrum that closely reproduces the input 21-cm signal. This figure demonstrates that SCF can successfully mitigate the artefacts from the PERIODIC missing channels in MWA, even within the delay-spectrum analysis framework.}
    \label{fig:Hann_GP_comparison_PERIODIC_delay}
\end{figure*}

The left panel of Figure~\ref{fig:Hann_GP_comparison_PERIODIC_delay}  corresponds to the case of EoR signal with smooth foreground. We see that SCF filters out the region $\tau < 0.4\,\mu {\rm s}$ beyond which both implementations of SCF effectively recovers the EoR 21-cm signal. We do not observe periodic leakage of the foreground into the power spectrum. Therefore, SCF enables the delay-spectrum analysis to remain viable even in the presence of PERIODIC flagging. However, as we noted earlier, SCF with a Hann window is not enough to remove the unsmooth foregrounds. The right panel illustrates the breakdown of Hann-SCF when the foreground is spectrally unsmooth.
We see that the Hann filter (blue dashed curve) fails to subtract the foregrounds adequately, and the residuals produce periodic ringing structures in delay space that completely obscure the 21-cm cosmological signal. The Bayes-SCF (orange solid curve) substantially suppresses the smooth component and yields a delay spectrum that is largely free of foreground contamination and closely matches the input EoR signal. 

Our analysis demonstrates that SCF can effectively mitigate the artefacts from the PERIODIC missing channels even when employing the delay-spectrum approach. 

% \balance
 
\bsp
\label{lastpage}
\end{document}